\documentclass[12pt, twocolumn]{aastex63}

\usepackage{amsmath}	% Advanced maths commands
\usepackage{amssymb}	% Extra maths symbols
\usepackage{graphicx}

\usepackage{color,comment}
\usepackage[normalem]{ulem}
\usepackage[caption=false]{subfig}
\usepackage{graphicx}
\usepackage{verbatim}
\usepackage{mathtools}
\newcommand{\red}[1]{{\textcolor{red}{#1}}}

\newcommand{\green}[1]{{\textcolor{green}{#1}}}

\begin{document}

\title{{\sc mirkwood:} Fast and Accurate SED Modeling Using Machine Learning}

\author[0000-0002-3645-4501]{Sankalp Gilda}
\affil{Department of Astronomy, University of Florida, 211 Bryant Space Science Center, Gainesville, FL, 32611, USA}
\author[0000-0003-4422-8595]{Sidney Lower}
\affil{Department of Astronomy, University of Florida, 211 Bryant Space Science Center, Gainesville, FL, 32611, USA}

\author[0000-0002-7064-4309]{Desika Narayanan}
\affil{Department of Astronomy, University of Florida, 211 Bryant Space Science Center, Gainesville, FL, 32611, USA}
\affil{University of Florida Informatics Institute, 432 Newell Drive, CISE Bldg E251 Gainesville, FL, 32611, USA}
\affil{Cosmic Dawn Centre at the Niels Bohr Institute, University of Copenhagen and DTU-Space, Technical University of Denmark}

\begin{abstract}
Traditional spectral energy distribution (SED) fitting codes used to derive galaxy physical properties are often uncertain at the factor of a few level owing to uncertainties in galaxy star formation histories and dust attenuation curves.  Beyond this, Bayesian fitting (which is typically used in SED fitting software) is an intrinsically compute-intensive task, often requiring access to expensive hardware for long periods of time. To overcome these shortcomings, we have developed {\sc mirkwood}: a user-friendly tool comprising of an ensemble of supervised machine learning-based models capable of non-linearly mapping galaxy fluxes to their properties. By stacking multiple models, we marginalize against any individual model's poor performance in a given region of the parameter space. We demonstrate \textsc{mirkwood}'s significantly improved performance over traditional techniques by  training it on a combined data set of mock photometry of z=0 galaxies from the \textsc{Simba}, \textsc{EAGLE} and \textsc{IllustrisTNG} cosmological simulations, and comparing the derived results with those obtained from traditional SED fitting techniques. \textsc{mirkwood} is also able to account for uncertainties arising both from intrinsic noise in observations, and from finite training data and incorrect modeling assumptions. To increase the added value to the observational community, we use Shapley value explanations (SHAP) to fairly evaluate the relative importance of different bands to understand why particular predictions were reached. We envisage \textsc{mirkwood} to be an evolving, open-source framework that will provide highly accurate physical properties from observations of galaxies as compared to traditional SED fitting.
\end{abstract}

\section{Introduction}\label{sec:introduction}
\subsection{Traditional SED Fitting}
Spectral energy distributions (SEDs) of galaxies are crucial in informing our understanding of the most fundamental physical properties, such as their redshifts, stellar population masses, ages, and dust content. The task of extracting these properties by fitting models of the stellar photometry and dust attenuation to observations of galaxies is commonly known as SED fitting, and is one of the most common ways of deriving physical properties of galaxies. In SED fitting, stellar populations are evolved with  over an assumed star formation history (SFH), with an assumed stellar initial mass function (IMF) and through an assumed dust attenuation curve, to produce a composite spectrum. This resulting spectrum is then compared to the galaxy photometric observation under consideration, and the input assumptions are varied until there is a reasonable match between the two. The best fit model is then used to infer the physical properties of the concerned galaxy (see recent reviews by \citealt{walcher2011fitting, conroy2013modeling}). 

This technique has been transformational in turning observed data into inferred galaxy physical properties. For instance, we now understand the evolution of the galaxy stellar mass function through z $\sim 7$ \citep{tomczak2014galaxy, grazian2015galaxy, leja2019a}, the diversity in shapes of galaxy SFHs \citep{papovich2011rising, leja2019b}, the nature of the star formation rate--stellar mass (SFR-M$^{\star}$) relation through z = 2 \citep{mobasher2008relation, iyer2018sfr}, the process of inside-out quenching \citep{jung2017evidence}, and the general shapes of galaxy dust attenuation curves at high-z \citep{salmon2016breaking}, all thanks to a combination of high quality data with SED fitting techniques. These developments have spurred -- and in turn have been accelerated by -- the development of modern SED fitting codes such as {\rm CIGALE} \citep{boquien2019cigale}, {\sc Prospector} \citep{leja_deriving_2017}, {\rm FAST} \citep{kriekfast2009}, {\rm BAGPIPES} \citep{bagpipes}, and {\rm MAGPHYS} \citep{da2008simplemagphys}. These codes rely on Bayesian optimization and Markov Chain Monte Carlo (MCMC) sampling methods to explore the large input base of computer simulated SED templates for close matches with an observed SED, and consequently output its physical properties.% \citep{leja_deriving_2017}.

At the same time, this traditional method of fitting SEDs has been known to introduce potential uncertainties and biases in the derived physical properties of galaxies. At the core of the issue are assumptions regarding the model stellar isochrones, spectral libraries, the IMF, the SFH, and the dust attenuation law. Each of these choices can dramatically impact the derived physical properties. For example, \cite{acquaviva2015simultaneous} evaluated the effects that different modelling assumptions have on the recovered SED parameters, and found that the incorrect parameterizations of the star formation history can significantly impact the derived physical properties.   This finding has been confirmed by the works of \citet{wuyts2009color, michalowski2014spatially, sobral2014stellar} and \citet{simha2014parametrising}, who have all found significant impact of SFH on the derived galaxy stellar mass in various observational datasets. \cite{pacifici_josh} found that the simple assumption of exponentially declining SFH, simple dust law, and no emission-line contribution fails to recover the true stellar mass - star formation rate relationship for star-forming galaxies. Similarly, \cite{iyer2017reconstruction} and \cite{lower_stellar_mass} found that fitting the SFH using simple functional forms leads to a bias of up to 70\% in the recovered total stellar mass. In addition, \cite{carnall2019vandels} demonstrated that such simple functional forms for SFH also impose strong priors on other physical parameters, which in turn bias our inference for fundamental cosmic relations, such as the stellar mass function (SMF) and the cosmic star formation rate density (CSFRD, \cite{ciesla2017sfr, leja_how_2019}), critical for answering crucial questions in the study of galaxy formation and evolution. Even with more complex SFH functions which have been able to better reproduce the CSFRD \citep{gladders2013imacs_lognormal}, individual stellar age estimates have been biased \citep{carnall2019vandels, leja2019a}. In a similar vein to SFH modeling, in their recent review \cite{salim2020dust} found that incorrect assumptions of the dust attenuation law, even within the commonly accepted family of locally-calibrated curves, can drive major uncertainties in the derived galaxy star formation rates.

To overcome some of these limitations of traditional SED fitting, two broad approaches have been introduced. The first has been to modify the SFH prior employed during SED fitting. This has been accomplished by either creating new functional, parametric forms for the SFH that result in lower biases in derived posteriors in physical parameters, for instance by being less subject to the `outshining bias' \citep{behroozi2013average, simha2014parametrising}, or by developing parameter-free (\emph{non-parametric}) descriptions of the SFH \citep{tojeiro2007, iyer2017reconstruction, iyer2019nonparametric, leja_how_2019}. The latter are those that allow for a number of flexible bins of star formation to vary in the SED modeling procedure in order to model the more complex star formation histories that likely represent real galaxies \citep{iyer2018sfr, leja2019new}. In \cite{lower_stellar_mass}, the authors used a subset of the hydrodynamical simulations used in this work to study the biases introduced by various commonly used parametric forms of SFH on the derived stellar masses, and found that some of the most commonly used formulations -- `constant', `delayed-$\tau$' and `delayed-$\tau$ + burst' -- dramatically under-predict the true stellar mass. They then compared the performance against that obtained by using a non-parametric SFH model \citep{leja_deriving_2017}, and found significantly better results. Even so, there were a large number of catastrophic outliers, with errors of up to 40\% (see Figure 3 in their work). This is because SED fitting is inherently a complex task involving several moving parts, and SFH is only one of several model assumptions that go in it. As already mentioned, for example, the assumed dust attenuation law is another factor that can significantly impact the derived galaxy properties (see Figure 2 in \citealt{salim2020dust}).

The upshot is that the uncertainties in the galaxy star formation history and star-dust geometry (as manifested in their dust attenuation curves) translate to significant uncertainties in the derived properties from traditional SED fitting techniques \citep{lower_stellar_mass, leja_deriving_2017, leja_how_2019, salim2020dust}. An alternative to SED fitting is to instead derive a mapping between the photometry of a galaxy and the underlying physical properties.   In this regard, machine learning techniques can serve as a potential alternative to traditional SED fitting.  In this paper, we demonstrate that machine learning is not only a suitable alternative, but potentially more accurate one.

\subsection{Machine Learning and SED Fitting}\label{sec:related_work}
To explore the complex and degenerate parameter space more efficiently, machine learning (ML) has emerged as an alternative tool for deriving the physical parameters of galaxies. Unlike traditional SED fitting procedures, including those utilizing novel non-parametric SFH priors, ML algorithms are able to directly learn the relationships between the input observations (either photometric of spectroscopic) and the galaxy properties of interest without the need for human-specified priors. With ML, the priors themselves become learnable. This is possible because ML algorithms learn from the entire population ensemble, learning not just the mapping between individual galaxy observations and physical properties, but also the distribution of properties.%Using a training data set consisting of samples (fluxes in different bands) with known output values (galaxy physical properties), they attempt to learn the underlying data-generating distribution (\emph{supervised learning}, Section \ref{sec:methods_intro_to_ml}).

 %(Sadeh et al. 2016; Ball et al. 2008; Hogan et al. 2015; Masters et al. 2015; more).
\cite{lovell2019learning}) trained convolutional neural networks (CNNs, \cite{krizhevsky2012imagenet_cnn}) to learn the relationship between synthetic galaxy spectra and high resolution SFHs from the \textsc{Eagle} \citep{schaye_2015_eagle} and \textsc{Illustris} \citep{vogelsberger2014introducingillustris} simulations. \cite{stensbo2017sacrificing} estimated specific star formation rates (sSFRs) and redshifts using broad-band photometry from SDSS (Sloan Digital Sky Survey, \cite{sdss_eisenstein2011}). \cite{surana2020predicting} used CNNs with multiband flux measurements from the GAMA (Galaxy and Mass Assembly, \cite{gama_driver2009}) survey to predict galaxy stellar mass, star formation rate, and dust luminosity. \cite{simet2019comparison} used neural networks trained on semi-analytic catalogs tuned to the CANDELS (Cosmic Assembly Near-infrared Deep Extragalactic Legacy Survey, \cite{grogin2011candels}) survey  to predict stellar mass, metallicity, and average star formation rate. Owing to their superior ability in capturing non-linearity in data, ML-based models have seen applications in almost every field of astronomical study, including identification of supernovae \citep{disanto2016}, classification of galaxy images %\red{add citations to papers by chris conselice's group who has done a lot here, as well as some student of brant robertsons who has some nice code called morpheus about this. also this paper: https://arxiv.org/pdf/2004.11981.pdf -- best to read these papers and look in their intros for further referencing as we really want to get the referencing complete}
\citep{ml_morphology0,ml_morphology1, hausen2020morpheus,ciprijanovic2020deepmerge,barchi2020machine}, and categorization of signals observed in a radio SETI experiment \citep{harp2019}. With the exponential growth of astronomical datasets, ML based models are slowly becoming an integral part of all major data processing pipelines \citep{siemiginowska2019astro2020}. \cite{baron2019machine} provides a comprehensive and current overview of the application of machine learning methodologies to astronomical problems.

\begin{table*}
    \caption{Summary statistics of the five galaxy properties in this paper, for all three hydrodynamical simulations.}
    \label{tab:datasetsummary}
    \centering
    \begin{tabular}{ccccc}
    \toprule
     & & {\sc Simba} &  {\sc Eagle} &  {\sc IllustrisTNG} \\ \tableline
     %& Min & (0.0, 2.0)  &  (0.0, 2.0) &  (0.0, 2.0) \\
     %& Max & (0.0, 2.0)  &  (0.0, 2.0) &  (0.0, 2.0) \\
    %$z$ & Mean & (0.0, 2.0)  &  (0.0, 2.0) &  (0.0, 2.0) \\
     %& Median & (0.0, 2.0)  &  (0.0, 2.0) &  (0.0, 2.0) \\
     %& Std.Dev. & (0.0, 0.0)  &  (0.0, 2.0) &  (0.0, 0.0) \\ \tableline
     & Min & 7.64  &  7.64 &  7.70 \\
     & Max & 12.15  &  12.01 &  12.19 \\
    $\log_{10}(\frac{\rm{M}^{\star}}{\rm{M}_{\odot}})$ & Mean & 8.91  & 8.71 &  8.82 \\
     & Median & 8.78  &  8.50 &  8.65 \\
     & Std.Dev. & 0.86  &  0.83 &  0.85 \\ \tableline
     & Min & 0.0  &  2.81 &  4.72 \\
     & Max & 8.63  &  10.5 &  10.26 \\
    $\log_{10}(1+\frac{\rm{M}_{\rm{Dust}}}{\rm{M}_{\odot}})$ & Mean & 5.39  &  6.25 &  6.82 \\
     & Median & 5.33  &  6.06 &  6.72 \\
     & Std.Dev. & 1.47  &  0.86 &  0.75 \\ \tableline
     & Min & -1.61  &  -1.05 &  -1.27 \\
     & Max & 0.08  &  0.30 &  0.17 \\
    $\log_{10}(\frac{\rm{Z}^{\star}}{\rm{Z}_{\odot}})$ & Mean & -0.86  &  -0.25 &  -0.49 \\
     & Median & -0.87  &  -0.24 &  -0.50 \\
     & Std.Dev. & 0.35  &  0.18 &  0.28 \\ \tableline
     & Min & 0.0  &  0.0 &  0.0 \\
     & Max & 1.35  &  1.14 &  1.57 \\
    $\log_{10}(1+\rm{SFR}_{100})$ & Mean & 0.10  &  0.05 &  0.09 \\
     & Median & 0.02  &  0.01 &  0.03 \\
     & Std.Dev. & 0.18  &  0.11 &  0.15 \\ \tableline
    \#Samples & & 1,797 & 4,697 & 10,073 \\ \tableline 
    \end{tabular}
\end{table*}

\subsection{This Paper}
\begin{figure*}
    \centering
    \includegraphics[width=0.99\textwidth]{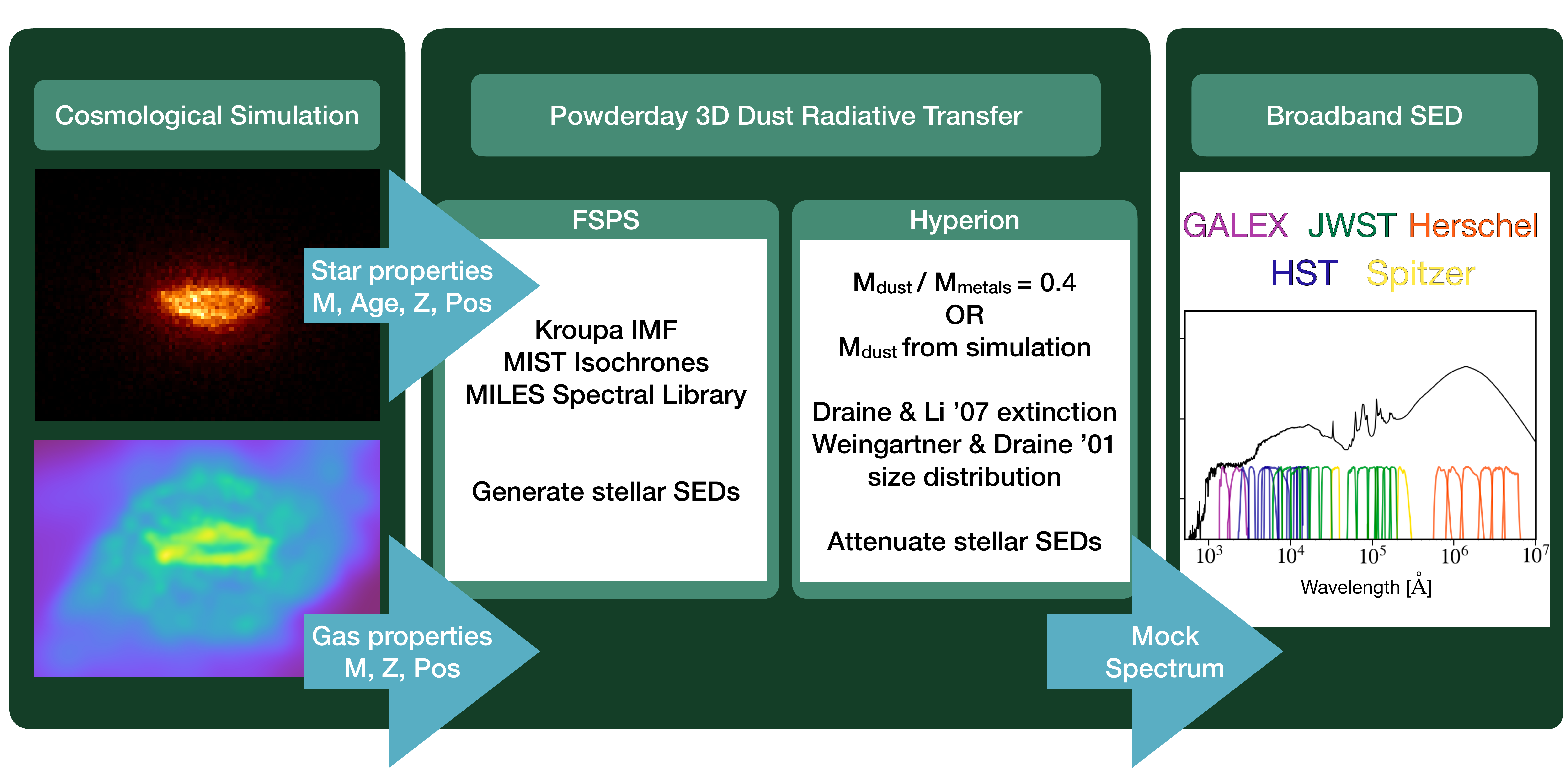}
    \caption{Data pipeline for each simulated galaxy. Once galaxies are selected from each simulation, we perform 3D dust radiative transfer with {\sc powderday}, which utilizes {\sc fsps} to generate the stellar SEDs and {\sc hyperion} to propagate the stellar SEDs through a dusty medium dependent. We then sample mock photometry in $33$ filters from the {\sc powderday} SED.} 
    \label{fig:pd_sed}
\end{figure*}

In this paper we introduce \textsc{mirkwood}, a fully ML-based tool for deriving galaxy physical parameters from multi-band photometry, while robustly accounting for various sources of uncertainty. We compare {\sc mirkwood}'s performance with that of the current state-of-the-art Bayesian SED fitting tool \textsc{Prospector}, by using data from galaxy formation simulations for which we have ground-truth values\footnote{While one could consider using observed photometric data of real galaxies instead, a glaring lack of well-defined ground truth physical parameters -- independent of any inference tool employed -- would make comparisons between results from \textsc{mirkwood} and any traditional SED fitting code difficult.}. To do this, we first create mock SEDs from three different cosmological hydrodynamical simulations (via the procedure discussed in detail in Section \ref{sec:data} and illustrated in Figure \ref{fig:pd_sed}) -- \textsc{Simba}, \textsc{Eagle}, and \textsc{IllustrisTNG}. We then fit GALEX, HST, Spitzer, Herschel, and JWST photometry of these mock SEDs to extract stellar masses, dust masses, stellar metallicities, and instantaneous SFRs (defined as the mean SFR in the past 100 Myrs from the time of observation) using the publicly available SED fitting software {\sc Prospector} \citep{leja_deriving_2017} with the state-of-the-art non-parametric SFH model from \cite{leja_deriving_2017, leja2019a} (Section \ref{sec:data}). We repeat this exercise using \textsc{mirkwood}, and compare both sets of derived galaxy properties to their true values from simulations.

The rest of this paper is organized as follows. In Section \ref{sec:data} we describe both our data and the traditional SED fitting technique in significant detail, including the three hydrodynamical simulations (Section \ref{sec:data_sims}), the method to select galaxies of interest from these simulations (Section \ref{sec:data_galaxyselection}), the dust radiative transfer process used to create SEDs (Section \ref{sec:data_radiativetransfer}), and finally the Bayesian code used to fit these SEDs to derive galaxy properties (Section \ref{sec:data_prospector}). In Section \ref{sec:methods_proposed_method} we describe in detail the SED modeling procedure using our proposed model \textsc{mirkwood}. We describe how \textsc{mirkwood} robustly estimates uncertainties in its predictions (uncertainty quantification, Section \ref{sec:methods_uncertainty}), and introduce metrics to quantify performance gains with respect to traditional SED fitting (Section \ref{sec:methods_metrics}). In Section \ref{sec:results}, we investigate the improvement in the determination of the four galaxy properties. Finally, in Section \ref{sec:discussion} we summarize our findings, and suggest possible improvements to both the proposed model and to areas of application. Throughout we assume a Planck 2013 cosmology with the following parameters: $\Omega_m = 0.30$, $\Omega_\lambda = 0.70$, $\Omega_b = 0.048$, $h = 0.68$, $\sigma_8 = 0.82$, and $n_s = 0.97$.

\section{Training on Cosmological Galaxy Formation Simulations}\label{sec:data}

The first step in a supervised ML code is to train the software on known results.  To do this, we employ three state-of-the-art cosmological galaxy formation simulations where we know the true physical properties.  We then generate mock SEDs from the galaxies in these simulations using stellar population synthesis models and dust radiative transfer; these mock SEDs allow our ML algorithms to develop a highly accurate mapping between input photometry and galaxy physical properties.

We train on a set of three galaxy formation simulations: {\sc Simba} \citep{dave_simba}, {\sc Eagle} \citep{schaye_2015_eagle,schaller_2015_eagle,mcalpine_2016_eagle_cat}, and {\sc IllustrisTNG} \citep{vogelsberger2014introducingillustris}. These models provide a large and diverse sample of galaxies with realistic growth histories, and free us from being dependent on any one set of galaxy model assumptions. In the following subsections, we describe each simulation and the galaxy formation models included in each suite as well as our galaxy selection method and computational techniques for generating mock SEDs for each galaxy. 

\subsection{Cosmological Galaxy Formation Simulations}\label{sec:data_sims}

\textbf{\textsc{Simba}} \citep{dave_simba}: The {\sc Simba} galaxy formation model is a descendent of the {\sc Mufasa} model, and uses the {\sc Gizmo} gravity and hydrodynamics code \citep{hopkins_2015_gizmo}. {\sc Simba} uses an H$_2$-based star formation rate (SFR), given by the density of H$_2$ divided by the local dynamical timescale. The H$_2$ density itself relies on the metallicity and local column density \citep{krumholz2014despotic}. The chemical enrichment model tracks $11$ elements from from Type II supernovae (SNe), Type Ia SNe, and asymptotic giant branch (AGB) stars with yields following \cite{nomoto_simba_sne_yields}, \cite{iwamoto_1999_sne_yields}, and \cite{Oppenheimer_2006_agb_yields}, respectively. Sub-resolution models for stellar feedback include contributions from Type II SNe, radiation pressure, and stellar winds. The two-component stellar winds adopt the mass-loading factor scaling from {\sc fire} \citep{angles-alcazar_FIRE} with wind velocities given by \cite{muratov_2015}. Metal-loaded winds are also included, which extract metals from nearby particles to represent the local enrichment by the SNe driving the wind. Feedback via active galactic nuclei (AGN) is implemented as a two-phase jet (high accretion rate ) and radiative (low accretion rate) model, similar to the model used in {\sc IllustrisTNG} described below. Finally, dust is modeled self-consistently, produced by condensation of metals ejected from SNe and AGB stars, and is allowed to grow and erode depending on temperature, gas density, and shocks from local Type II SNe \citep{qi_dust,li2020}. The tunable parameters in {\sc Simba} were chosen to reproduce the M$_{BH}$-$\sigma$ relation and the galaxy stellar mass function (SMF) at redshift z $=0$. For our analysis, we use a box with $25/h$ Mpc side length with $512^3$ particles, resulting in a baryon mass resolution of $1.4\times$ $10^6 \: \mathrm{M}_{\odot}$.

\textbf{Eagle} \citep{crain_2015_eagle, schaye_2015_eagle, schaller_2015_eagle, mcalpine_2016_eagle_cat}: The Evolution and Assembly of GaLaxies and their Environments ({\sc eagle}) suite of cosmological simulations is based on a modified version of the smoothed particle hydrodynamics (SPH) code {\sc gadget} 3 \citep{springel_2005_gadget}. Star formation occurs via gas particles that are converted into star particles at a rate that is pressure dependent and gas temperature and density limited, following \cite{schaye_2004_eagle_sfr} and \cite{schaye_dallavecchia_2008_eagle_sfr}. The chemical enrichment model tracks $9$ independent elements generated via winds from AGB stars, winds from massive stars, and Type II SNe following \cite{Wiersma_2009_eagle_yields}, \cite{Portinari_1998_eagle_yields}, and \cite{Marigo_2001_eagle_yields}, respectively. Stellar feedback from Type II SNe heats the local ISM, delivering fixed jumps in temperature and a fraction of energy from the SNE, and similarly, a single feedback mode for AGN injects thermal energy proportional to the BH accretion rate at a fixed efficiency into the surrounding gas. Like {\sc simba}, the parameters in {\sc eagle} have been tuned to reproduce the z $=0.1$ galaxy SMF. For our analysis, we use a $50/h$ Mpc box size with $752^3$ particles, resulting in a baryon mass resolution of $1.81\times$ $10^6 \: \mathrm{M}_{\odot}$. 
    
\textbf{IllustrisTNG} \citep{weinberger_2017_tng, pillepich_2018_tng1, pillepich_2018_tng2}: {\sc IllustrisTNG} is an updated version of the {\sc illustris} project, based on the {\sc arepo} \citep{springel_2010_arepo} magneto-hydrodynamics code. Star formation occurs in gas above a given density threshold, at a rate following the Kennicutt-Schmidt relation \citep{schmidt_1959_sfr, kennicutt_1989_sfr}. As stars evolve and die, nine elements are tracked via enrichment from Type II SNe and AGB stars following the models and yields of \cite{Wiersma_2009_eagle_yields}, \cite{Portinari_1998_eagle_yields}, and \cite{nomoto_simba_sne_yields}. Star formation also drives galactic scale winds. These hydrodynamically decoupled wind particles are injected isotropically with an initial speed that scales with the the local 1D dark matter velocity dispersion. The galactic winds carry additional thermal energy to avoid spurious artifacts where the wind particles hydrodynamically recouple with the gas. Like {\sc simba}, {\sc IllustrisTNG} features a two-mode feedback model for AGN: thermal energy is injected into the surrounding ISM at high accretion rates, while BH-driven winds are produced at low accretion rates. Tunable parameters were chosen in {\sc IllustrisTNG} to match observations including the z $=0$ stellar mass function, the star formation rate density, and the stellar mass-halo mass relation at z $=0$. We use the TNG100 box with a baryonic mass resolution of $1.4\times$ $10^6 \: \mathrm{M}_{\odot}$. 

\begin{table}[!htbp]
\begin{tabular}{llc}
\tableline
Instrument    & Filter & Effective Wavelength (AA) \\ \tableline
GALEX         & FUV    & 1549                      \\
              & NUV    & 2304                      \\ \tableline
              & F275W  & 2720                      \\
              & F336W  & 3359                      \\
              & F475W  & 4732                      \\
              & F555W  & 5234                      \\
              & F606W  & 5780                      \\
HST/WFC3      & F814W  & 7977                      \\
              & F105W  & 10431                     \\
              & F110W  & 11203                     \\
              & F125W  & 12364                     \\
              & F140W  & 13735                     \\
              & F160W  & 15279                     \\ \tableline
Spitzer/IRAC  & Ch1    & 35075                     \\ \tableline
              & Blue   & 689247                    \\
Herschel/PACS & Green  & 979036                    \\
              & Red    & 1539451                   \\ \tableline
              & F070W  & 7007                      \\
              & F090W  & 8980                      \\
JWST/NIRCam   & F115W  & 11487                     \\
              & F150W  & 14942                     \\
              & F200W  & 19791                     \\
              & F277W  & 27466                     \\ \tableline
              & F560W  & 56151                     \\
              & F770W  & 75911                     \\
              & F1000W & 99230                     \\
JWST/MIRI     & F1130W & 127677                    \\
              & F1280W & 113035                    \\
              & F1500W & 150078                    \\
              & F1800W & 179390                    \\ \tableline
Spitzer/MIPS  & 24 um  & 232096                    \\ \tableline
              & PSW    & 2428393                   \\
Herschel/SPIRE & PMW   & 3408992                   \\
              & PLW    & 4822635                   \\ \tableline
\end{tabular}
\caption{Table of the 33 filters used to extract photometry from the synthetic {\sc powderday} spectra.\label{table:filters}}
\end{table}

\subsection{Galaxy Selection}\label{sec:data_galaxyselection}
Galaxies from each simulation were identified with their respective friends-of-friends (FOF) algorithms. We use the publicly available galaxy catalogues for {\sc eagle} \citep{mcalpine_2016_eagle_cat} and {\sc IllustrisTNG} \citep{nelson_tng_cat} in which subhalo structures are identified by a minimum of 32 particles with the {\sc subfind} algorithm \citep{springel_2001_subfind, dolag_2009_subfind}. For {\sc simba}, we have employed  {\sc caesar}\footnote{https://github.com/dnarayanan/caesar} \citep{caesar}, which identifies halos and galaxies in snapshots based on the number of bound stellar particles (a minimum of 32 particles defines a galaxy). To ensure our galaxy sample is robust across the three simulations, we select galaxies from the full subhalo populations that have (i) a stellar mass above the minimum mass found in the z $=0$ {\sc simba} snapshot ($4.4\times$ $10^7 \: \mathrm{M}_{\odot}$) and (ii) have a nonzero gas mass. Due to computational limits, we randomly sample $\sim$10,000 galaxies from the $\sim$85,000 previously identified {\sc IllustrisTNG} galaxy sample, though we ensure that this subsample matches the intrinsic stellar mass function of the entire sample. After these selections, we have a total of $1797$, $4697$, and $10000$ from the {\sc Simba}, {\sc Eagle}, and {\sc IllustrisTNG} simulations respectively for redshift $z=0$.% For redshift $z=2$, an additional $17,020$ galaxies are included in the training set. 

\subsection{Mock SEDs from 3D Radiative Transfer}\label{sec:data_radiativetransfer}

Our analysis relies on realistic mock galaxy SEDs as input to predict physical properties including stellar mass and SFR. To produce these synthetic SEDs, we perform full 3D dust radiative transfer (RT) for each galaxy in our subsamples. We use the radiative transfer code {\sc powderday}\footnote{https://github.com/dnarayanan/powderday} \citep{pd_2020} to construct the synthetic SEDs by first generating with {\sc fsps} \citep{fsps_1, fsps_2} the dust free SEDs for the star particles within each cell using the stellar ages and metallicities as returned from the cosmological simulations. For these, we assume a \cite{kroupa_initial_2002} stellar IMF and the {\sc mist} stellar isochrones \citep{mist_1, mist_2}. These {\sc fsps} stellar SEDs are then propagated through the dusty ISM. Here, differences in the RT model arise as dust is implemented differently in each simulation. {\sc eagle} and {\sc IllustrisTNG} do not have a native dust model and so we assume a constant dust mass to metals mass ratio of $0.4$ \green{\citep{dwek_1998_dust_to_metals, vladilo_1998_dust_to_metals, watson_2011_dust_to_metals}}. For {\sc simba}, the diffuse dust content is derived from the on-the-fly self-consistent model of \cite{qi_dust}. From there, the dust is treated identically between simulations: it is assumed to have extinction properties following the carbonaceous and silicate mix of \cite{draine_infrared_2007}, that follows the \cite{weingartner_draine} size distribution and the \cite{draine_03} renormalization relative to hydrogen. We assume $R_{\rm V} \equiv A_{\rm V}/E(B-V) = 3.15$. We do not assume further extinction from sub-resolution birth clouds. Polycyclic aromatic hydrocarbons (PAHs) are included following the \citet{robitaille_pahs} model in which PAHs are assumed to occupy a constant fraction of the dust mass (here, modeled as grains with size $a<20$ \AA) and occupying $5.86\%$ of the dust mass. The dust emissivities follow the \citet{draine_infrared_2007} model, though are parameterized in terms of the mean intensity absorbed by grains, rather than the average interstellar radiation field as in the original Draine \& Li model. The radiative transfer propagates through the dusty ISM in a Monte Carlo fashion using {\sc hyperion} \citep{robitaille_hyperion}, which follows the \cite{lucy_rt} algorithm in order to determine the equilibrium dust temperature in each cell. We iterate until the energy absorbed by $99\%$ of the cells has changed by less than $1\%$.  Note, a result of these calculations is that while we assume the \cite{draine_infrared_2007} extinction curve in every cell, the effective {\it attenuation} curve is a function of the star-dust geometry, and therefore varies from galaxy to galaxy \citep{attenuation_araa}.

The result of the {\sc powderday} radiative transfer is the UV - FIR spectrum for each galaxy. From there, we sample broadband photometry in $33$ filters, ensuring robust SED coverage. These filters are shown in Table \ref{table:filters}. Our training set for {\sc mirkwood} includes photometry at signal-to-noise ratios (SNR) of 2, 10, and 20.  

\subsection{Prospector SED Fitting}\label{sec:data_prospector}

We compare our results from {\sc mirkwood} to traditional galaxy SED fitting using the state-of-the-art SED fitting code, {\sc Prospector} \citep{2019ascl.soft05025J, leja_deriving_2017, leja_how_2019}\footnote{https://github.com/bd-j/prospector}. {\sc Prospector} wraps {\sc fsps} stellar modeling with {\sc dynesty} \citep{dynesty} Bayesian inference to provide estimates of galaxy properties given models forms for the galaxy star formation history (SFH), metallicity, and dust content. For the galaxy SFH, we use the nonparametric linear piece-wise function as described in \cite{leja_how_2019} with 6 time bins spaced equally in logarithmic time. The stellar metallicity is tied to the inferred stellar mass of the galaxy, constrained by the \cite{gallazzi_ages_2005} M*-Z relation from SDSS DR7 data. The dust attenuation model follows \cite{kc_law}, in which the strength of the 2175 \AA \ bump is tied to the powerlaw slope. Dust emission follows the \cite{draine_infrared_2007} model. Prior distributions for model components approximately follow those outlined in \cite{lower_stellar_mass} and have been shown to reasonably reproduce {\sc simba} galaxy properties including stellar mass and SFR. 

We provide the same mock {\sc powderday} broadband SEDs to {\sc prospector} as {\sc mirkwood} with the same SNRs. In all comparisons to {\sc mirkwood}, we report the median value of the posterior distribution from the {\sc prospector} fit for each galaxy property. 

\section{Proposed Method}\label{sec:methods_proposed_method}

We now present implementation details of our proposed method. {\sc mirkwood} is a \emph{supervised} learning algorithm, meaning it requires a \emph{training set} consisting of a pairs of inputs and outputs, to be able to learn the mapping from the former to the latter. Each input consists of a set of discrete \emph{features}, or independent variables, whereas the outputs -- also referred to as \emph{labels} -- are the dependent variables. The algorithm learns this mapping by minimizing a user-specified \emph{loss function}.  This is similar to the minimization of the mean squared error commonly employed when fitting data in linear regression analysis. In this work, each input $\mathbf{x}$ is a mock SED vector consisting of flux values in 35 bands, and the associated galaxy properties (redshift, stellar mass, dust mass, stellar metallicity, and instantaneous star formation rate) form a vector output of length 5. While certain types of ML algorithms are able to map each input vector to an output vector consisting of multiple labels (for example, neural networks), this is not possible with {\sc mirkwood}. Hence we associate with each input only one galaxy property at a time, and use the model five times. Let us denote by $\mathbf{x}$ an input SED vector, by $y$ a galaxy property, by $n$ the number of SED samples, and by $d$ the number of bands/filters. Then the data set on hand can be mathematically specified as $\mathcal{D}=\left\{\left(\mathbf{x}_{i}^{1:d}, y_{i}\right): i=1, \ldots, n\right\}$. Our aim is to learn a data-driven mapping from the set of all $\mathbf{x}_{i}$s to the set of all $ y_i$s, to be able to predict $y_\mathrm{new}$ for a new, unseen observation $\mathbf{x_\mathrm{new}}$. To solve this \emph{regression} problem\footnote{Classification tasks are those where the goal is to predict discrete labels, e.g. identifying whether an object is a hotdog or not a hotdog. In regression problems we aim to prediction continuous labels, e.g. predicting the price of a stock tomorrow given its price over the past year.}, we employ the \textsc{NGBoost} algorithm \citep{ngboost}. Unlike most commonly used ML algorithms and libraries such as Random Forests \citep{randomforests}, Randomized Trees \citep{extremelyrandomizedtrees}, XGBoost \citep{xgboost} and Gradient Boosting Machines \citep{lightgbm}, \textsc{NGBoost} enables us to easily work in a probabilistic setting, and corresponding to every input galaxy SED, output both a measure of the central tendency (i.e. $\mu$), and a measure of dispersion (i.e. $\sigma$) for the galaxy property of interest. We thus use {\sc NGBoost} as the backbone of {\sc mirkwood}, and use the hyper-parameters suggested by \cite{ren2019rongba_ngboost}\footnote{\cite{ren2019rongba_ngboost} found that their new set of hyperparameters result in faster convergence than the default values supplied by the developers at \url{https://github.com/stanfordmlgroup/ngboost/blob/master/ngboost/ngboost.py}.}. We make the assumption that samples are drawn from Gaussian distributions. % therefore its probability density function is given by $\mathrm{PDF}(y_i) = (2\pi \sigma_{\rm a}^2)^{-1/2} \exp{\frac{-(y_i - \mu)}{\sigma_{\rm a}^2}}\mathcal{N}$
Our loss function is the \emph{negative likelihood function} \citep{ngboost}.

\begin{figure*}
    \centering
    \includegraphics[width=.80\textwidth]{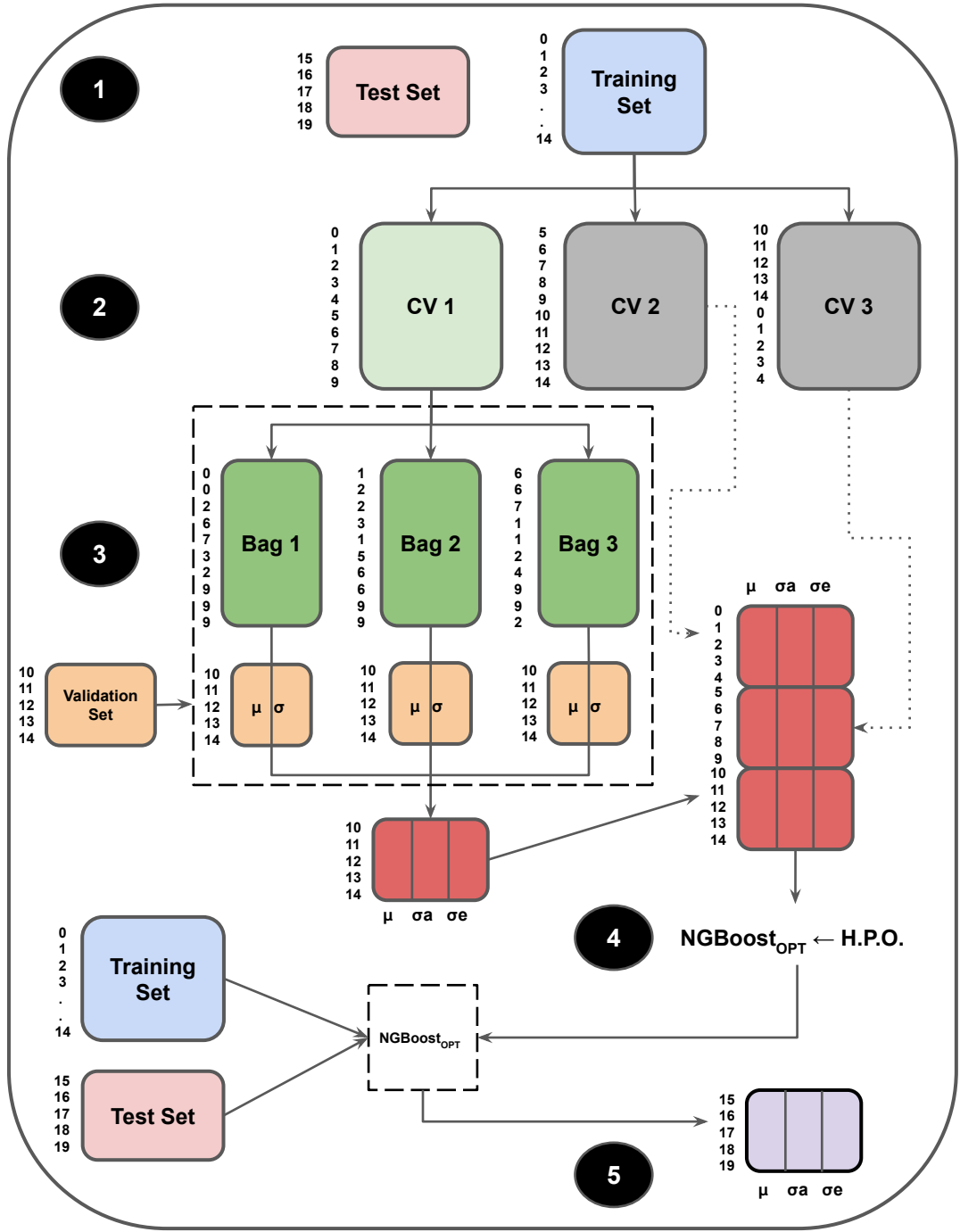}
    \caption{\textbf{Training pipeline for {\sc mirkwood} in five sequential steps.} %In what follows, we refer to the test subset -- for which we desire to obtain predictions -- as $\mathcal{D_{TEST}}$; the training subset -- on which the \textsc{NGBoost} algorithm is trained -- as $\mathcal{D_{TRAIN}}$; and the input data set consisting of both of these as $\mathcal{D}$. Given a sample of galaxy flux values as input, a `prediction' consists of a mean ($\mu$) and a standard deviation ($\sigma$) for the desired galaxy property. 
    \emph{\textbf{Step 1:}} We divide the input data set $\mathcal{D}$ into $N_{\rm CV,1} = 4$ cross-validation folds, at any point referring to the collation of $N_{\rm CV,1}-1$ of them as $\mathcal{D_{TRAIN}}$, and the remaining fold as $\mathcal{D_{TEST}}$. We repeat this $N_{\rm CV,1}$ times to cover all samples in $\mathcal{D}$, but depict only one such iteration here for illustration. \emph{\textbf{Step 2:}} We sub-divide $\mathcal{D_{TRAIN}}$ into $N_{\rm CV,2} = 3$ CV folds. Same as before, $N_{\rm CV,2}-1$ folds are collated while the remaining fold, referred to as the validation set $\mathcal{V}$, is set aside. \emph{\textbf{Step 3:}} Each of the $N_{\rm CV,2}$ CV folds is used to create $N_{\rm bags} = 3$ `bags' by randomly shuffling its data and picking the same number of samples with replacement. An \textsc{NGBoost} model with a given set of hyperparameters is copied over $N_{\rm bags}$ times, and each copy is trained on one `bag' of data. The predictions of all $N_{\rm bags}$ models on $\mathcal{V}$ are recorded and averaged according to Equation \ref{eqn:main_averaging_eqn} to obtain the mean, epistemic, and aleatoric uncertainties for $\mathcal{V}$. \emph{\textbf{Step 4:}} This process is repeated 100 times to find better performing hyperparameters for \textsc{NGBoost}; we refer to thus optimized model as {\sc NGBoost$_{\rm{OPT}}$}. \emph{\textbf{Step 5:}} Finally, we train {\sc NGBoost$_{\rm{OPT}}$} on $\mathcal{D_{TRAIN}}$ and predict the output for $\mathcal{D_{TEST}}$.} %the predictions on $\mathcal{V}$ are used to \emph{post hoc} modify the predictions on $\mathcal{D_{TEST}}$ as described in Section ().}
    \label{fig:mirkwood_pipeline}
\end{figure*}

Consider the case where we have samples from all three simulations. As a reminder, we have 1,797 samples from {\sc Simba}, 4,697 from {\sc Eagle}, and 10,073 from {\sc IllustrisTNG}. %The corresponding figures at z=2 are 1,961, 5,734 and 9,324.
To predict any given galaxy property for each of the $n=1,797$ {\sc Simba} %($n=33,586$)
galaxies, we proceed as follows. First, we fix the parameters of the {\sc NGBoost} model to those recommended by \cite{ren2019rongba_ngboost}. Then, we divide the $n$ samples into $N_{\rm{CV,1}}=5$ distinct subsets (or \emph{folds}) based on \emph{stratified sampling} -- $y_i$s in each subset are picked such that their distribution closely matches the distribution of the parent sample $y_i$s. The first four subsets taken together form the \emph{training set}, $\mathcal{D_{TRAIN}}$, while the fifth, left-out fold for which we assume output labels are unavailable, is called the \emph{test set}, $\mathcal{D_{TEST}}$. This is the first step show in Figure \ref{fig:mirkwood_pipeline}. We then sub-divide $\mathcal{D_{TRAIN}}$ into $N_{\rm{CV,2}} = 5$ subsets, again in a stratified fashion. The model is trained cyclically on the combination of any distinct four sub-subsets -- referred to as $\mathcal{D_{TRAIN_{TRAIN}}}$, while its predictions on the samples in the left-out sub-subset, $\mathcal{D_{TRAIN_{VAL}}}$, are recorded and stored. Instead of training directly on $\mathcal{D_{TRAIN_{TRAIN}}}$, we use create  $N_{\rm bags} = 48$ `bags' by randomly shuffling the data and picking the same number of samples \emph{with replacement}. The {\sc NGBoost} model initialized earlier is trained on all bags, and the respective predictions on  $\mathcal{D_{TRAIN_{VAL}}}$ are averaged. The exact method of averaging is expounded upon in Section \ref{sec:methods_uncertainty}. After $N_{\rm{CV,2}}$ rounds we have $N_{\rm{CV,2}}$ non-overlapping $\mathcal{D_{TRAIN_{VAL}}}$ data sets containing distinct samples, which taken together constitute $\mathcal{D_{TRAIN}}$. These together constitute the second and third steps in the pipeline. For each of the 1,797 %$14,770$ 26,869
samples, we compare the predicted galaxy properties with their ground truth values, and note the average negative-log likelihood, $\mathcal{NLL_{\rm{VAL}}}$. We repeat this exercise multiple times, on each iteration tuning {\sc NGBoost}'s hyperparameters so that the $\mathcal{NLL_{\rm{VAL}}}$ is as low as possible. With the optimized set of {\sc NGBoost} hyperparameters thus obtained, we train {\sc mirkwood} on the entire training set $\mathcal{D_{TRAIN}}$, and record its predictions on $\mathcal{D_{TEST}}$. This process of \emph{hyperparameter optimization} is the fourth step in the pipeline. Finally, this process is repeated $N_{\rm{CV,1}}-1$ more times, to obtain predictions on all $n$ samples -- this is the fifth and final step in the pipeline. The entire process is carried out in a \emph{chained} fashion -- the model is first trained to use galaxy flux values to predict stellar mass,, then the predicted masses in conjunction with the original flux values are used to predict dust mass, and so on; this is explained pictorially in Figure \ref{fig:chaining}. We know that galaxy properties can be strongly degenerate (for instance, the well-known age-reddening-metallicity degeneracy); by chaining, we `hack' our model to leverage the inter-dependencies between these parameters. To predict SFR$_{100}$, the galaxy SFR average over the last 100 Myr, for an unseen sample $x_{\textrm{new}}$, for instance, we will use \textsc{NGBoost} %the model that was trained on samples split stratifically based on SFR$_{100}$, and
to first predict galaxy stellar mass, then galaxy dust mass, then stellar metallicity, and finally SFR$_{100}$.

\subsection{Uncertainty Quantification}\label{sec:methods_uncertainty}
Real observations differ from simulations in a number of ways.
\begin{itemize}
    \item It is rarely the case that observers have access to all filters in a survey for every observed galaxy. For example, one might encounter a situation where they are interested in determining a galaxy's mass from its photometry, but do not have access to $K$ and $H$ band information, thus making their task difficult.
    \item Different filters have differing sensitivities and noise properties, resulting in non-uniform measurement noises across observation bands. Even in an ideal case where observations in UV, optical and IR are equally robust, there is an irreducible Poisson measurement error.
    \item While modern hydrodynamical simulations have demonstrated a number of successes, there is an inherent, irreducible \emph{domain gap} between their approximation of reality, and actual reality. Imperfect modeling assumptions both about known physical processes, and arising from our sheer lack of knowledge of them, are primary contributors to this gap. Some specific reasons are modeling star formation with a sub-resolution model, divergence of modeled stellar mass functions from observed ones, and incomplete understanding of AGN feedback processes.
    \item Any supervised ML algorithm is only as good as the data on which it is trained. It might well be the case that an observed galaxy is in a completely different parameter space than any of the galaxies in our training set. %For example, in Figure () we consider the case where \textsc{mirkwood} is trained on only \textsc{Simba} galaxies, and is asked to predict stellar mass for galaxies in \textsc{Eagle} and \textsc{IllustrisTNG}. As can be seen, the model performance is severely lacking in the non-overlapping part of the histograms in Figure ()(). 
    \item Even if some particular model, say \textsc{NGBoost}, with a given set of hyperparameters, is able to faithfully and satisfactorily predict galaxy properties, it is very well possible that there exist other models which can also make predictions that are just as accurate. It is also possible that for our chosen model/set of models, a different data pre-processing scheme, choice of hyperparameters (as mentioned in Section \ref{sec:methods_proposed_method}, the chosen set of hyperparameters is locally optimal but is not guaranteed to be globally optimal), or choice of scheme for balancing classes might result in better predictions.
\end{itemize}

Owing to these factors, a highly desirable behavior of any predictive model would be to not only return a point prediction (per galaxy property under consideration), but also some quantity conveying the model's belief or confidence in its output \citep{kendall2017uncertainties, choi2018uncertainty}. We can break down such \emph{predictive uncertainty} into two parts--\emph{aleatoric} and \emph{epistemic}  \cite{kendall2017uncertainties}\footnote{The word epistemic comes from the Greek ``episteme", meaning ``knowledge"; epistemic uncertainty is ``knowledge uncertainty". Aleatoric comes from the Latin ``aleator", meaning ``dice player"; aleatoric uncertainty is the ``dice player's" uncertainty \citep{gal_thesis}}. Aleatoric uncertainty captures the uncertainty in the data generating process -- for example, measurement noise in filters, or the inherent randomness of a coin flipping experiment. It cannot be reduced by collecting more training data\footnote{We know that a fair coin has a 0.5 probability of landing heads, yet we cannot assert with 100\% certainty the outcome of the next flip, no matter how many times we have already flipped the coin.}; however, it can be reduced by collecting more informative features\footnote{Uncertainty in galaxy mass estimation in the absence of H and K band information can be drastically reduced by collecting high quality data in these bands.}. On the other hand, epistemic uncertainty models the ignorance of the predictive model, and can indeed be explained away given a sufficiently large training dataset\footnote{If we are given a biased coin where we do not know the probability $p$ of turning up heads, we can determine it to arbitrary precision by flipping the coin an arbitrarily large number of times and counting the number of times it turns up heads.}. Aleatoric uncertainty is thus \emph{irreducible} or \emph{statistical}, epistemic being its \emph{reducible} or \emph{systematic} cousin. Aleatoric uncertainty covers first three of the five issues above, while epistemic uncertainty subsumes the last two points. High aleatoric uncertainty can be indicative of noisy measurements or missing informative features, while high epistemic uncertainty could be a clue that the observed galaxy is very different that any of the galaxies the model was trained on (such a galaxy is referred to as being \emph{out of distribution}).

As previously discussed, the aim for a supervised ML algorithm is to learn the latent data generating mechanism that maps each input sample $\mathbf{x}_{i}$ to its corresponding output label $y_{i}$. This can be represented as follows:%We compute probability density function for every input. Here we
\[
y_i=f(\mathbf{x}_i)+\eta_{i}
\]
where $f(\cdot)$ is the \emph{unknown} latent function and a measurement error $\eta$ follows a standard normal distributions distribution with a variance $\sigma_{\eta,i}^{2},$ i.e., $\eta_i \sim \mathcal{N}\left(\mu=0, \sigma_{\eta,i}^{2}\right)$. The variance of $\eta_i$ corresponds to the aleatoric uncertainty, i.e., $\sigma_{\eta,i}^{2}=\sigma_{a,i}^{2}$ where we will denote $\sigma_{a,i}^{2}$ as aleatoric uncertainty associated with the observation $\mathbf{x}_{i}$. Similarly, epistemic uncertainty will be denoted as $\sigma_{e,i}^{2}$. Since $f(\cdot)$ is unknown, by training on the \emph{training set} the model attempts to approximate, or guess, this function as closely as possible. Suppose that we train $\hat{f}(\mathbf{x})$ to approximate $f(\mathbf{x})$ from the samples in the input data set, $\mathcal{D}$. Then, we can see that
%\[
%\sigma_{e}^{2}=\mathbb{E}\|f(\mathbf{x})-\hat{f}(\mathbf{x})\|^{2}
%\]
%Then, we can see that
\[
\begin{aligned}
\mathbb{E}\|y_i-\hat{f}(\mathbf{x}_i)\|^{2} &=\mathbb{E}\|y_i-f(\mathbf{x}_i)+f(\mathbf{x}_i)-\hat{f}(\mathbf{x}_i)\|^{2} \\
&=\mathbb{E}\|y_i-f(\mathbf{x}_i)\|^{2}+\mathbb{E}\|f(\mathbf{x}_i)-\hat{f}(\mathbf{x}_i)\|^{2} \\
&=\sigma_{a,i}^{2}+\sigma_{e,i}^{2},
\end{aligned}
\]
where $\mathbb{E}\|\cdot\|$ is the expectation operator. This indicates the total predictive variance is the sum of aleatoric uncertainty and epistemic uncertainty.

The \textsc{NGBoost} algorithm that forms the backbone of {\sc mirkwood} is naturally able to output both the mean $\mu_i$ and the aleatoric uncertainty $\sigma_{a,i}^2$ for a given observation $\mathbf{x}_i$. To obtain the epistemic uncertainty $\sigma_{e,i}^2$, we use \emph{bootstrapping}. This involves creating multiples copies of the same training dataset, each time picking samples \emph{with replacement}, training the same model on all copies, and averaging the results. The goal is to simulate a scenario where a model only has access to limited samples -- and hence limited parameter space -- and to study how this impacts its predictive ability (see Figure \ref{fig:mirkwood_pipeline}, especially \emph{\textbf{Step 3}}). That is, given an \textsc{NGBoost} model, the predicted label for an observation $\mathbf{x}_{i}$ is represented as:
\begin{align*}
p(\mathbf{y} \mid \mathbf{x}_{i}) &= \mathcal{N}\left(\mu_{i}, \sigma_{a,i}^2\right),
\end{align*}

Creating $N_{\rm bags}$ bootstrapped bags essentially creates $N_{\rm bags}$ copies of the above equation. To obtain final summary statistics, we combine them as
\begin{align}
\hat{\mu}_i {=}\sum_{j=1}^{N_{\rm bags}}\mu_{i,j};\;\hat{\sigma}_{i,a}^2 {=} \sum_{j=1}^{N_{\rm bags}} \sigma_{i,a,j}^{2};\;\hat{\sigma}_{i,e}^2 {=} \sum_{j=1}^{N_{\rm bags}} (\mu_{i,j} {-} \hat{\mu}_i)^2
\label{eqn:main_averaging_eqn}
\end{align}

The upshot of this exercise is that the errors modeled by {\sc mirkwood} encapsulates both the propagated noise due in the input samples, and the uncertainty arising due to the finite size of the training set.  The inclusion of the latter source of uncertainty represents a significant advancement over the current state-of-the-art \citep[e.g.][]{lovell2019learning,acquaviva2015simultaneous}, and helps mitigate the over-confidence in predictions that is typical of machine learning models.

\begin{table*}
    \caption{Summary of machine learning terms and variables appearing in this work.}
    \label{tab:meaningsofterms}
    \centering
    \begin{tabular}{p{0.15\linewidth} p{0.80\linewidth}}
    \toprule
    Term & Meaning \\ \tableline
    \emph{Bagging} & Bootstrapped Aggregation. A technique of building several models at a time by randomly sampling with replacement, or bootstrapping, from the original data set. Each bag may or may not be the same size as the parent training data set. \\ 
    \emph{HPO} & Hyperparameter optimization. The task of finding a well performing set of hyperparameters for any given ML algorithm. \\
    \emph{Chaining} & Given multiple outputs of interest (dependent variables), the technique of running the regression model in sequence to exploit correlations among them. \\
    \emph{Cross validation} & The task of dividing the training set into non-overlapping `folds' for the purpose of hyperparameter optimization. \\
    $\mathcal{D}$ & Input training set comprising of all samples from {\sc Simba}, {\sc Eagle}, and {\sc IllustrisTNG}. Each sample consists of input (independent variables)-output(dependent variables) pairs.  \\
    $\mathcal{D_{TRAIN}}$ & Training set, derived from $\mathcal{D}$. {\sc mirkwood} is trained on this, and used for inference on $\mathcal{D_{TEST}}$. \\
    $\mathcal{D_{TRAIN_{TRAIN}}}$ & Training set, derived from $\mathcal{D_{TRAIN}}$. Used for HPO. \\
    $\mathcal{D_{TRAIN_{VAL}}}$ & Validation set, derived from $\mathcal{D_{TRAIN}}$. Used for HPO. \\
    $\mathcal{D_{TEST}}$ & Test set. \\ 
    $N_{\rm{CV,1}}$ & Number of CV folds that is used to produce $\mathcal{D_{TRAIN}}$ and $\mathcal{D_{TEST}}$ from $\mathcal{D}$. \\
    $N_{\rm{CV,2}}$ & Number of CV folds that is used to produce $\mathcal{D_{TRAIN_{TRAIN}}}$ and $\mathcal{D_{TRAIN_{VAL}}}$ from $\mathcal{D_{TRAIN}}$. \\
    $\sigma_{e}$ & Epistemic uncertainty. \\
    $\sigma_{a}$ & Aleatoric uncertainty. \\
    $f(\cdot)$ & Functional representation of {\sc mirkwood}. This acts on a given input vector $\mathbf{x}$ to produce the output $y$. \\
    $\mu$ & The mean of a predicted galaxy property, using one bag of training data. \\
    $\hat{\mu}$ & The mean of $\mu$s predicted by all $N_{\rm{bags}}$ bags. \\ \tableline
    \end{tabular}
\end{table*}

\subsection{Performance Metrics}\label{sec:methods_metrics}
As we have discussed, for each input sample {\sc mirkwood} outputs three values: $\mu$, $\sigma_{a}$, and $\sigma_{e}$. As the same time, for each input, {\sc Prospector} outputs two values: $\mu$, and $\sigma$. We quantify the performance of each model by comparing the predicted galaxy properties against the true ones known from simulations. We divide this task into two parts -- comparing just the predicted mean \emph{deterministic predictions} of each galaxy property with its ground truth value, and judging the quality of the entire predicted PDF  (probability distribution function; \emph{probabilistic predictions}) by comparing it against the ground truth value.

\subsubsection{Metrics for Deterministic Predictions}\label{sec:metrics_det}
For determining the `goodness' of the predicted means for the galaxy properties, we adopt the following three performance criteria:
\begin{itemize}
    \item Normalized root-mean-square error (RMSE):
    \begin{align*}
        \mathrm{NRMSE}=\sqrt{\frac{1}{n} \sum_{i=1}^{n}\left(\hat{\mu}(i)-Y(i)\right)^{2}}
    \end{align*}
    \item Normalized mean absolute error (NMAE):
    \begin{align*}
        \mathrm{NMAE}=\frac{1}{n} \sum_{i=1}^{n}\left|\hat{\mu}(i)-Y(i)\right|
    \end{align*}
    %\item Normalized mean absolute percentage error (NMAPE):
    %\begin{align}
    %    \mathrm{NMAPE}=\frac{1}{n} \sum_{i=1}^{n} \frac{\left|\tilde{y}(i)-y(i)\right|}{\max _{i \in 1, \mathrm{n}]} y(i)} \times 100 \%,
    %\end{align}
    \item Normalized bias error:
    \begin{align*}
        \mathrm{NBE} = \frac{1}{n} \sum_{i=1}^{n} \hat{\mu}(i)-Y(i)
    \end{align*}
\end{itemize}
where $Y(i)$ and $\hat{\mu}(i)$ are the true and predicted values of any of the five galaxy properties for the $i^{th}$ sample, while $n$ is the total number of samples. A perfect prediction would result in NRMSE, NMAE, and NBE of 0. These values can be seen in the top-left plots in Figures \ref{fig:mass} through \ref{fig:sfr}.

\subsubsection{Metrics for Probabilistic Predictions}\label{sec:metrics_prob}
Probabilistic predictions are assessed by constructing Confidence Intervals (CI) and using them in evaluation criteria. In this work, we adopt average coverage area (ACE) and interval sharpness (IS) as our metrics. $\alpha$ is the significance level -- the probability that one will mistakenly reject the null hypothesis when in fact it is true. Specifically, since in this paper we are interested in $+/- 1\sigma$ predictions (CI=68.2\%), we choose $\alpha = 1-\rm{CI} = 0.318$. For {\sc mirkwood} results, $\sigma = \sqrt{\sigma_{a}^2 + \sigma_{e}^2}$, while for {\sc Prospector} results, there is only one $\sigma$ per prediction. For a given sample $i$, if the prediction of any galaxy property, from either {\sc mirkwood} or {\sc Prospector}, is \{$\mu(i)$; $\sigma(i)$\}, then the lower and upper confidence intervals  $L_\alpha(i)$ and $U_\alpha(i)$ are given by $\mu(i) - \sigma(i)$ and $\mu(i) + \sigma(i)$, respectively. $l_\alpha(i)$ and $u_\alpha(i)$ correspond to the $u-$values for these: $l_\alpha(i) = 0.5 - \rm{CI}/2 = 0.159$, $u_\alpha(i) = 0.5 + \rm{CI}/2 = 0.841$. $y(i)$ corresponds to the $u-$value of the ground truth value $Y(i)$. If the predicted mean $\hat{\mu}(i)$ for a galaxy property exactly matches $Y(i)$, then $y_\alpha(i)=0.5$; if the predicted mean overshoots or undershoots the ground truth value, then $y_\alpha(i) > 0.5$ or $y_\alpha(i) < 0.5$, respectively.

\begin{itemize}
    \item Average Coverage Error (ACE): %ACE is an indicator to appraise the reliability of prediction interval, which has the following equation:
    \begin{align*}
        \mathrm{ACE}_{\alpha}=\frac{1}{n} \sum_{i=1}^{n} c_{\alpha}(i) \times 100 \%-100 \times(1-\alpha) \%,
    \end{align*}
    where $c_i$ is the indicative function of coverage:
    \begin{align*}
        c_{\alpha}(i)=\left\{\begin{array}{l}
1, Y(i) \in\left[L_{\alpha}(i), U_{\alpha}(i)\right] \\
0, Y(i) \notin\left[L_{\alpha}(i), U_{\alpha}(i)\right]
\end{array}\right.
    \end{align*}
    ACE is effectively the ratio of target values falling within the confidence interval to the total number of predicted samples.
    \item Interval Sharpness (IS):
    \begin{align*}
        \mathrm{IS}_{\alpha}=\frac{1}{n} \sum_{t=1}^{n}\left\{\begin{array}{ll}
-2 \alpha\Delta_{\alpha}(i)\\-4\left[l_{\alpha}(i)-y(i)\right], & y(i)<l_{\alpha}(t) \\
-2 \alpha\Delta_{\alpha}(i)\\-4\left[y(i)-u_{\alpha}(i)\right], & y(i)>u_{\alpha}(t) \\
-2 \alpha\Delta_{\alpha}(i), & \mathrm{otherwise}\end{array}\right.
    \end{align*}
    where $\Delta_{\alpha}(i) = u_{\alpha}(i)-l_{\alpha}(i) = 0.682$. As a reminder, corresponding to $+/- 1\sigma$ intervals, $\alpha=0.318$, $l_{\alpha}(i)=0.159$, and $u_{\alpha}(i)=0.841$.
\end{itemize}

From the equation of ACE, a value close to zero denotes the high reliability of prediction interval. As an infinitely wide prediction interval is meaningless, IS is another indicator contrary to the coverage rate of interval, which measures the accuracy of probabilistic forecasting. $\mathrm{IS}_{\alpha}$ is always $<0$; a great prediction interval obtains high reliability with a narrow width, which has a small absolute value of IS. 

\begin{figure*}
    \centering
    \includegraphics[width=0.98\textwidth]{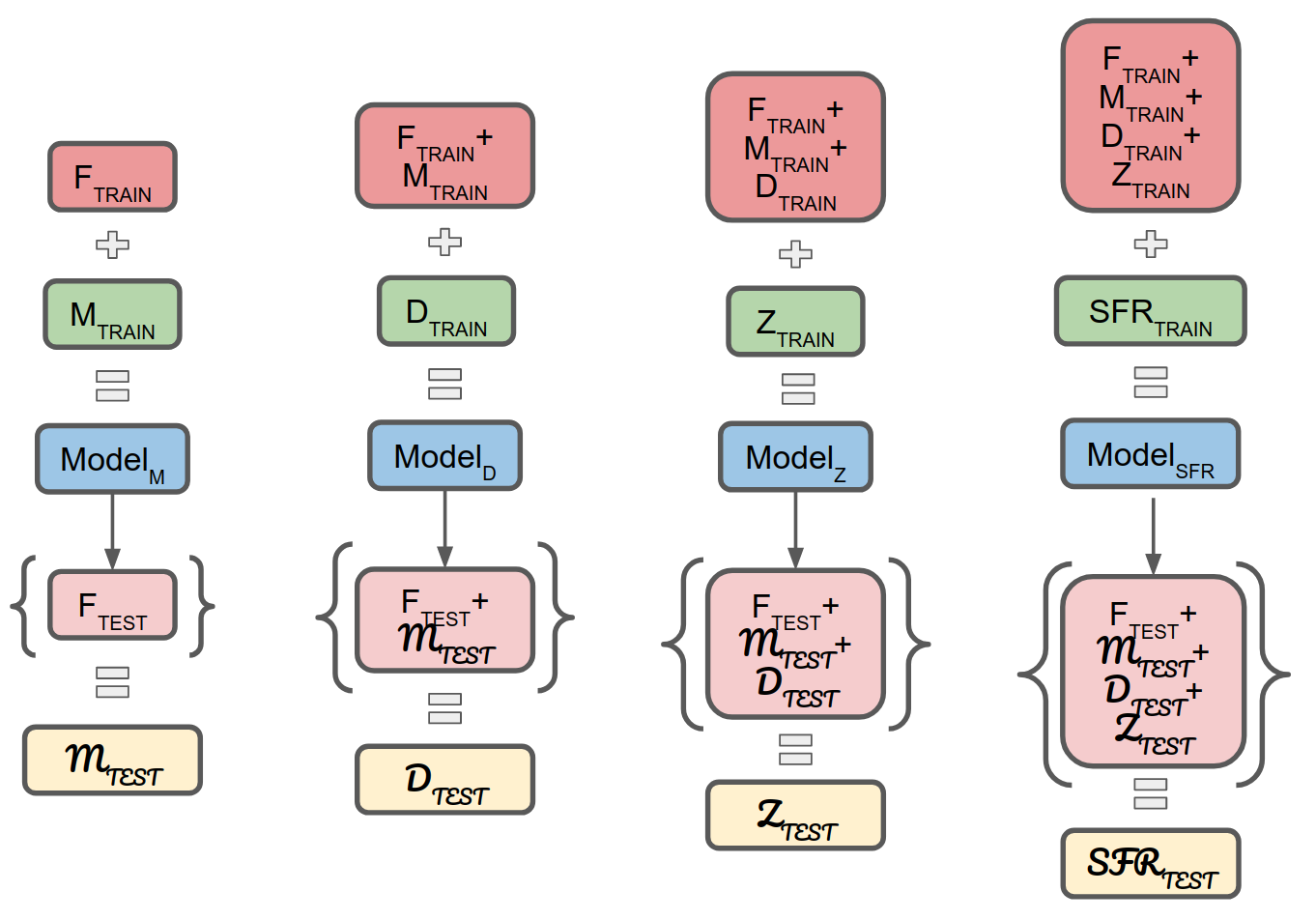}%chaining.pdf
    \caption{Chained multi-output regression. From left to right, we train models to sequentially predict galaxy stellar mass, dust mass, stellar metallicity, and SFR$_{100}$. At the first step, we train {\sc mirkwood} on fluxes (as input) and stellar mass (as output) in the training set $\mathcal{D_{TRAIN}}$, and use it to predict stellar mass for samples in the test set ($\mathcal{D_{TEST}}$) using their respective flux values as inputs. In the second step, we append the true stellar masses of samples in $\mathcal{D_{TRAIN}}$ with their fluxes, and append the predicted stellar masses of the samples in $\mathcal{D_{TEST}}$ to theirs. A new model is then trained on the extended training set, and used to predict dust mass mass for samples in the test set. We continue thus until all five properties have been successfully predicted for all samples in the original $\mathcal{D_{TEST}}$.}
    \label{fig:chaining}
\end{figure*}

\section{Results}\label{sec:results}

\begin{figure*}%[!htbp]
\gridline{\leftfig{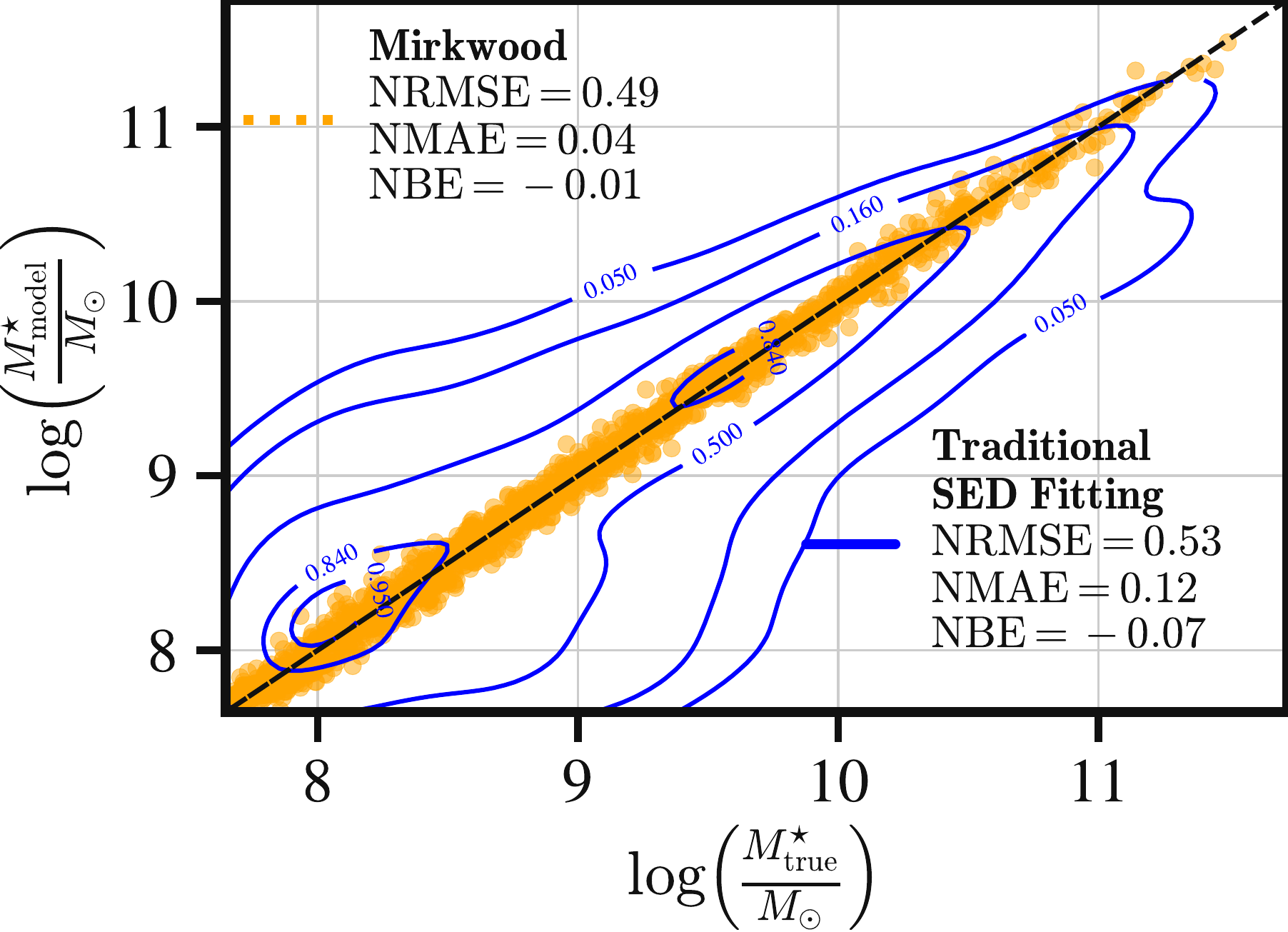}{0.48\textwidth}{(a)}
          \rightfig{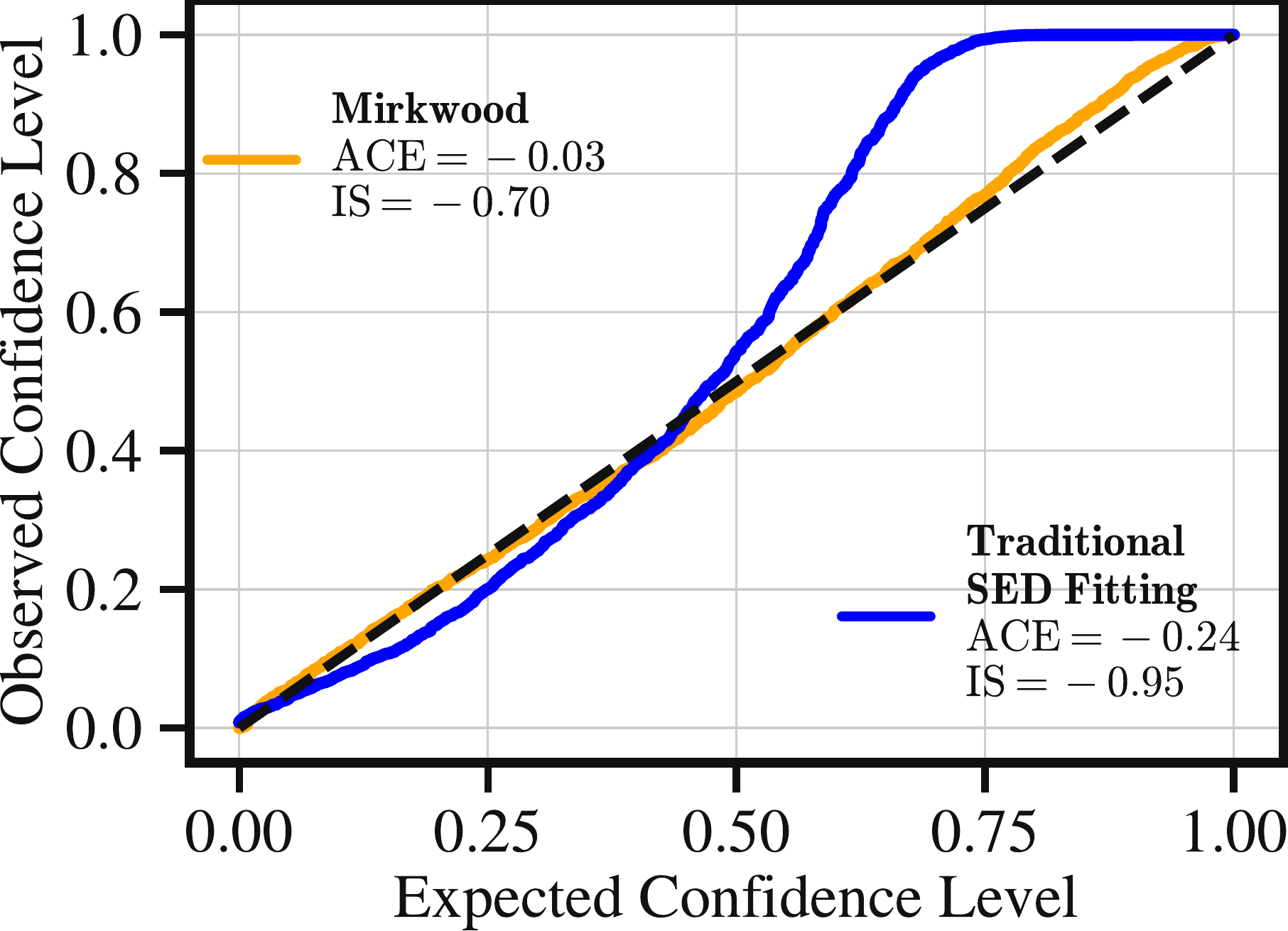}{0.48\textwidth}{(b)}
          }
%\gridline{\fig{SHAP_Mass_snr=5_20201218.pdf}{0.98\textwidth}{(c)}}
\gridline{\leftfig{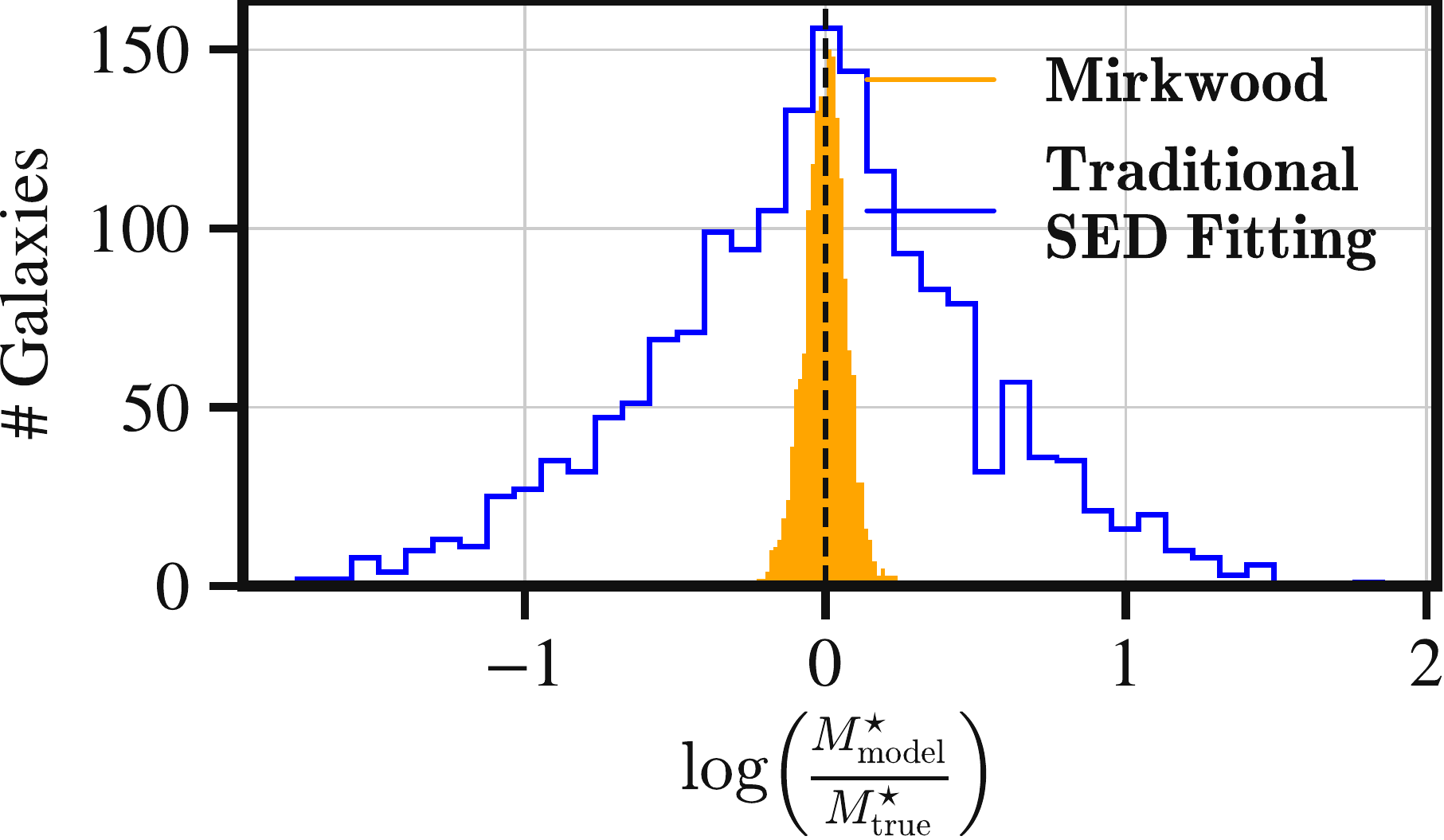}{0.48\textwidth}{(c)}
          \rightfig{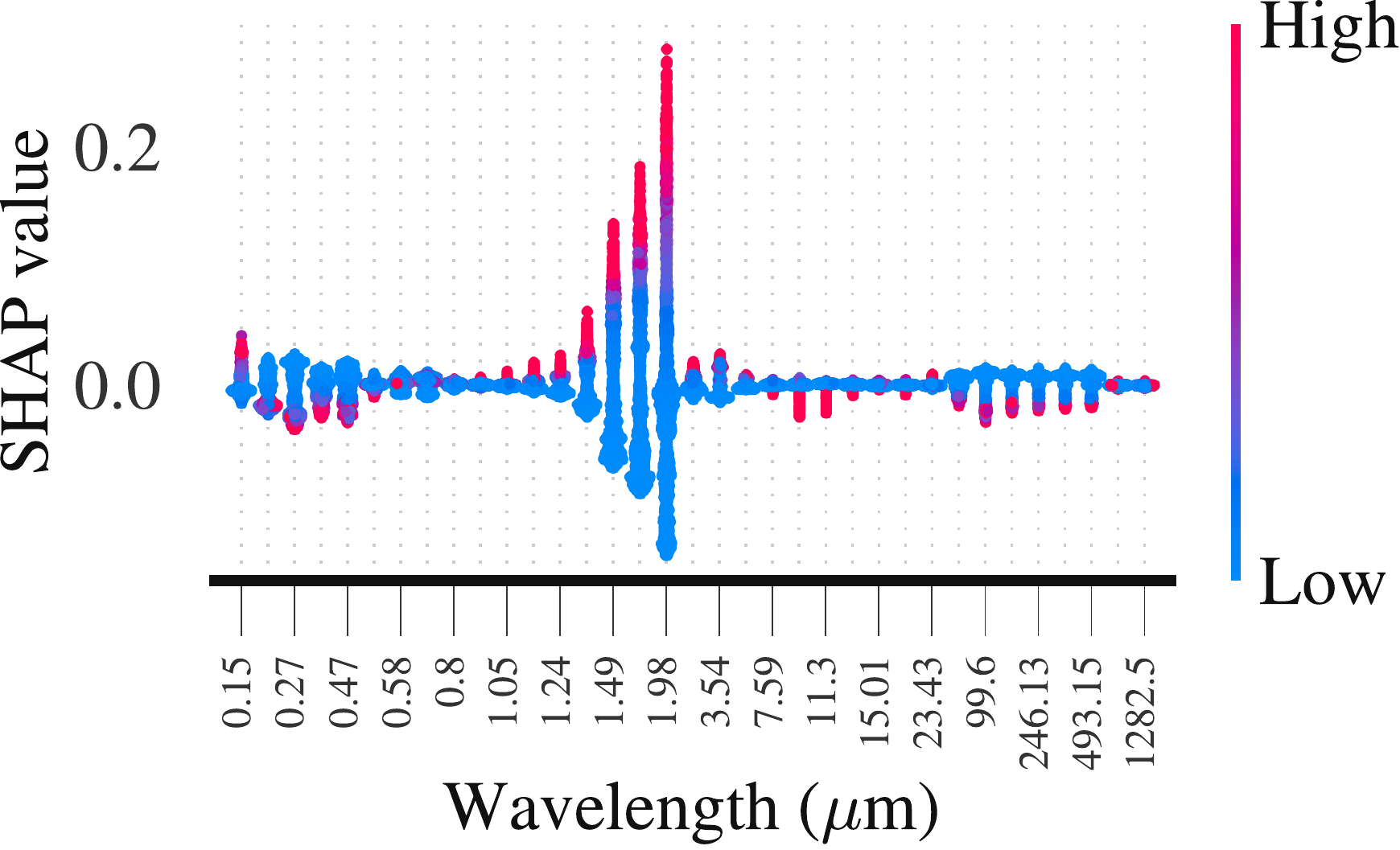}{0.48\textwidth}{(d)}
          }
\caption{Comparison of predicted v/s true stellar mass -- in terms of both means and uncertainties -- for samples from the {\sc Simba} simulation. \textbf{(a), (c) Predicted mean stellar mass from {\sc mirkwood} v/s {\sc Prospector}:} We can see both qualitatively and quantitatively (via the three metrics normalized root mean square error, normalized mean absolute error, and normalized bias error) that predictions from {\sc mirkwood} are a significant step-up from traditional SED-based results. As a reminder, the smaller the three metrics, the better the predictions. For clearer visualization, we have used scatter points for results from {\sc mirkwood}, and contour lines for results from {\sc Prospector}. In subplot \textbf{(c)}, we employ histograms to demonstrate the radically different bias and variance in predictions from the two models. \textbf{(b) Predicted aleatoric uncertainty in stellar mass from {\sc mirkwood} v/s {\sc Prospector}:} We compare the predicted aleatoric uncertainty from {\sc mirkwood} with the aleatoric uncertainty from traditional SED fitting. Both qualitatively and quantitatively -- by employing the metrics average coverage error, and interval sharpness -- we see that predictions from {\sc mirkwood} are superior. As explained in Section \ref{sec:metrics_prob}, these probabilistic metrics such as ACE and IS are essential to evaluate predicted uncertainties, something that is outside the purview of traditional deterministic metrics such as those used in subplot \textbf{(a)} and expounded upon in Section \ref{sec:metrics_det}. \textbf{(d) Importance of different bands in predicting stellar mass from {\sc mirkwood}:} SHAP values are a way to obtain contribution of each feature (predictive variable) in predicting a model's output. They measure how reactive the output is to changes in the input, and thus they can be thought of as a type of sensitivity measure; see Section \ref{sec:results} and Appendix \ref{sec:shapley_values} for detailed discussion. Here we see that K and H bands are the most important predictors of stellar mass, with higher flux values (red color) corresponding to higher mass values. \label{fig:mass}}
\end{figure*}

\begin{figure*}%[!]
\gridline{\leftfig{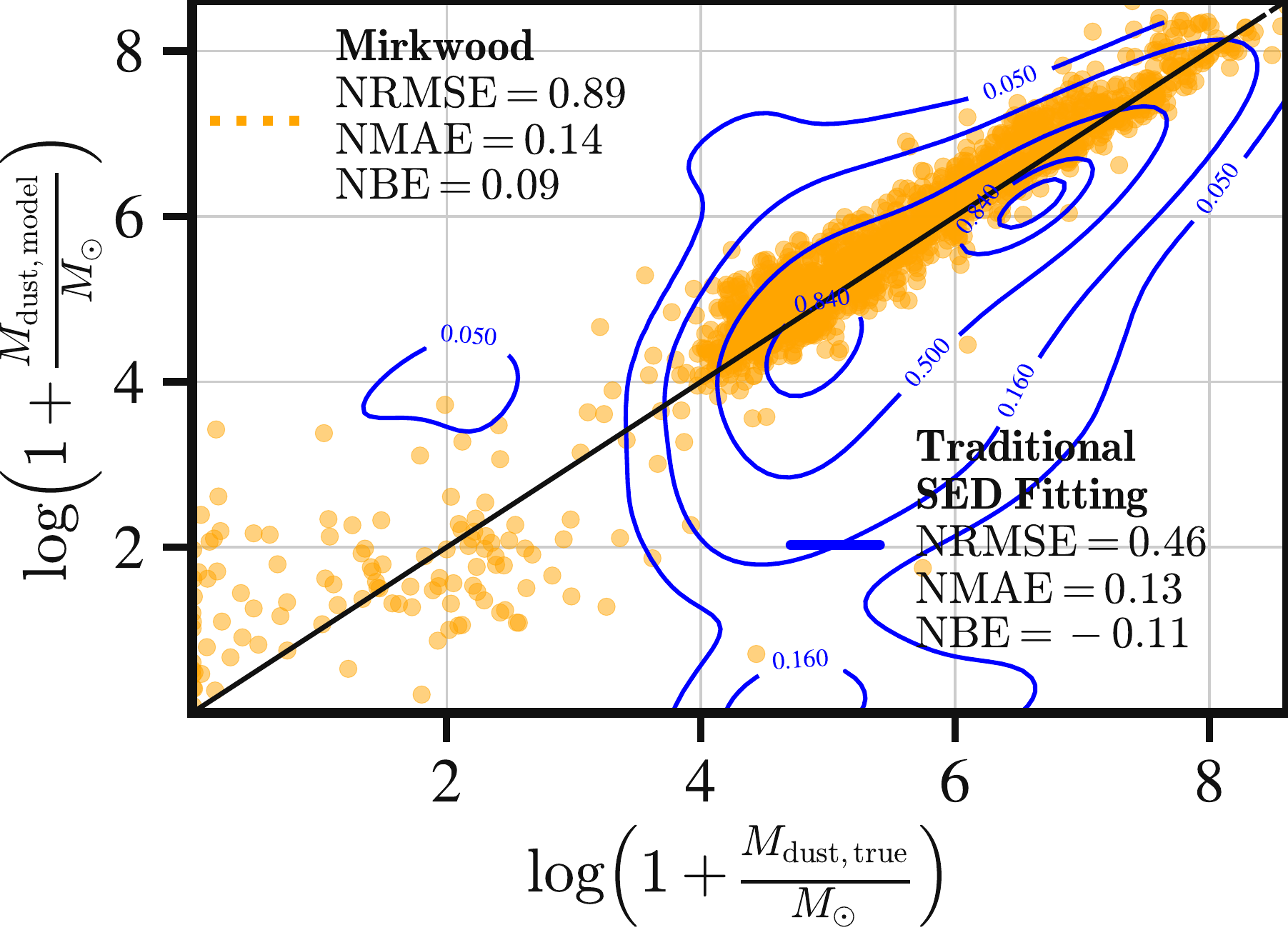}{0.48\textwidth}{(a)}
          \rightfig{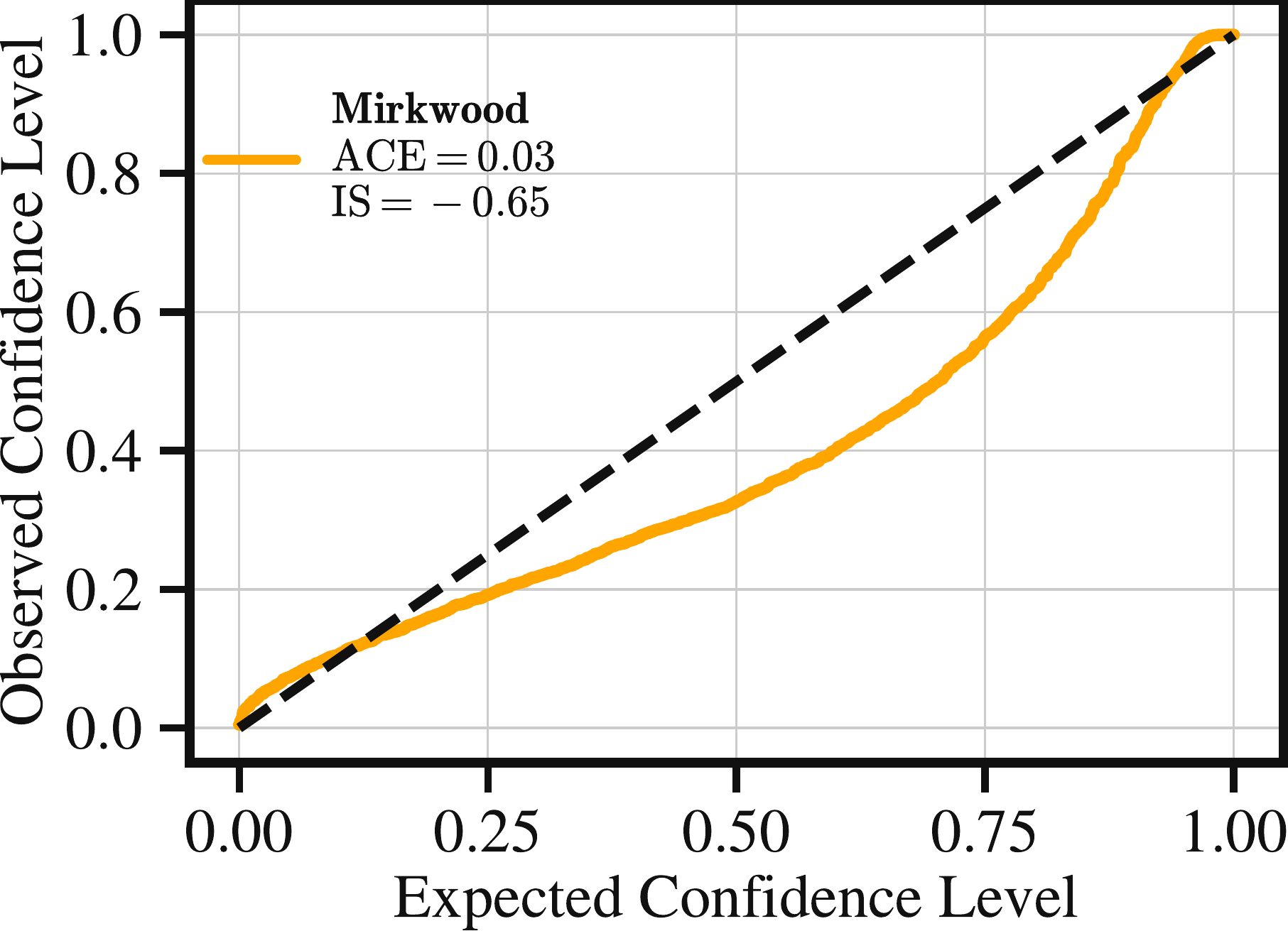}{0.48\textwidth}{(b)}
          }
\gridline{\leftfig{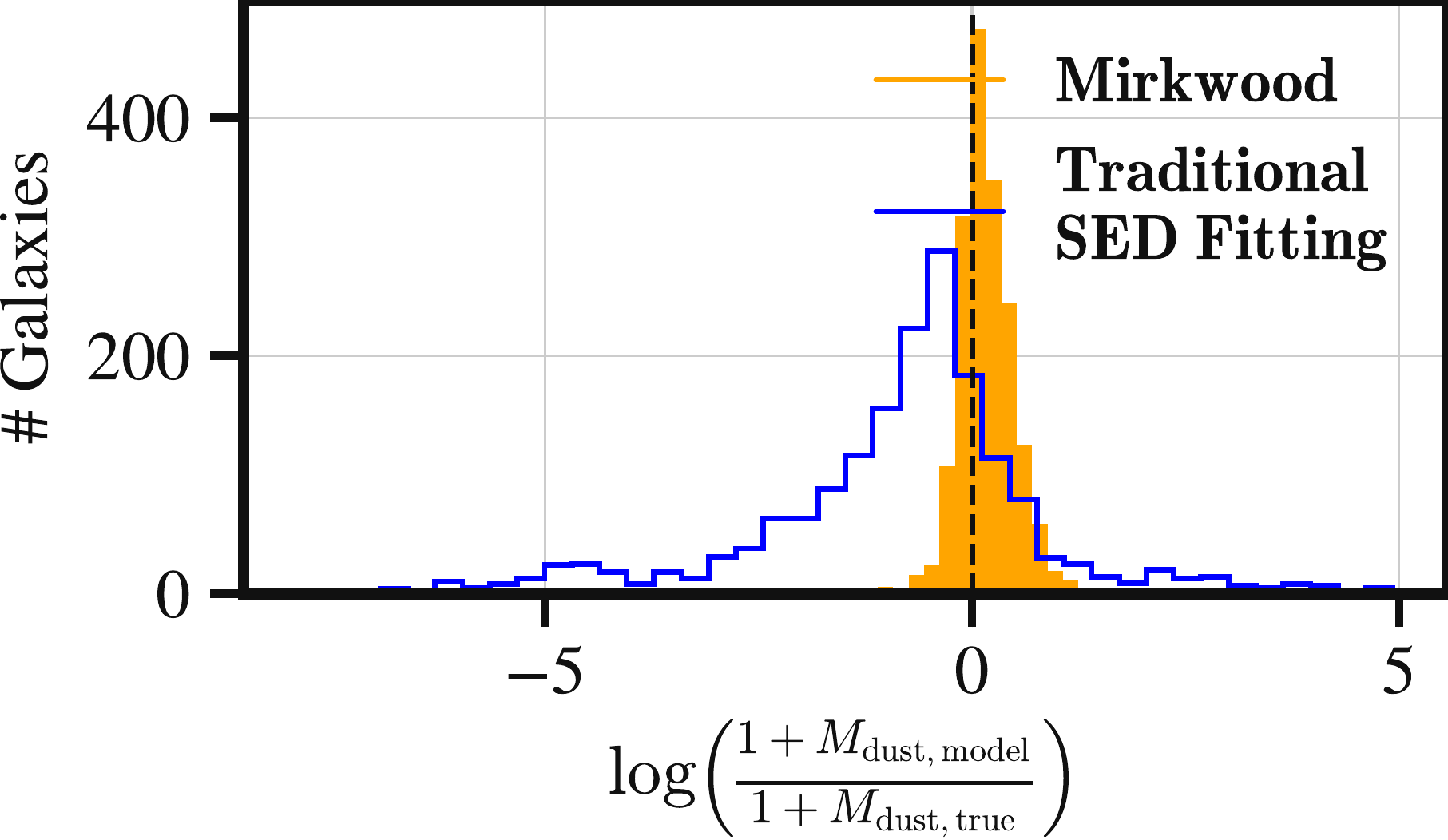}{0.48\textwidth}{(c)}
          \rightfig{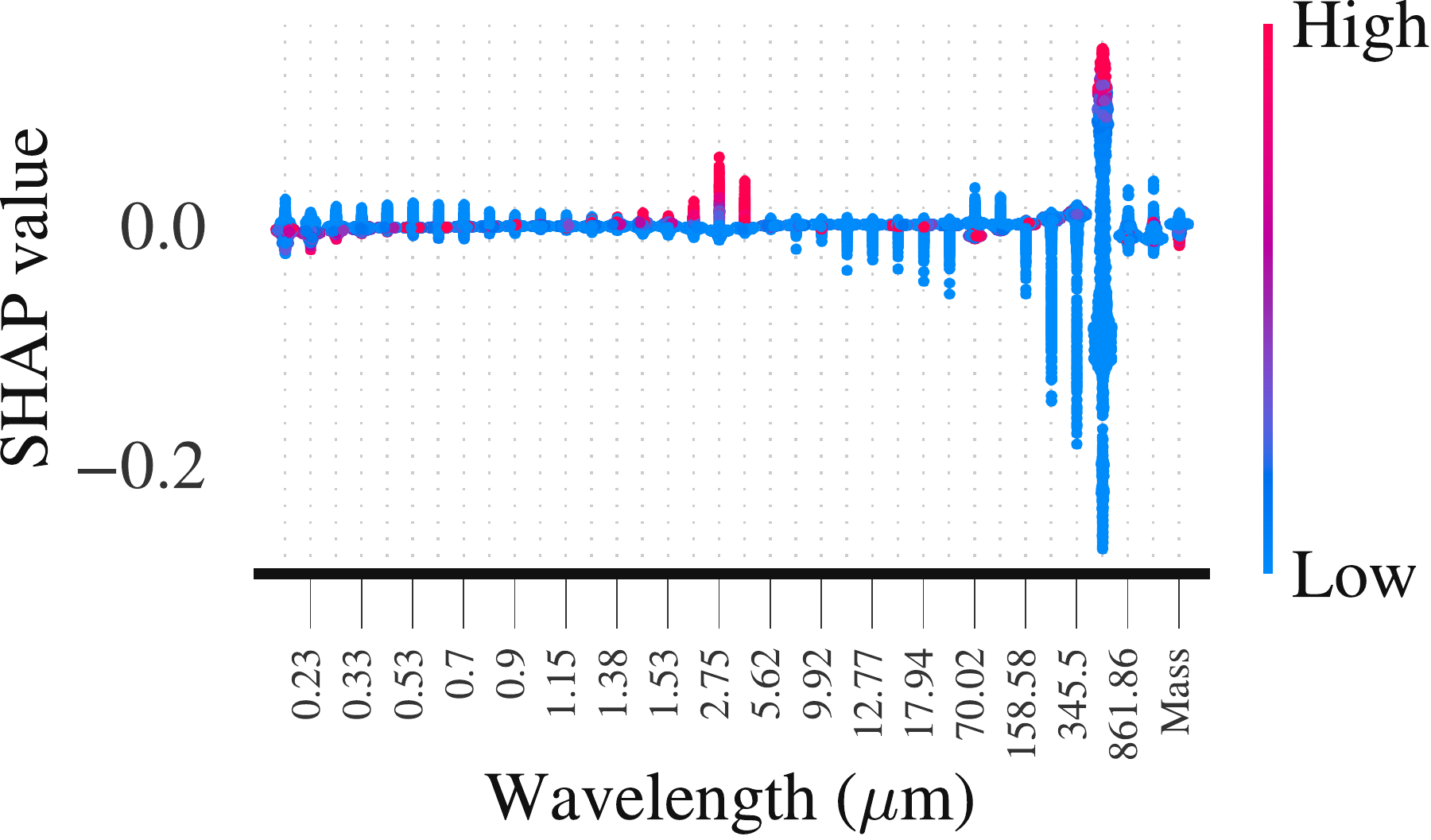}{0.48\textwidth}{(d)}
          }
%\gridline{\leftfig{Epis_z0_Dust_snr=5_20201218.pdf}{0.48\textwidth}{(e)}
%          \rightfig{Epis_z2_Dust_snr=5_20201218.pdf}{0.48\textwidth}{(f)}
%          }
\caption{Very similar to Figure \ref{fig:mass}, this figure compares predicted v/s true dust mass -- in terms of both means and uncertainties -- for samples from the {\sc Simba} simulation. \textbf{(a), (c) Predicted dust mass from {\sc mirkwood} v/s {\sc Prospector}:} Just like in Figure \ref{fig:mass}, we see that predictions from {\sc mirkwood} are a significant step-up from traditional SED-based results. \textbf{(b) Predicted aleatoric uncertainty in dist mass from {\sc mirkwood}:} Unlike in Figure \ref{fig:mass}, we only plot the calibration curve for aleatoric uncertainties from {\sc mirkwood}, since in this work we do not have access to uncertainties in dust mass from {\sc Prospector}. \textbf{(d) Importance of different bands in predicting dust mass from {\sc mirkwood}:} We see that the NIR bands are the most predictive of dust mass, with only a tiny contribution from predicted mass. This tracks with our understanding of dust mass properties. \label{fig:dustmass}}
\end{figure*}

\begin{figure*}%[!htbp]
\gridline{\leftfig{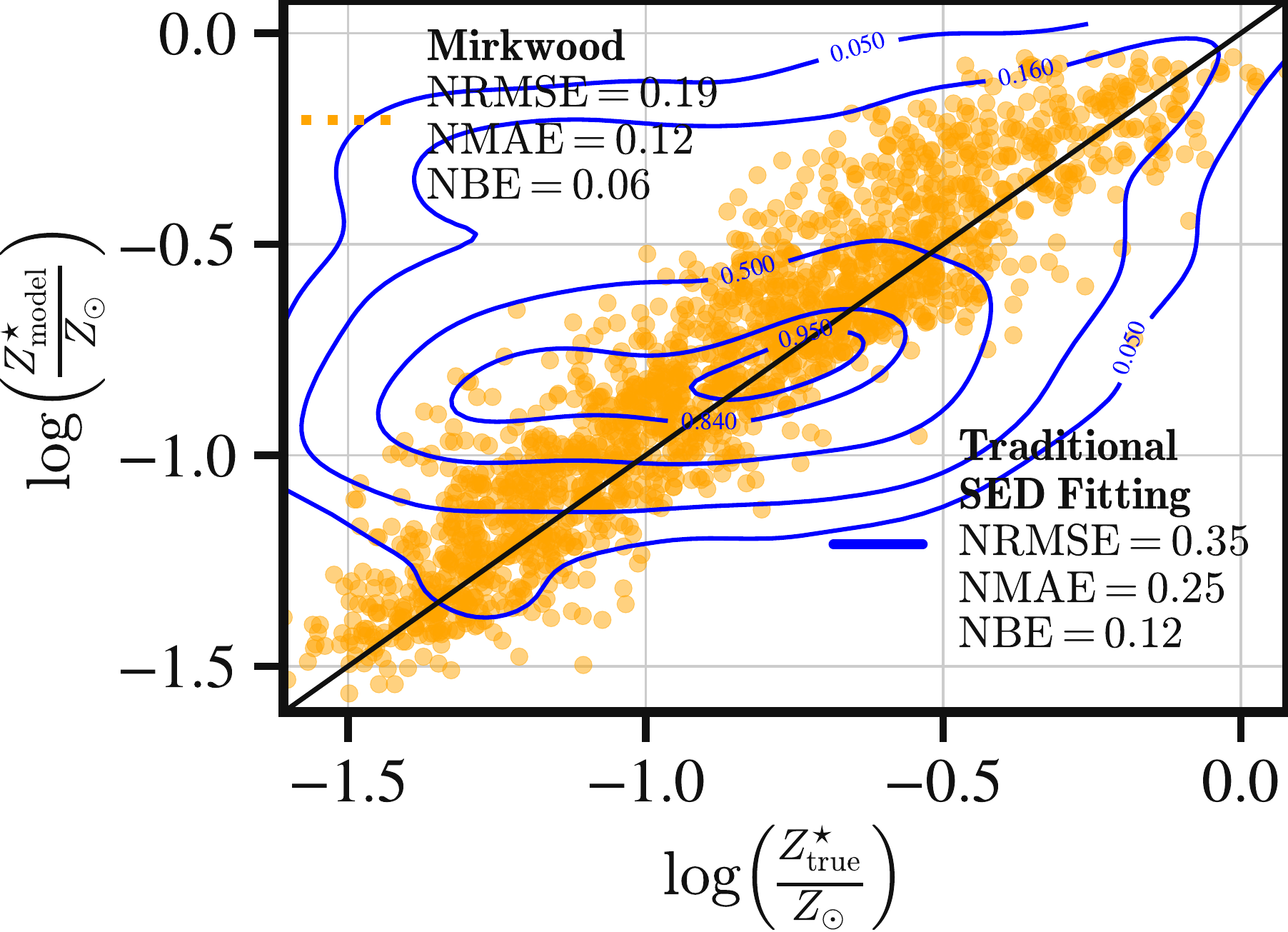}{0.48\textwidth}{(a)}
          \rightfig{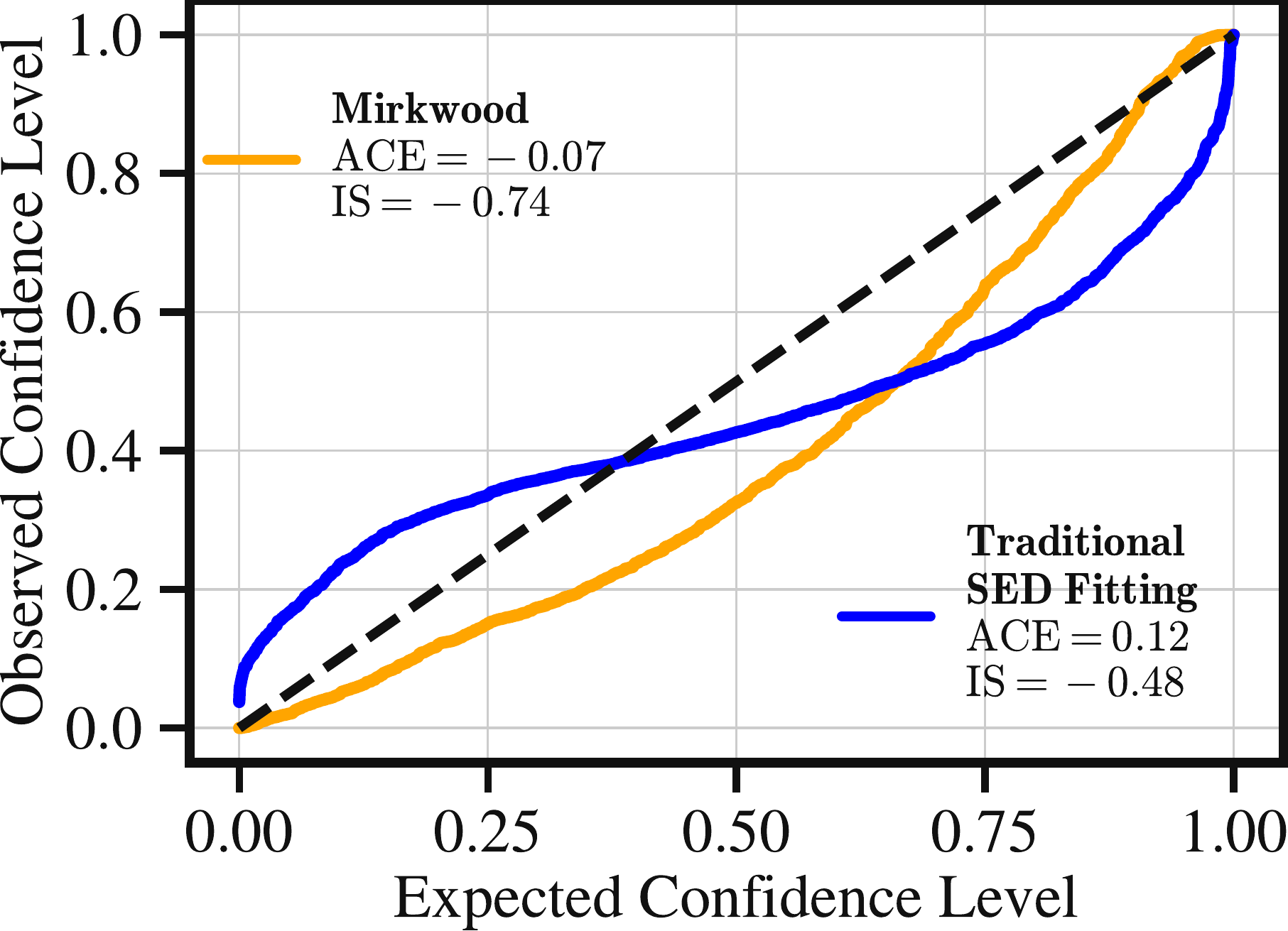}{0.48\textwidth}{(b)}
          }
\gridline{\leftfig{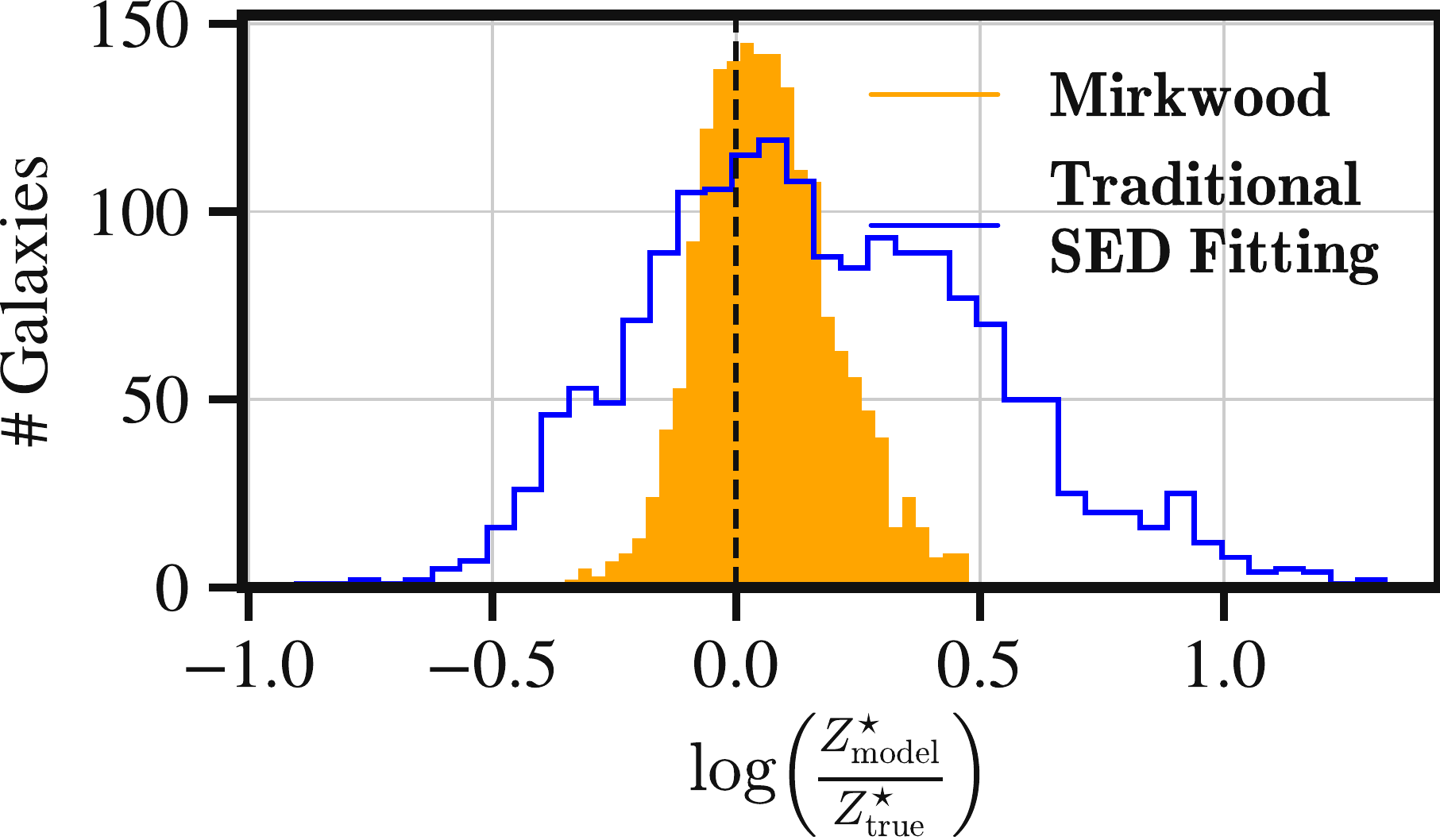}{0.48\textwidth}{(c)}
          \rightfig{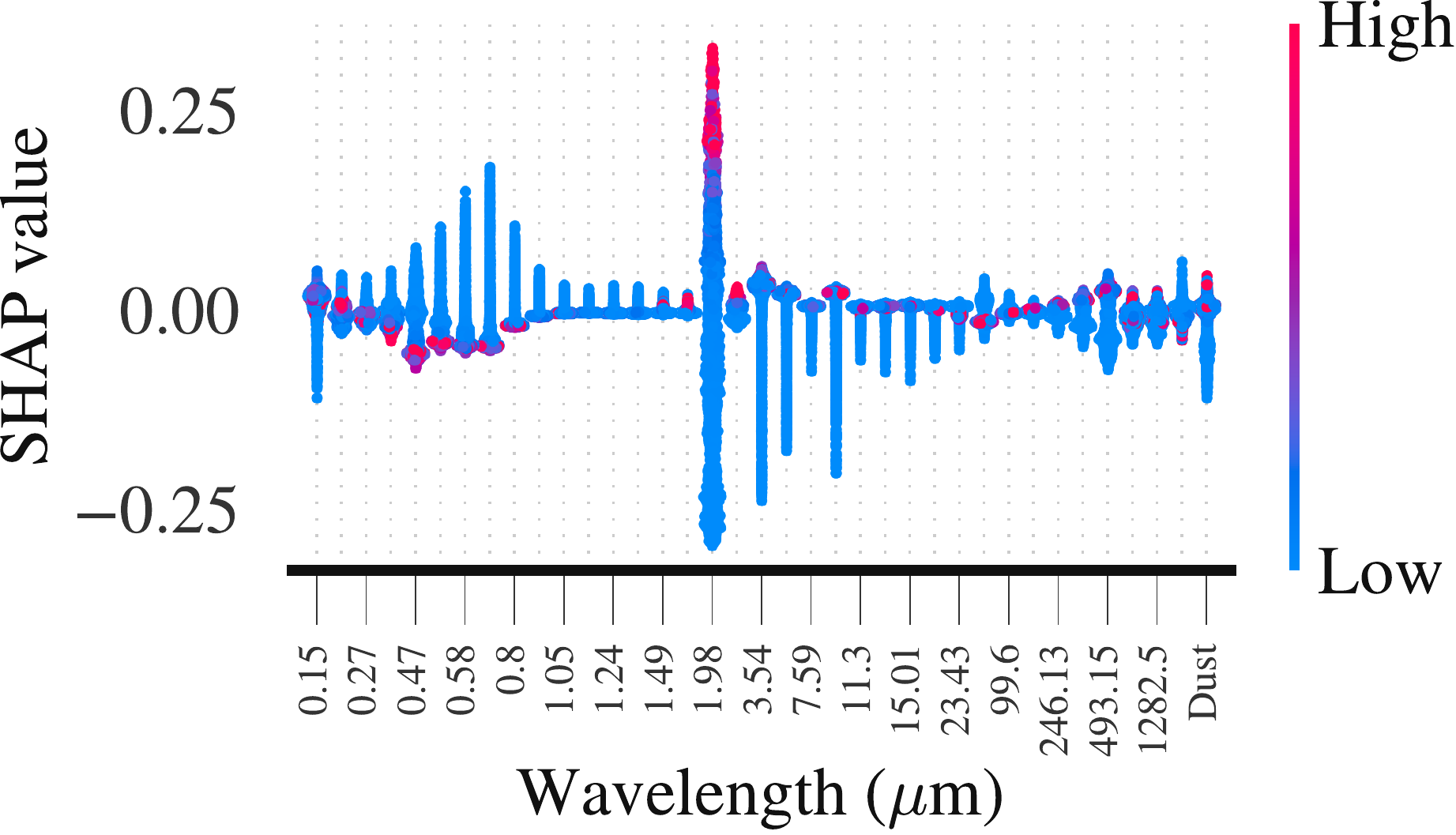}{0.48\textwidth}{(d)}
          }
%\gridline{\leftfig{Epis_z0_Z_snr=5_20201218.pdf}{0.48\textwidth}{(e)}
%          \rightfig{Epis_z2_Z_snr=5_20201218.pdf}{0.48\textwidth}{(f)}
%          }
\caption{Same as Figures \ref{fig:mass} and \ref{fig:dustmass}, but for instantaneous star formation rate metallicity Z. \label{fig:metallicity}}
\end{figure*}

\begin{figure*}[!tbp]
\gridline{\leftfig{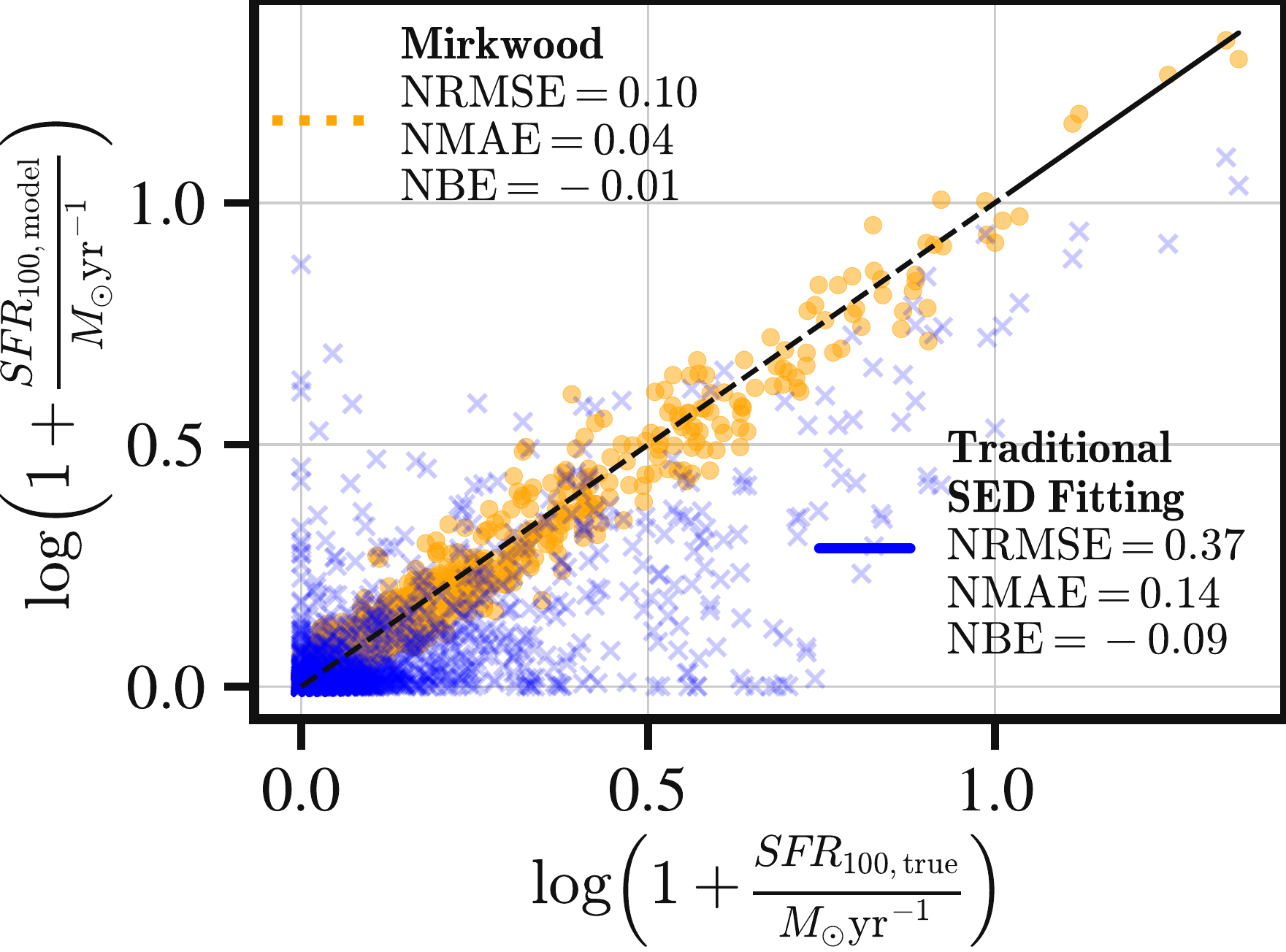}{0.48\textwidth}{(a)}
          \rightfig{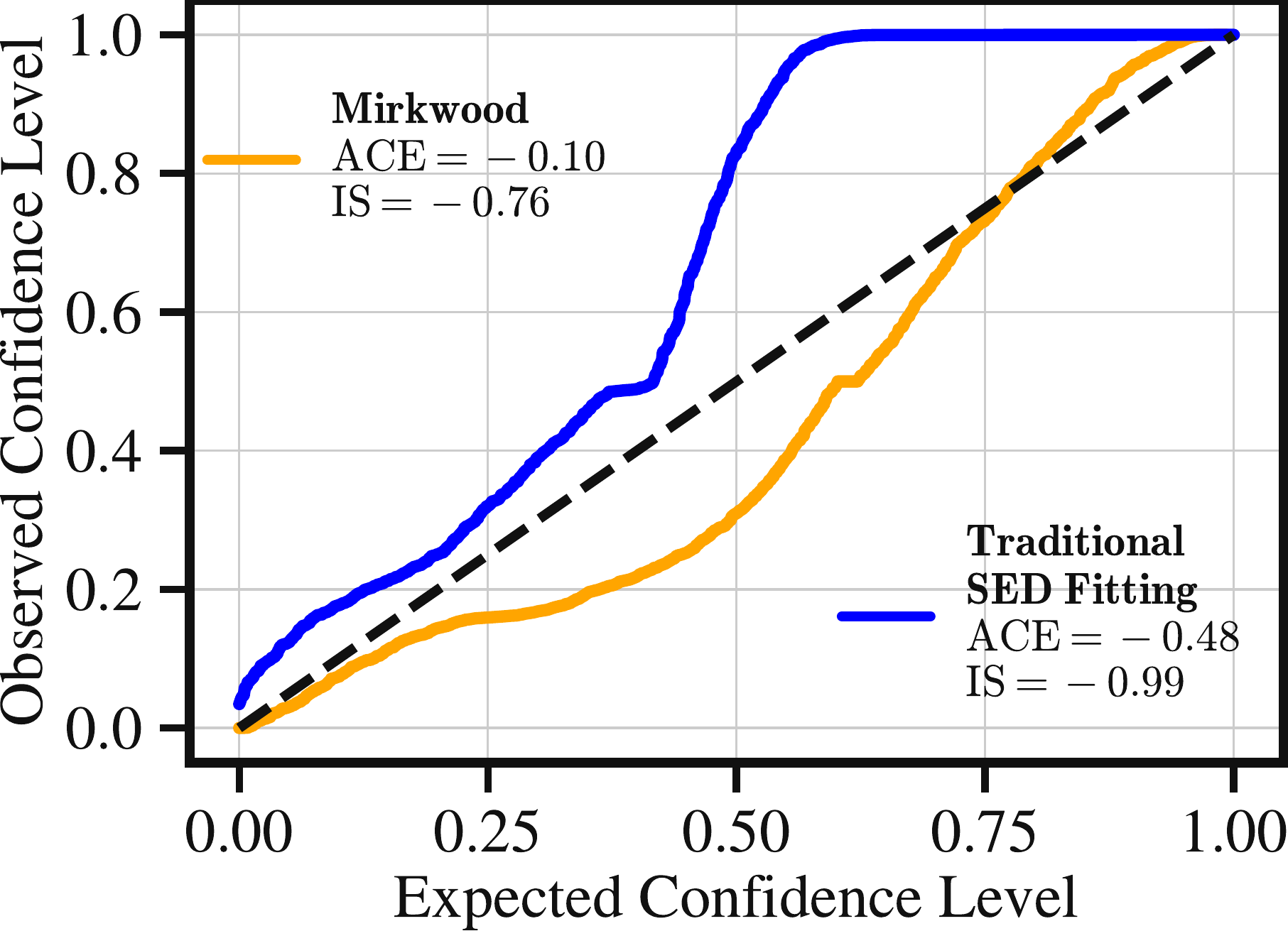}{0.48\textwidth}{(b)}
          }
\gridline{\leftfig{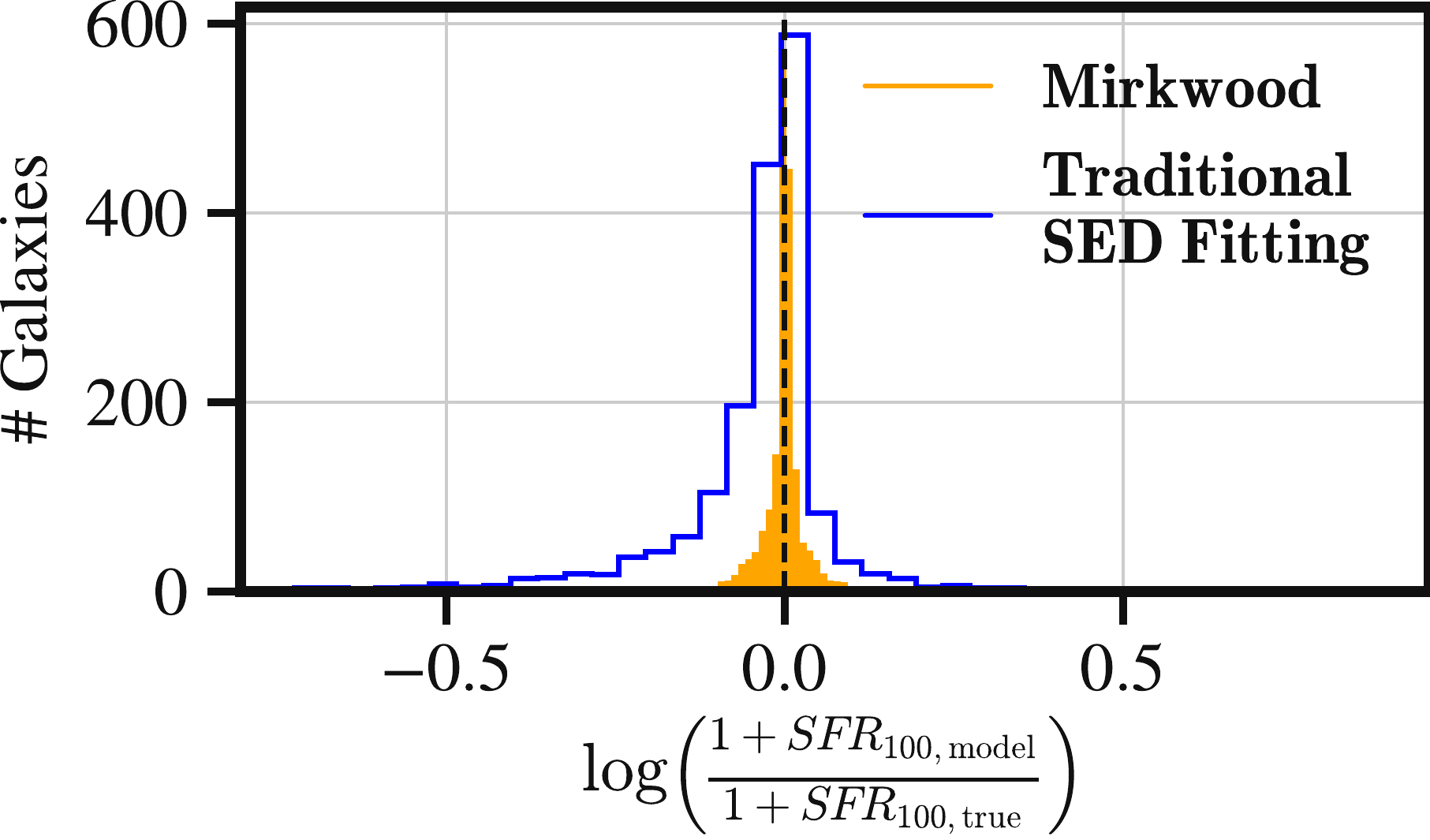}{0.48\textwidth}{(c)}
          \rightfig{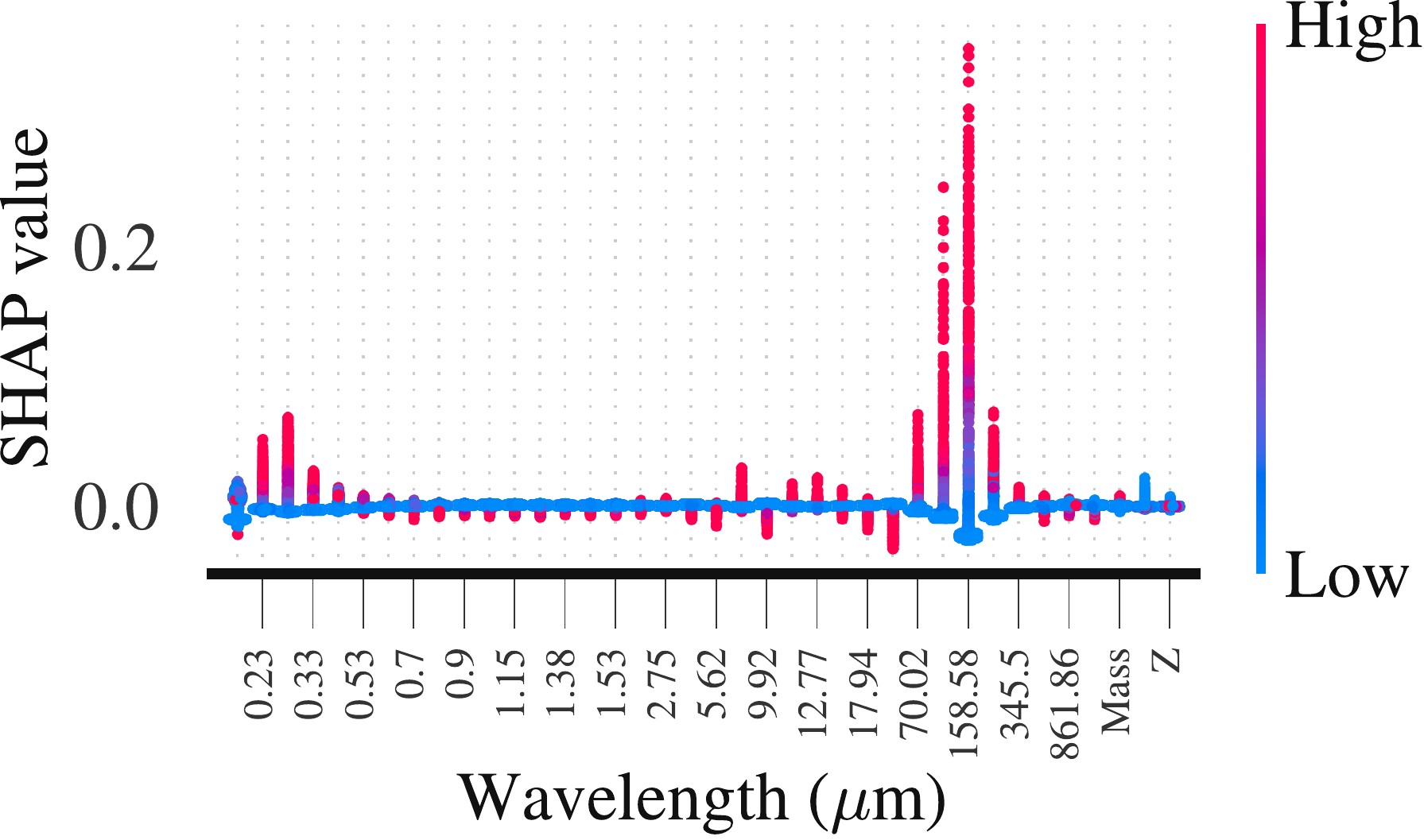}{0.48\textwidth}{(d)}
          }
%\gridline{\leftfig{Epis_z0_SFR_snr=5_20201218.pdf}{0.48\textwidth}{(e)}
%          \rightfig{Epis_z2_SFR_snr=5_20201218.pdf}{0.48\textwidth}{(f)}
%          }
\caption{Same as Figures \ref{fig:mass}, \ref{fig:dustmass}, and \ref{fig:metallicity} but for instantaneous star formation rate SFR$_{100}$. \label{fig:sfr}}
\end{figure*}

\iffalse
\begin{figure*}[!htb]
\gridline{\fig{Barplot_Mass_snr=5_20201101.pdf}{0.98\textwidth}{(a)}}
\gridline{\fig{Barplot_Mass_snr=5_20201101.pdf}{0.98\textwidth}{(b)}}
\gridline{\fig{Barplot_Mass_snr=5_20201101.pdf}{0.98\textwidth}{(c)}}
\gridline{\fig{Barplot_Mass_snr=5_20201101.pdf}{0.98\textwidth}{(d)}}
\caption{b}
\label{fig:stackedbarplots}
\end{figure*}
\fi

In order to test our proposed model for SED fitting, we compare against fits from the Bayesian SED fitting software {\sc prospector}.  {\sc prospector} and {\sc mirkwood} are given the same information in order to infer galaxy properties. This includes broadband photometry in $35$ bands with 5\%, 10\%, and 20\% Gaussian uncertainties (respectively SNRs = 20, 10, and 5). We present here the results from both methods for several galaxy properties including stellar mass, dust mass, star formation rate averaged over the last 100 Myr (SFR$_{100}$), and stellar metallicity.

In Figures \ref{fig:mass} through \ref{fig:sfr}, we present {\sc mirkwood's} predictions when trained on the full dataset comprising of simulations from {\sc Simba}, {\sc Eagle}, and {\sc IllustrisTNG}, for SNR=5\footnote{Corresponding plots for SNR=10 and SNR=20, along with our complete code, are available online at \url{https://github.com/astrogilda/mirkwood}}. The training set contains all 10,073 samples from {\sc IllustrisTNG}, 4,697 samples from {\sc Eagle}, and 359 samples from {\sc Simba} selected by stratified 5-fold cross-validation (see Section \ref{sec:methods_proposed_method} for details on implementation), while the test set contains the remaining 1,438 samples from {\sc Simba}. After making inference on all test splits, we collate the results, thus successfully predicting all four galaxy properties for all 1,797 samples from {\sc Simba}. Each predicted output for a physical property contains three values -- the mean $\mu$, the aleatoric uncertainty $\sigma_{\rm a}$, and the epistemic uncertainty $\sigma_{\rm e}$. The first type of uncertainty quantifies the inherent noise in the data, while the second quantifies the confidence of the model in the training data (see Section \ref{sec:methods_uncertainty} for detailed description). The results from traditional SED fitting are a subset of the total training data.% ; we only fit samples from {\sc Simba} due to computational constraints. Thus this dataset consists of a total of 12,287 samples, out of which (\red{MISSING NUMBER HERE}) are from {\sc Simba}. %14,258 samples -- 12,287 from all three simulations at z=0, and the rest from {\sc Simba} at z=2.
We present a brief overview of the data sets in Table \ref{tab:datasetsummary}.

In subplots (a) of Figures \ref{fig:mass} through \ref{fig:sfr}, we plot the recovered physical properties from {\sc mirkwood} and {\sc prospector} against the true values from the simulations for each of the four modeled properties: stellar mass, dust mass, metallicity, and star formation rate. In each case, the blue contours represent the physical properties derived from traditional SED fitting, while the orange points denote the results from {\sc mirkwood}. Without exception, {\sc mirkwood} provides superior predictions, as evidenced both by a visual inspection, and via a more quantitative approach by using three different metrics (detailed in Section \ref{sec:methods_metrics}). 

%\red{I know you're trying to be concise here, but you probably need to explain this to a dolt astronomer who knows very little about ML [like me...] and walk the reader through carefully.  There's no point in plotting it if an astronomer can't understand it (and, I don't think from this explanation they will be able to).  Remember the audience is not ML: it's an OIR observer who may switch from SED fitting to this model.}
In subplots(b), we plot `calibration diagrams' \citep{crude_probability_calibration}. %\red{Desika leaving off here please leave this note so i know where to re-join}
On the y-axis are the inverse-CDFs (cumulative distribution functions) for predictions from both {\sc mirkwood} and {\sc Prospector}. These are commonly used in ML literature -- the plotting function takes as input a vector consisting of the ground truth value, predicted mean, and predicted standard deviation, and outputs a value between 0 and 1 (see \cite{probability_calibration1} for detailed instructions on constructing these diagrams). We repeat this for all samples in the test set $\mathcal{D_{TEST}}$, sort the resulting vector in ascending order, and plot the curve on the y-axis. For {\sc mirkwood}, the uncertainty used is $\sigma_{\rm a}$, while {\sc Prospector} does not differentiate between the two types of uncertainties and outputs a single value per galaxy sample. A 1:1 line on this curve indicates a perfectly calibrated set of predictions, with deviations farther from it indicative of poorer calibration. Both visually and quantitatively by using the two probabilistic metric (Section \ref{sec:methods_metrics}) average coverage error and interval sharpness, we ascertain that {\sc mirkwood} outperforms traditional SED fitting.

In subplots (c) and (d), we plot the histograms for predicted means, and SHAP summary plots for samples in $\mathcal{D_{TEST}}$, respectively. %, for redshifts 0 and 2 respectively.
The histograms in subpolots (c) enable us to visualize both bias and variance in predictions from {\sc mirkwood}, and compare the results to those from {\sc Prospector}. As is readily apparent, both these quantities are significantly smaller for predictions from our ML-based model than from traditional SED fitting, for all four galaxy properties.
We explain in detail how SHAP values are calculated in Section \ref{sec:shapley_values}. In brief, for a given filter (dotted vertical lines in subplots (d)), an absolute value farther from 0 denotes a larger impact of that feature in affecting the model outcome. SHAP values can thus be thought of as a type of sensitivity -- it controls how reactive the output is to changes in the input (here flux in the filter). Red and blue indicate high and low ends of values per feature/filter. Thus for a given feature, SHAP values $> 0$ with a red hue indicate that the galaxy property increases as that feature increases. Similarly, SHAP values $> 0$ with a blue hue indicate that the galaxy property decreases as the feature value decreases. As an example, in panel (d) of Figure~\ref{fig:mass}, the largest correlations with the $M^{\star}$ of a galaxy are, quite naturally, the $\sim 1-2 \mu$m range.% \red{desika leaving off here}

%In subplots (e) and (f), we plot epistemic errors from {\sc mirkwood} against the most informative feature derived from their respective subplots (c) and (d). Over-plotted on these are the scaled histograms for the same feature, from the training and test sets, at these two redshifts\footnote{To be able to plot values differing by orders of magnitude on the same plot, we convert both the $x$ and $y$ axis to log-scale; the addition of the unit vector ensures that small values remain close to 0 after conversion and do not turn negative.}. $\sigma_{\rm e}$ informs us of the degree of confidence the model has in its predictions in a given part of the parameter space; by plotting it against the most informative filter derived from the SHAP plot, we can get a first order visual insight into the model's decision making process. In all Figures from \ref{fig:mass} through \ref{fig:sfr}, we see a general trend that epistemic uncertainties increase when when the informative features become sparse, just as we would intuitively expect.

In Table \ref{tab:stackedbartable}, we delineate all five metrics from Section \ref{sec:methods_metrics} to quantitatively compare {\sc mirkwood}'s performance with that of traditional SED fitting, across different levels of noise. As we would expect, the results from both models improve with increasing SNR. Across every metric and galaxy property, {\sc mirkwood} demonstrates excellent performance. We also note that the difference in their relative performances increases with increasing noise level. Below a certain SNR, traditional SED fitting stops producing meaningful results (as can also be seen clearly from subplots (a) in Figures \ref{fig:mass} through \ref{fig:sfr}
), whereas our ML-based approach is still able to extract signal from the noisy data.

\begin{figure*}%[!htbp]
\gridline{\fig{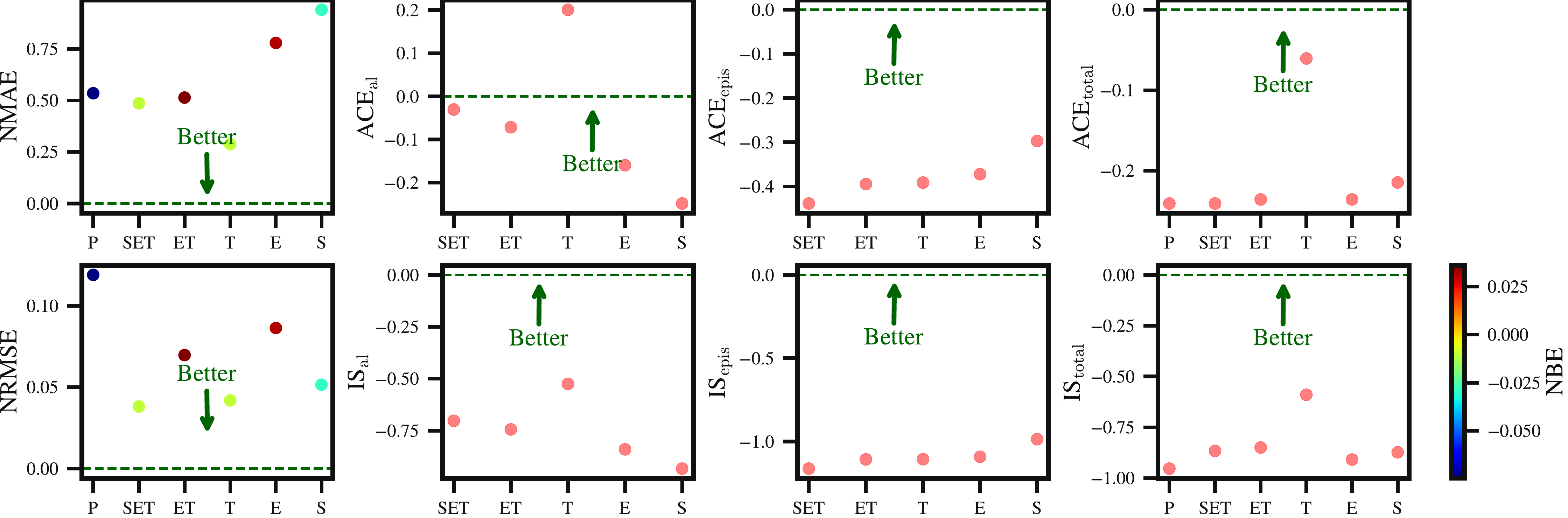}{0.88\textwidth}{(a) SNR = 5}}
\gridline{\fig{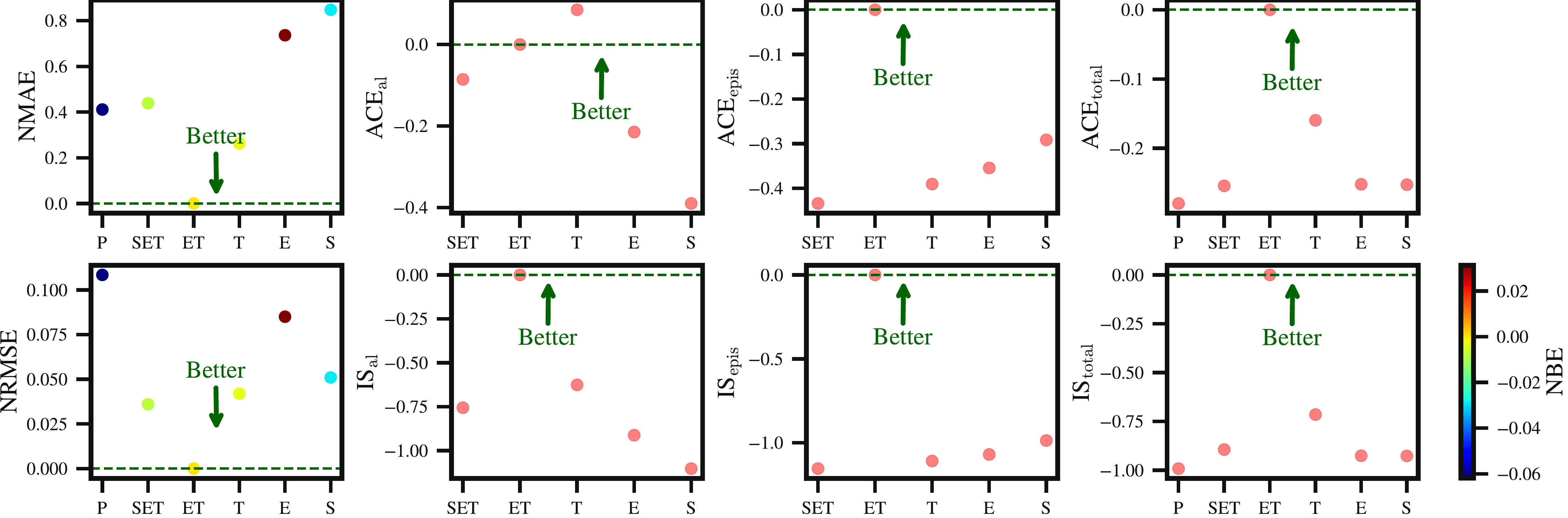}{0.88\textwidth}{(b) SNR = 10}}
\gridline{\fig{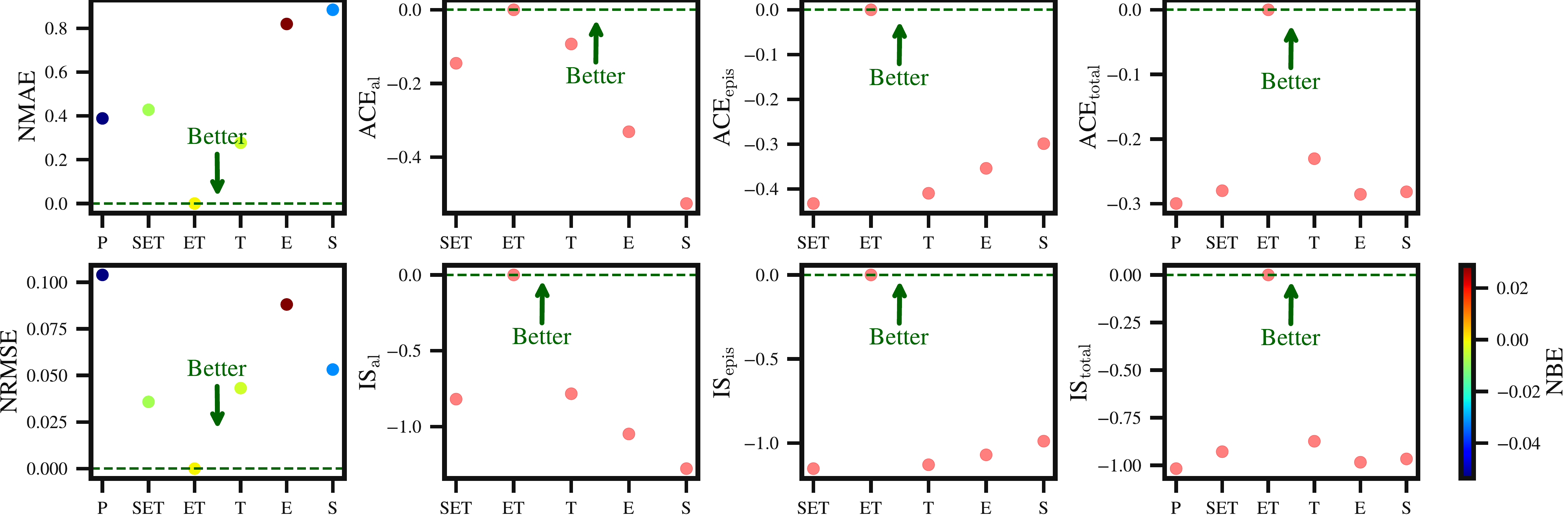}{0.88\textwidth}{(c) SNR = 20}}
\caption{\textbf{Impact of training set on {\sc mirkwood}'s performance on predicting galactic stellar mass:} Using the three deterministic metrics %(normalized root mean squared error (NRMSE), normalized mean absolute error (NMAE), and normalized bias error (NBE))
and two probabilistic metrics %(average coverage error (ACE) and interval sharpness (IS))
introduced in \ref{sec:methods_metrics}, we evaluate the impact that different training sets have on predictions of stellar mass from {\sc mirkwood} on galaxy samples from {\sc Simba}. `P' stands for results from {\sc prospector} (traditional SED fitting), `SET' for training set consisting of all {\sc Simba}, {\sc Eagle}, and {\sc IllustrisTNG}, and so on. The test set is always {\sc Simba}. We separate the aleatoric error from the epistemic uncertainty to provide a more granular visualization. As is apparent, at all noise levels, {\sc mirkwood} outperforms traditional SED fitting. Aleatoric uncertainty is generally the lowest when the training set is as diverse as possible, while epistemic uncertainty is always the lowest when the training set samples are drawn from the same simulation as the test set samples ({\sc Simba}). \label{fig:mass_snr_comparison}}
\end{figure*}

\begin{figure*}%[!htbp]
\gridline{\fig{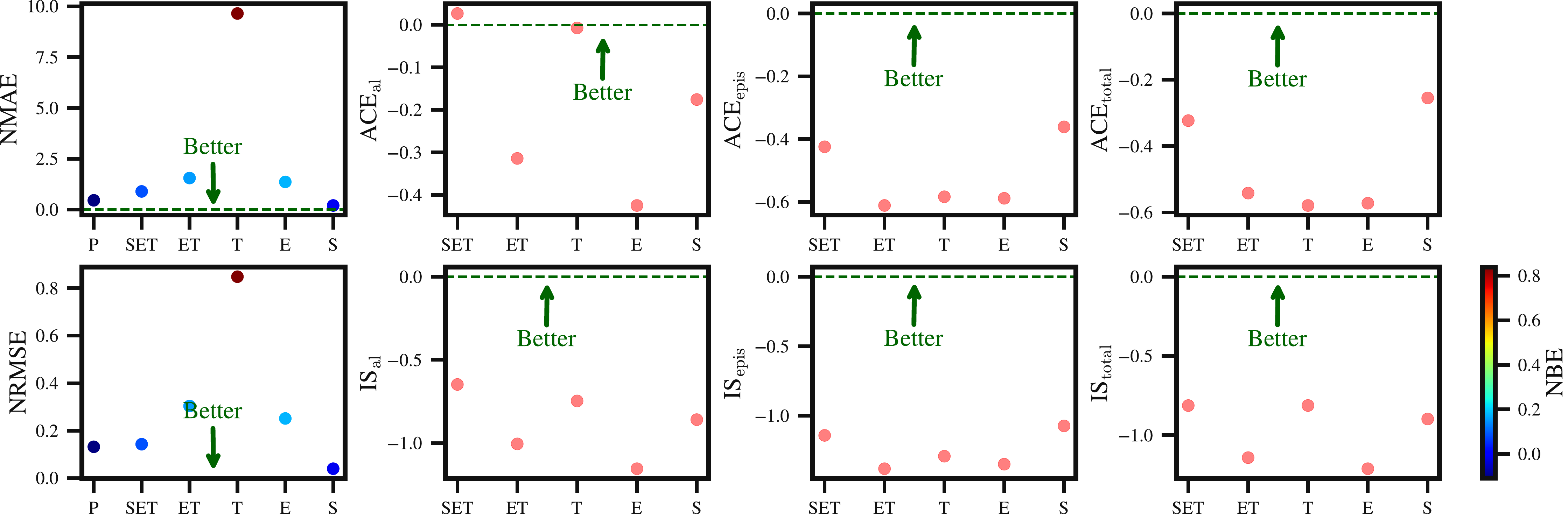}{0.98\textwidth}{(a) SNR = 5}}
\gridline{\fig{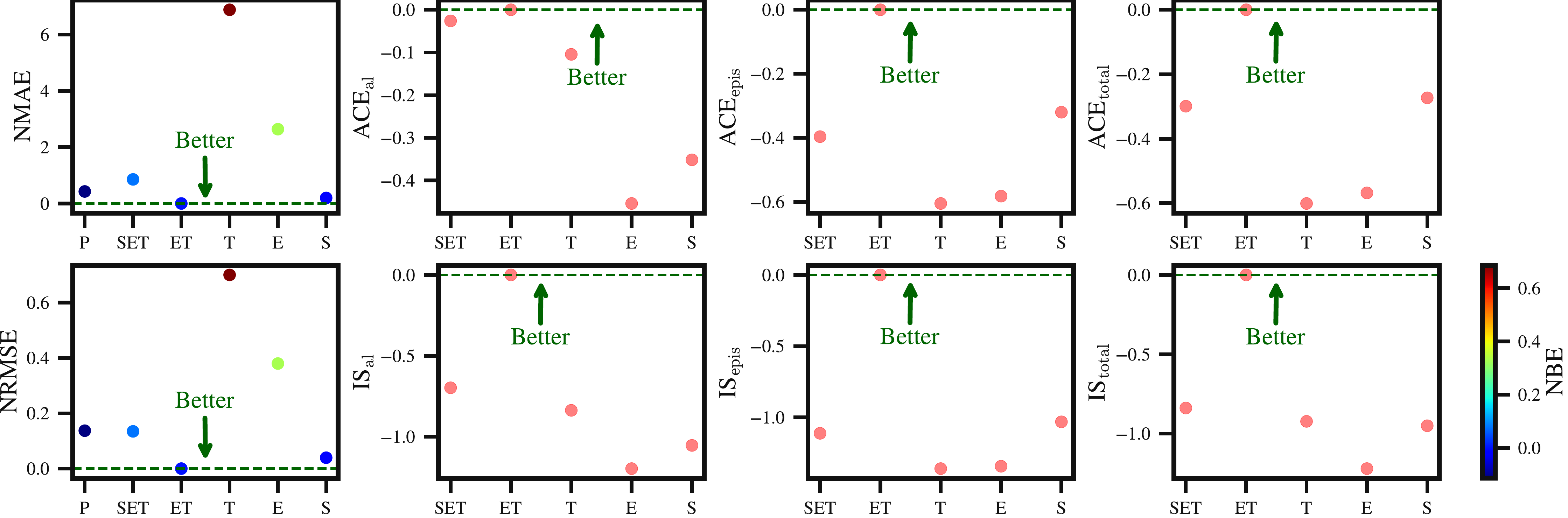}{0.98\textwidth}{(b) SNR = 10}}
\gridline{\fig{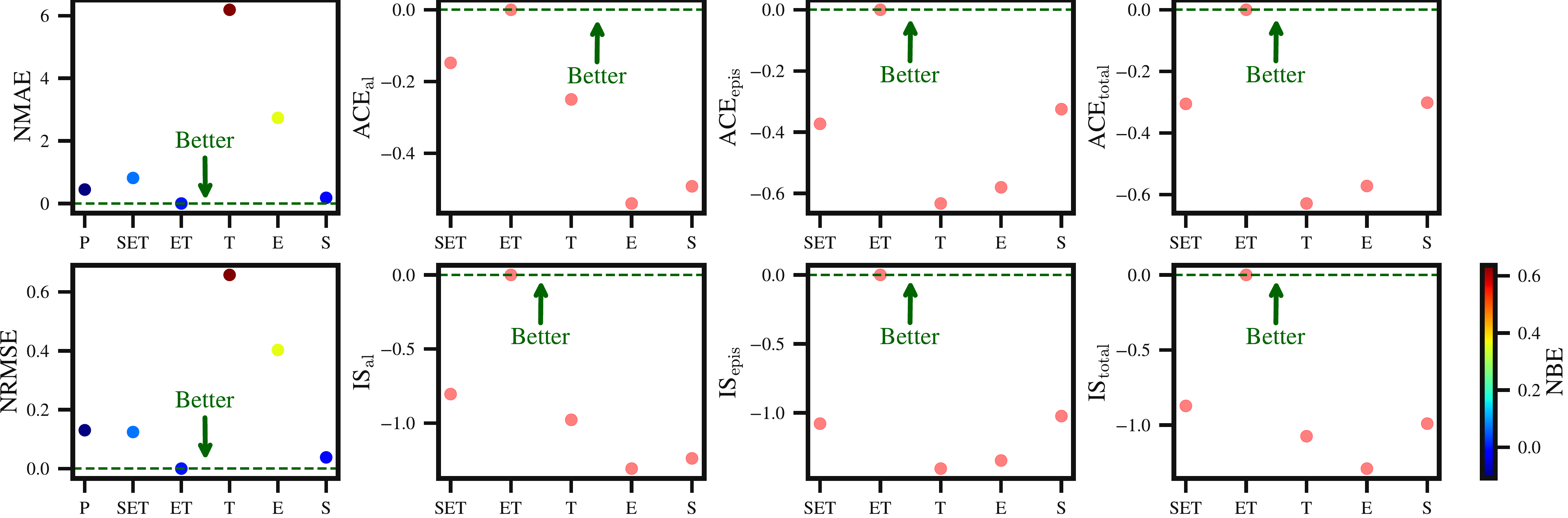}{0.98\textwidth}{(c) SNR = 20}}
\caption{Same as Figure \ref{fig:mass_snr_comparison}, but for dust mass. \label{fig:dustmass_snr_comparison}}
\end{figure*}

\begin{figure*}%[!htbp]
\gridline{\fig{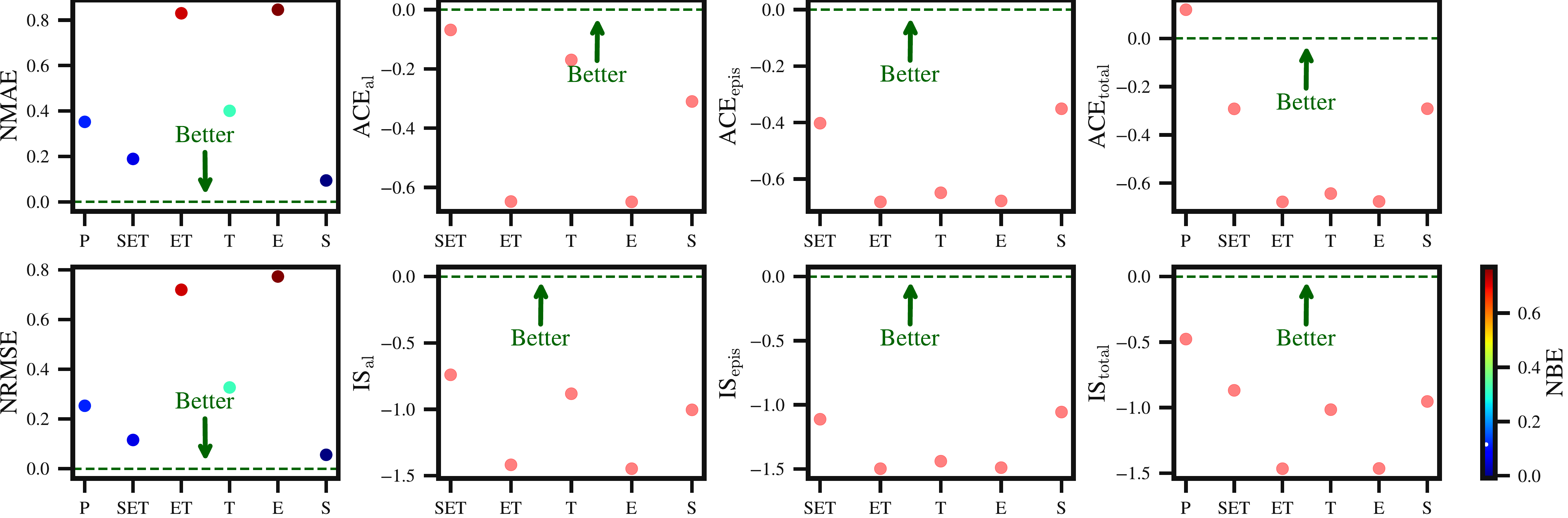}{0.98\textwidth}{(a) SNR = 5}}
\gridline{\fig{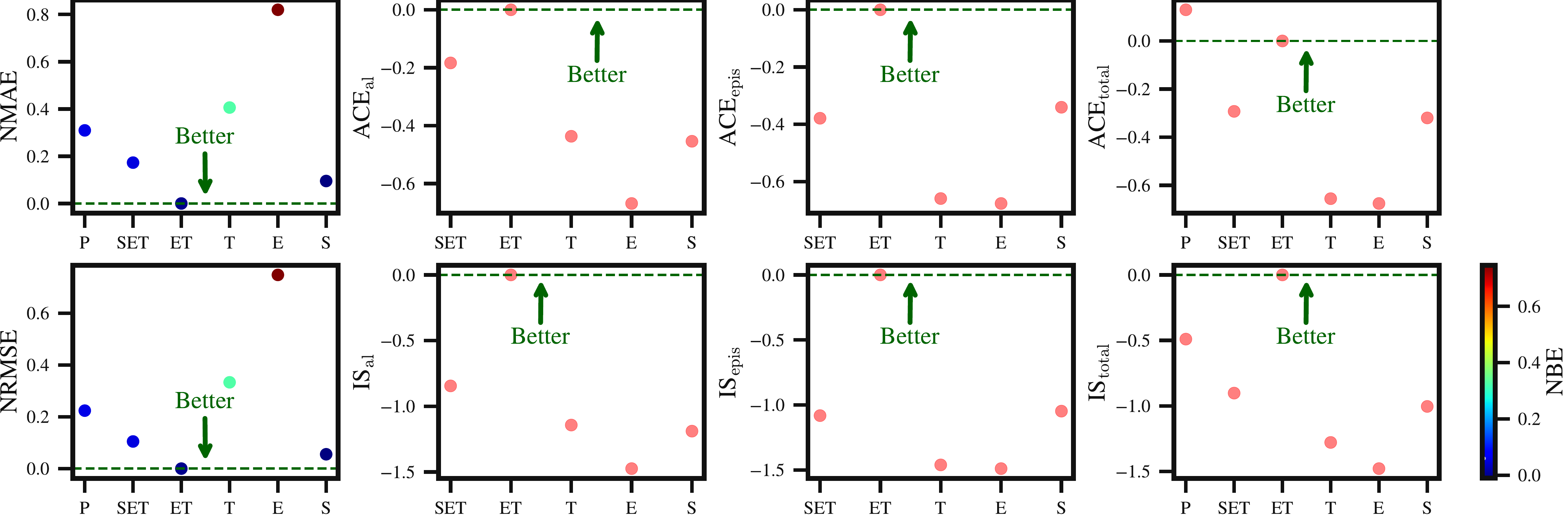}{0.98\textwidth}{(b) SNR = 10}}
\gridline{\fig{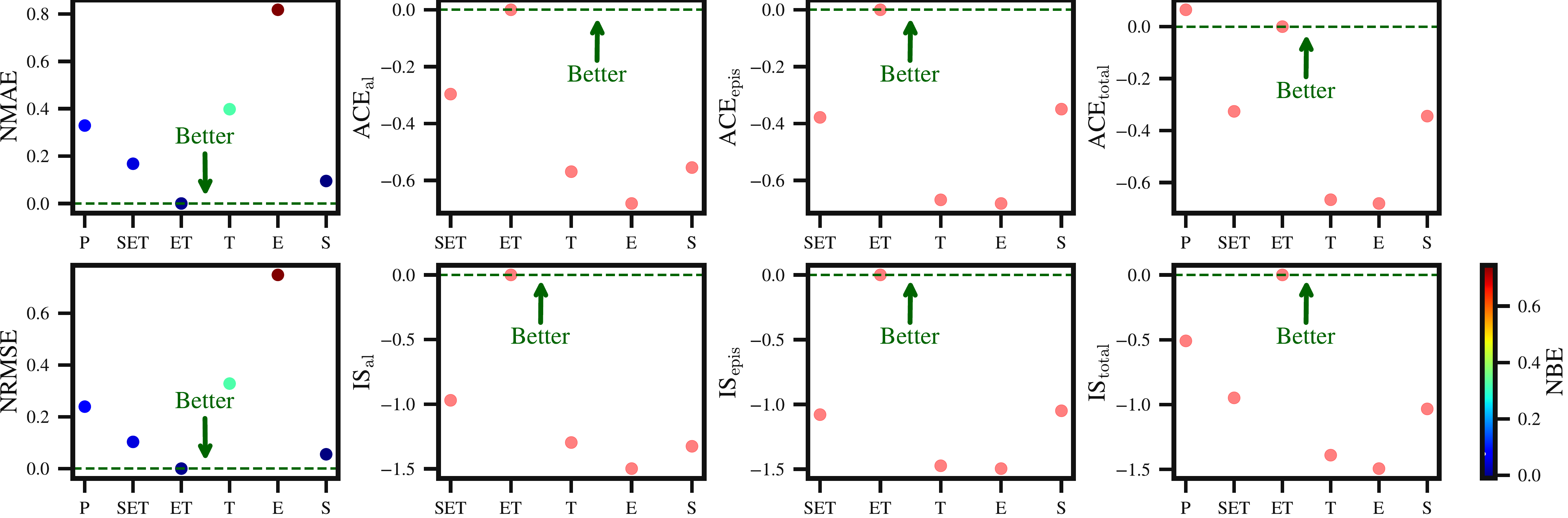}{0.98\textwidth}{(c) SNR = 20}}
\caption{Same as Figures \ref{fig:mass_snr_comparison} and \ref{fig:dustmass_snr_comparison}, but for metallicity Z. \label{fig:met_snr_comparison}}
\end{figure*}

\begin{figure*}%[!htbp]
\gridline{\fig{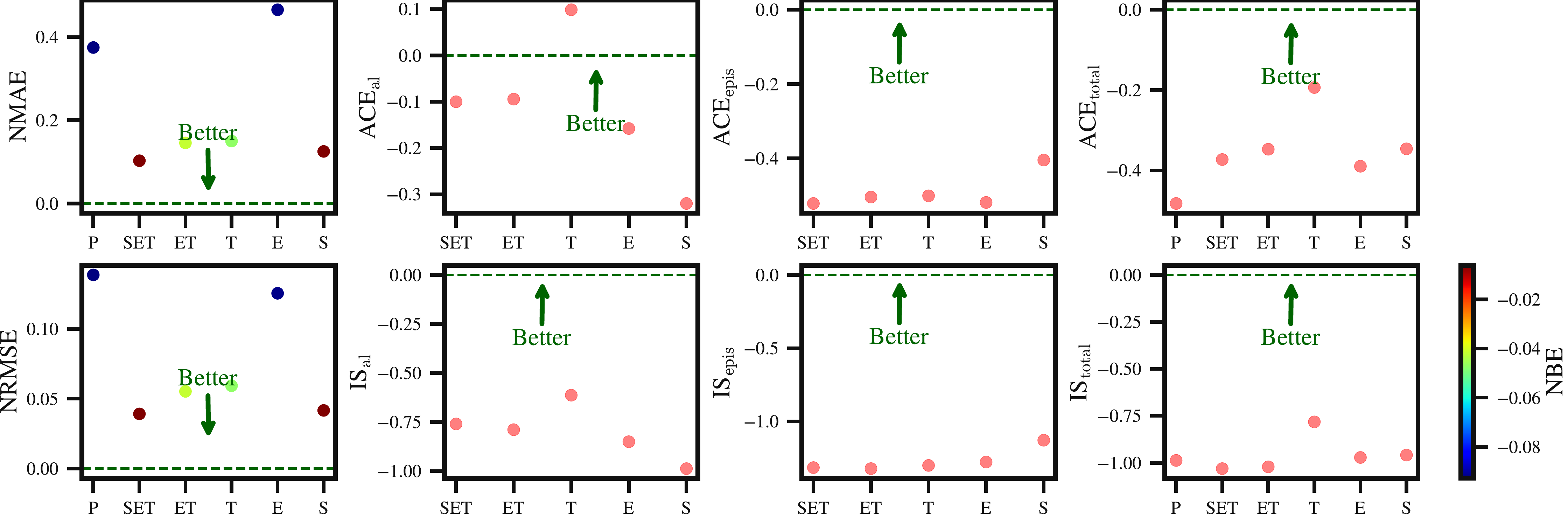}{0.98\textwidth}{(a) SNR = 5}}
\gridline{\fig{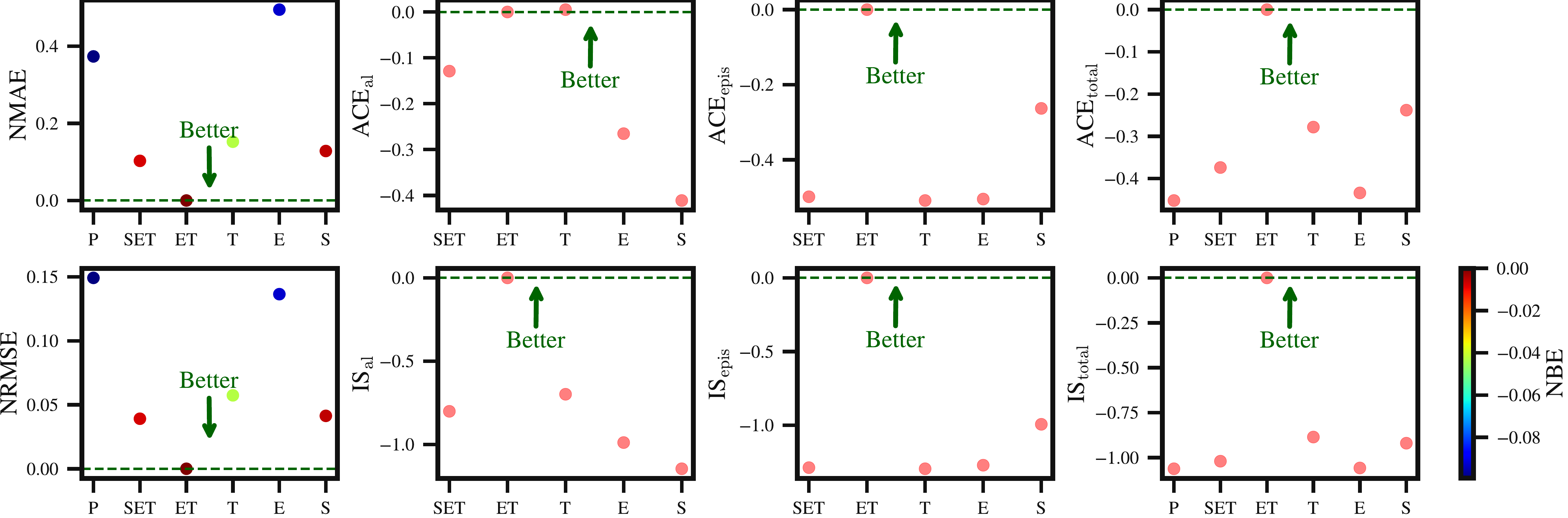}{0.98\textwidth}{(b) SNR = 10}}
\gridline{\fig{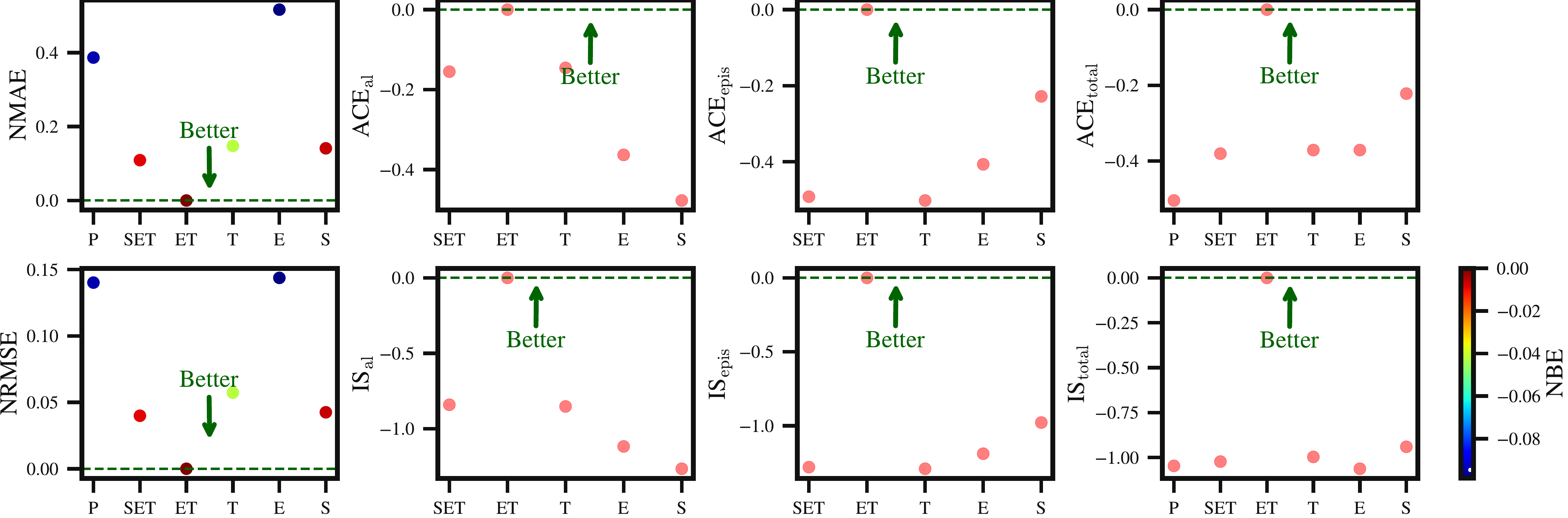}{0.98\textwidth}{(c) SNR = 20}}
\caption{Same as Figures \ref{fig:mass_snr_comparison}, \ref{fig:dustmass_snr_comparison} and \ref{fig:met_snr_comparison} but for instantaneous star formation rate SFR$_{100}$.\label{fig:sfr_snr_comparison}}
\end{figure*}

Finally, in Figures \ref{fig:mass_snr_comparison} through \ref{fig:sfr_snr_comparison}, we demonstrate the impact of different training sets on predictions from {\sc mirkwood}, for all three SNRs of 5, 10, and 20. In all figures, `P' stands for results from {\sc Prospector}, `SET' for results from {\sc mirkwood} when the training dataset contains samples from {\sc IllustrisTNG}, {\sc Eagle}, and {\sc Simba}, `ET' for {\sc Eagle} and {\sc IllustrisTNG}, `T' for only {\sc IllustrisTNG}, `E' for only {\sc Eagle}, and `S' for only {\sc Simba}. For `SET' and `S', we use stratified 5-fold cross validation to select non-overlapping samples from {\sc Simba} in the training set. As expected, we obtain the best predictions, as quantified by various deterministic and probabilistic metrics, when the training set consists of samples from all three simulations. This ensures the most diverse coverage in parameter space, and hence offers the highest generalization power. We also notice that prediction quality improves as the SNR improves, something that one would intuitively expect.

\section{Discussion and Future Work}\label{sec:discussion}
%https://arxiv.org/pdf/1905.08996.pdf
%To field-test {\sc mirkwood}, we plan to use it with data from the GAMA survey. This
 Our work shows that it is possible to reproduce galaxy property measurements with high accuracy in a purely data-driven way by using robust ML algorithms that do not have any knowledge about the physics of galaxy formation or evolution. The biggest caveat here is that the training set must be closely representative of the data of interest, i.e. observations of real galaxies. We also demonstrate the necessity of carefully accounting for both the noise in the data and sparsity of the training set in any given parameter-space of interest. By combining galaxy simulations with disparate physics, we offer a way to better train our model for a real observation. Across all SNRs, we show a considerable improvement over state-of-the-art SED fitting using {\sc Prospector} and non-parametric SFH model, both in terms of prediction quality and compute requirements. For comparison, {\sc mirkwood} ran for $\sim$ 24 hours on the first author's home workstation with 12 cores, to derive all four galaxy parameters for $\sim$ 1700 {\sc SIMBA} galaxies; {\sc prospector} required about the same time, but 3,000 cores on the University of Florida HiPerGator supercomputing system. Another notable result is that we are able to extract accurate galaxy properties from photometries with SNRs as low as 5. All these contributions considered together imply that scientifically interesting galaxy properties can be accurately and quickly measured on future datasets from large-scale photometric surveys.

Several avenues for future research remain open. The crux of our follow-up efforts will focus on development of {\sc mirkwood} to a higher level of complexity and precision. We identify the following areas for improvement:
\begin{enumerate}
    \item Extensive hyperparameter optimization: In this work we use {\sc NGBoost} models with hyperparameters only slightly modified from the prescription of \cite{ren2019rongba_ngboost}. This is an easy avenue for gaining performance boost with respect to reconstruction of galaxy properties -- in future work we will leverage modern Bayesian hyperparameter optimization libraries to conduct extensive HPO to eke out maximum performance from our models.
    \item Managing missing observations: Most real-life observations of all galaxies do not contain observations in the same set of filters, nor is the SNR of the data in each filter the same. Currently, {\sc mirkwood} cannot easily accommodate these complications, since {\rm NGBoost} is unable to natively handle missing feature values. In future work we will explore other boosted tree models, such as {\rm CatBoost} \citep{prokhorenkova2018catboost}, to extract galactic properties from real observations.
    \item Synergistic spectroscopy: While spectroscopic data are considerably more expensive to obtain, there exist targets where photometry and spectra are available simultaneously \citep{carnall2019vandels}. Modeling both modes of data simultaneously can increase the SNR available to any model, and help significantly in reducing uncertainties in derived galaxy properties, thus enabling accurate re-construction of higher resolution ones such as a galaxy's entire star formation history. We will explore inclusion of a convolutional neural network to this end, similar to the work of \cite{lovell2019learning}.
    \item Probability calibration: This is a post-processing step that has been shown to improve predictions, measured both by deterministic and probabilistic metrics. ML algorithms are often over-confident, and unable to assess the degree of their fallacy -- if given a completely novel input, most commonly used ML algorithms are likely to predict a wrong label for it, instead of just saying, `I do not know the correct answer.' Probability calibration is widely used in the machine learning community \citep{measuring_calibration_in_deep_learning, probability_calibration1, crude_probability_calibration, lakshminarayanan_probability_calibration0}, albeit its adoption in astronomy has been slow. We expect probability calibration would result in more sensible uncertainties in the second through fourth columns of Figures \ref{fig:mass_snr_comparison} through \ref{fig:sfr_snr_comparison}. Specifically, we expect that calibrated aleatoric, epistemic, and total uncertainties would mean the lowest ACE and IS values for the `SET' column, which is not consistently the case in this paper.
    \item Randomized chaining: Here we have considered one particular way of chaining galaxy properties: only fluxes are used when predicting stellar mass, fluxes and predicted stellar mass are synergistically used to predict dust mass, and so on all the way till SFR$_{100}$ (see Figure \ref{fig:chaining} for details). However, given the interdependence of all properties, a more justified approach would be to consider multiple such orderings and average the result. For instance, to predict stellar mass, $\rm{model}_1$ will be trained on just the galaxy fluxes, $\rm{model}_2$ on fluxes + metallicity, $\rm{model}_3$ on fluxes + dust mass + SFR$_{100}$, and all such permutations. Finally, the predicted stellar mass results will be averaged. We expect this would average out any biases introduced by chaining in any particular order, and we plan to explore this line of research in more detail.
    \item Redshifts as both input and output: We plan on expanding the training set considerably by incorporating photometries from galaxies a range of redshifts to enable extraction of properties of real galaxies. In this sense, redshift would be the fifth output variable. On the other hand, it might very well be the case that for a given set of samples, high-confidence spectroscopic redshift is available, but photometry data is present only in a few bands. In such a case, it would be sensible to use redshift as an input rather than an output. We plan on upgrading our network so that {\sc mirkwood} can easily take a redshift as either a fixed input, or take a user-given prior for redshift which it will then aim to improve.
    \item Semi-supervised learning: In an effort to reduce the `domain gap' --  the difference between the training set(s) and real life data -- several semi-supervised algorithms have been developed and used. These can smoothly combine a computer simulated set of data with high-confidence real life-observations, such as spectra, to more closely approximate reality.
    \item Outlier detection and elimination: In its current implementation, {\sc mirkwood} is unable to identify and reject outlying inputs. These can result from any of the several data processing pipelines that are used to create scientifically interesting data products can fail and output nonsensical numbers, for example a flux absurdly high as $10^{252}$. This issue becomes even more urgent when dealing with semi-supervised learning, where data from real-life observations form part of the training set. In future we plan on exploring algorithms such as variational auto-encoders \citep{ghosh2016robust_variationalbayesianinference,zellner1988optimal_variationalbayesianinference, gilda2020astronomical_neurips_rvae} to this end.
    %\red{it may be valuable to note that we assume a particular input set of models in our training, including a set of IMF, isochrones and stellar spectra assumptions that require a full retraining of the models and regeneration of the SEDs to accomodate that.} \green{true, but I am not sure how to mention this. like, how do we propose to improve this in future work...?}
\end{enumerate}

\subsection{On SED Modeling Priors and Assumptions}
%\green{SL: feel free to move this section to where you see fit, i don't know if it should go before / after future work.}

A fundamental aspect of SED modeling is the choice of model for each component (e.g. SFH or dust attenuation) and the subsequent priors used to constrain the models. In this work, we have used {\sc prospector} to fit the synthetic SEDs of galaxies from three cosmological simulations, following the model assumptions outlined in \cite{lower_stellar_mass}. There, the model components, namely the non-parametric SFH, were optimized to result in the best possible estimates for stellar mass, mass-weighted stellar age, and recent SFR for the {\sc simba} simulation at $z=0$. However, it is conceivable that these optimizations do not hold true for {\sc eagle} or {\sc illustris TNG}. %, nor even {\sc simba} at redshift $z=2$.
For instance, in recent work by \cite{iyer_2020_div_sfh}, the above cosmological simulations produce significantly different SFHs depending on the different subgrid models for, e.g., star formation and stellar feedback, which modulate the short and long-timescale variations of a galaxy's SFH. Thus, using the same SFH model to describe the galaxy SFHs from these simulations may not be justified, but is ultimately outside the scope of this paper to fully tune the SFH models for each simulation. Similarly, in future work, we are working to implement physically motivated priors to further improve the results from traditional SED fitting by including variable dust attenuation curves. %, focusing on the impact of a variable dust attenuation curve on inferred galaxy properties. 

One advantage of {\sc mirkwood} in this regard is the freedom from choosing a particular SFH model, dust attenuation law, or set of priors to infer galaxy properties from broadband photometry. Combining the input from three cosmological simulations as a training set for {\sc mirkwood} removes the simplifications of modeling galaxy SEDs with relatively simple analytic functions as in traditional SED modeling. This method of inferring galaxy properties can also overcome the biases that affect traditional SED fitting, such as the outshining of older stellar populations by younger, massive stars which generally causes stellar masses to be under-estimated for star forming galaxies \citep{Papovich_2001_outshining} and the generally poor reconstruction of early SFHs due to the non-uniform sensitivity to variations in star formation as a function of time (i.e.that SED fitting is more sensitive to recent star formation and much information about star formation beyond a few Gyr before the observation is lost)  \citep{iyer2019nonparametric}.  

\subsection{Caveats to the Mirkwood Training Set}
%\green{SL: my attempt at compiling the 'however...'s about the limitaitons of the current mirkwood training set}

Though {\sc mirkwood} is trained on a wealth of data from three moderate volume cosmological simulations, we are not yet free from broad assumptions pertaining to stellar population synthesis modeling. {\sc mirkwood} relies on synthetic galaxy SEDs generated by post-process 3D dust radiative transfer modeling via {\sc hyperion} and {\sc fsps}. In practice, we have assumed a universal initial mass function (IMF) for all galaxies as well as one library of stellar spectra and isochrones. Although this method is similar to traditional SED fitting, our training set is tied to these particular models. To understand the magnitude of importance of these assumptions on inferring galaxy properties, we would need to regenerate the synthetic galaxy SEDs for each combination of IMFs, spectral libraries, and isochrones, representing millions of hours of CPU time in addition to the time required to retrain {\sc mirkwood} on each set of SEDs for each redshift. Given the scale this problem requires, this retraining is ultimately outside the scope of this work. 

%\red{This paragraph kind of comes out of nowhre, and I suggest axing it \sout{Finally, any new algorithm must be validated against known results from literature. Since we have developed {\sc mirkwood} to appeal to observers, we will use it to derive physical properties of galaxies from DR3 of the GAMA survey \citep{gama_driver2009}. In a recent work, \cite{bellstedt2020galaxy} formulated a novel SFH prior, and derived SFH and metallicty posteriors of $\sim 7000$ galaxies in the nearby universe. They were able to recreate the cosmic star formation rate density much better than using other parametric formulations of the SFH. We plan on using this same dataset and comparing our results with the GAMA Team.}}

\begin{table*}[!htpb]
    \caption{Comparative performance of {\sc mirkwood} v/s {\sc Prospector} across different noise regimes. The five metrics are the normalized root mean squared error (NRMSE), normalized mean absolute error (NMAE), normalized bias error (NBE), average coverage error (ACE), and interval sharpness (IS). In any given cell, the tuple of values consists of the metric of interest from {\sc mirkwood} and {\sc Prospector}, respectively. A value of `nan' represents lack of predictions from {\sc Prospector}. %Since traditional SED fitting hasn't been used to derive the redshift, the second values in all tuples in the `Redshift' field are `nan's. Similarly,
    We do not have predicted error bars from {\sc Prospector} for dust mass, hence ACE and IS values corresponding to this property are `nan's.}
    \label{tab:stackedbartable}
    \centering
    \begin{tabular}{ccccccc}
    \toprule
    & SNR &  NRMSE &  NMAE & NBE & ACE & IS \\ \tableline
    %& 5  &  (0.0, nan) &  (0.0, nan) &  (-0.0, nan) &  (0.317, nan) &  (0.0, nan) \\
    %Redshift & 10 &  (0.0, nan) &  (0.0, nan) &  (-0.0, nan) &  (0.317, nan) &  (0.0, nan) \\
    %& 20 &  (0.0, nan) &  (0.0, nan) &   (0.0, nan) &  (0.317, nan) &  (0.0, nan) \\ \tableline
    & 5  &  (\textbf{0.198}, 1.003) &  (\textbf{0.123}, 1.091) &  (\textbf{-0.042}, -0.528) &  \textbf{(-0.002}, -0.497) &  (\textbf{0.001}, 0.005) \\
    Mass & 10 &    (\textbf{0.165}, 1.0) &  (\textbf{0.118}, 1.088) &  (\textbf{-0.035}, -0.518) &  (\textbf{-0.021}, -0.502) &  (\textbf{0.001}, 0.004) \\
    & 20 &  (\textbf{0.155}, 1.002) &  (\textbf{0.115}, 1.117) &  (\textbf{-0.041}, -0.479) &  (\textbf{-0.066}, -0.482) &  (\textbf{0.001}, 0.033) \\ \tableline
    & 5  &   (\textbf{0.48}, 0.996) &  (\textbf{0.339}, 0.998) &  (\textbf{-0.219}, -0.905) &   (\textbf{0.003}, nan) &  (\textbf{0.001}, nan) \\
    Dust Mass & 10 &  (\textbf{0.456}, 0.996) &  (\textbf{0.332}, 0.998) &  (\textbf{-0.209}, -0.905) &  (\textbf{-0.033}, nan) &  (\textbf{0.001}, nan) \\
    & 20 &  (\textbf{0.475}, 1.263) &  (\textbf{0.336}, 1.212) &  (\textbf{-0.215}, -0.679) &  (-0.076, nan) &  (0.001, nan) \\ \tableline
    & 5  &  (\textbf{0.062}, 0.544) &   (\textbf{0.06}, 0.478) &  (\textbf{-0.011}, -0.297) &  (\textbf{-0.024}, 0.046) &  (\textbf{0.041}, 0.301) \\
    Metallicity & 10 &  (\textbf{0.058}, 0.534) &  (\textbf{0.055}, 0.464) &   (\textbf{-0.01}, -0.275) &  (\textbf{-0.032}, 0.041) &  (\textbf{0.036}, 0.295) \\
    & 20 &  (\textbf{0.056}, 0.547) &  (\textbf{0.052}, 0.487) &   (\textbf{-0.01}, -0.229) &  (\textbf{-0.063}, 0.036) &  (\textbf{0.032}, 0.302) \\ \tableline
    & 5  &  (\textbf{0.241}, 0.907) &   (\textbf{0.205}, 0.99) &  (\textbf{-0.069}, -0.687) &  (\textbf{0.074}, -0.557) &  (7.314,\textbf{ 0.001}) \\
    SFR & 10 &   (\textbf{0.329}, 0.91) &  (\textbf{0.226}, 0.992) &   (\textbf{-0.09}, -0.686) &  (\textbf{0.048}, -0.564) &  (\textbf{1.937}, 0.001) \\
    & 20 &  (\textbf{0.277}, 1.988) &  (\textbf{0.215}, 2.911) &   (\textbf{-0.078}, 1.437) &  (\textbf{0.035}, -0.547) &    (\textbf{0.006}, 0.2) \\ \tableline
    \end{tabular}
\end{table*}

\section*{Acknowledgements}
SG is grateful to Dr. Sabine Bellstedt (GAMA collaboration, University of Western Australia) for several fruitful discussions, Prof. Viviana Aquaviva (Department of Physics, CUNY NYC College of Technology) for an in-depth review of the first draft, and Dr. Joshua Speagle (University of Toronto) for his helpful comments. D.N. acknowledges support from the US NSF via grants AST-1715206 and AST-1909153, and from the Space Telescope Science Institute via grant HST-AR15043.001-A. S.L. was funded in part by the NSF via AST-1715206.

\bibliography{bib}
\bibliographystyle{aasjournal}

\appendix
\section{Shapley Values}\label{sec:shapley_values}
Consider a dataset $\textbf{X}$ with $N$ features and $S$ samples $\textbf{X}=\left\{X^{(1)}_{1:S}, \ldots, X^{(N)}_{1:S}\right\},$ and a continuous, real-valued target variable vector $Y_{1:S}$. Given an instance (observation) $x \in \textbf{X} \in \textsc{R}^{1:N},$ a model $v$ forecasts the target as $v(x)$. In addition to this forecasting, we would also like to know if we can attribute the departure of the prediction $v(x)$ from the average value of the target variable $\bar{y} = \bar{Y}_{1:S}$ to each of the features (predictor variables) in the observed instance $x$? In a linear model, the answer is trivial :$v(x)-\bar{y}=\beta_{1} x^{(1)}+\cdots+\beta_{N} x^{(N)}$ (see Slide 13 in \citealt{prado_shapleyvalues}). We would like to do a similar local decomposition for any (non-linear) ML model. %\footnote{As explained in \S\ref{sec:introduction} and evident from Figure \red{YOU'RE MISSING A FIGURE HERE}{}, non-linear models provide much better predictions than linear ones.}
Shapley values answer this \emph{attribution} problem through game theory \citep{shapley}. Specifically, they originated with the purpose of resolving the following scenario--a group of differently skilled participants are all \emph{cooperating} with each other for a collective reward; how should this reward be \emph{fairly} divided amongst the group? In the context of machine learning, the participants are the features of our input dataset $\textbf{X}$ and the collective \emph{payout} is the \emph{excess} model prediction (i.e., $v(x) - \bar{y}$). Here, we use Shapley values to calculate how much each individual feature \emph{marginally} contributes to the model output--both for the dataset as a whole (\emph{global explainability}) and for each instance (\emph{local explainability}).

While a goal of this paper is to explain the contributions of narrow- and broad-bands to various galaxy properties, for the sake of simplicity we chose to explain Shapley values using a toy example. Let us say that we operate a small hedge fund, and our team consists of three people: \emph{\underline{A}lice}, \emph{\underline{B}ob}, and \emph{\underline{C}arol}. On average, they manage to make approximately \$100,000 everyday ($\bar{y} = 100$ (thousand)). At the end of the year, we would like to distribute bonuses to our team members in direct proportion to their contribution to beating the average daily profit. Let us consider one day in particular, when owing partly to market volatility and partly to luck, today they make \$120,000 (i.e., excess income = 20 (thousand) dollars). Below, we delineate the process of determining each person's fair contribution to this number.

\begin{enumerate}
    \item First, we consider the $2^{N=3}$ possible coalitions (interactions) between the players (features), where some players may not participate (i.e., remain at their average value), and other may participate (depart from their average value). See Table \ref{table:coalitions}.
    \item Second, we use the coalitions table to compute the marginal contribution of each player conditional on the contribution of every other player. The marginal impact of changing \emph{A} after changing \emph{B} may differ from the marginal impact of changing \emph{B} after changing \emph{A}. Accordingly, we must account for all possible $N! = 6$ sequences of conditional effects. See Table \ref{table:marginal_conditional_contribution}.
    \item Finally, the \emph{Shapley value} of a player is calculated as the average conditional marginal contribution of that player across all the possible $N!$ ways of conditioning the predictive model $v$. See last row of Table \ref{table:marginal_conditional_contribution} and Equation \ref{eq:shapley_main}.
\end{enumerate}

\begin{table}%[h!]
\centering
\begin{tabular}{|c|c|c|c|}
\hline $\text{Alice} \neq \overline{\text{Alice}}$ & $\text{Bob} \neq \overline{\text{Bob}}$ & $\text{Carol} \neq \overline{\text{Carol}}$ & $v(...) \left(= v[x \mid \ldots]\right)-\bar{y}$ \\
\hline 0 & 0 & 0 & $v(\varnothing) (= v[x \mid 000])-\bar{y} = 0$ \\
\hline 0 & 0 & 1 & $v(C) (= v[x \mid 001])-\bar{y} = 10$ \\
\hline 0 & 1 & 0 & $v(B) (= v[x \mid 010])-\bar{y} = 5$ \\
\hline 0 & 1 & 1 & $v(B,C) (= v[x \mid 011])-\bar{y} = 7$ \\
\hline 1 & 0 & 0 & $v(A) (= v[x \mid 100])-\bar{y} = 2$ \\
\hline 1 & 0 & 1 & $v(A,C) (= v[x \mid 101])-\bar{y} = 8$ \\
\hline 1 & 1 & 0 & $v(A,B) (= v[x \mid 110])-\bar{y} = 10$ \\
\hline 1 & 1 & 1 & $v(A,B,C) (= v[x \mid 111])-\bar{y} = \textcolor{orange}{20}$ \\
\hline
\end{tabular}
\caption{In a model with 3 features, there are $2^{3}=8$ possible interactions. For each interaction, we compute the departure of the model's forecast from its baseline (average value). We encode as `1' a feature that is not at its average value (it forms part of a \emph{coalition}), and `0' a feature that is at its average value. Thus $v(A,B)=v(B,A)$, $v(B,C)=v(C,A)$, $v(A,C)=v(C,A)$, and $v(A,B,C)=v(B,A,C)=v(A,C,B)$.}
\label{table:coalitions}
\end{table}

\iffalse
%similar plot to the one in /home/sgilda/Downloads/SSRN-id3637020.pdf, page 9
An attribution of the departure of a model's prediction, $f[x],$ from its average value, $\bar{y}$ Because addition is commutative, the sequence of the effects does not alter the result. Hence,
in a linear model, the attribution is invariant to the sequence of effects.
\fi

%\subsection{Coalitional Game Theory}

\begin{table}%[h!]
\centering
\begin{tabular}{|l|l|l|l|l|}
\hline Combination & \multicolumn{1}{|c|} {Alice} & \multicolumn{1}{|c|} {Bob} & \multicolumn{1}{|c|} {Carol} & Total \\
\hline $\mathrm{A}, \mathrm{B}, \mathrm{C}$ & $v(\mathrm{A})-v(\varnothing) = 2$ & $v(\mathrm{A}, \mathrm{B})-v(\mathrm{A}) = 8$ & $v(\mathrm{A}, \mathrm{B}, \mathrm{C})-v(\mathrm{A}, \mathrm{B}) = 10$ & \textcolor{orange}{20}\\
\hline $\mathrm{A}, \mathrm{C}, \mathrm{B}$ & $v(\mathrm{A})-v(\varnothing) = 2$ & $v(\mathrm{A}, \mathrm{C}, \mathrm{B})-v(\mathrm{A}, \mathrm{C}) = 12$ & $v(\mathrm{A}, \mathrm{C})-v(\mathrm{A}) = 6$ & \textcolor{orange}{20} \\
\hline $\mathrm{B}, \mathrm{A}, \mathrm{C}$ & $v(\mathrm{B}, \mathrm{A})-v(\mathrm{B}) = 5$ & $v(\mathrm{B})-v(\varnothing) = 5$ & $v(\mathrm{B}, \mathrm{A}, \mathrm{C})-v(\mathrm{B}, \mathrm{A}) = 10$ & \textcolor{orange}{20} \\
\hline $\mathrm{B}, \mathrm{C}, \mathrm{A}$ & $v(\mathrm{B}, \mathrm{C}, \mathrm{A})-v(\mathrm{B}, \mathrm{C}) = 13$ & $v(\mathrm{B})-v(\varnothing) = 5$ & $v(\mathrm{B}, \mathrm{C})-v(\mathrm{B}) = 2$ & \textcolor{orange}{20} \\
\hline $\mathrm{C}, \mathrm{A}, \mathrm{B}$ & $v(\mathrm{C}, \mathrm{A})-v(\mathrm{B}) = 3$ & $v(\mathrm{C}, \mathrm{A}, \mathrm{B})-v(\mathrm{C}, \mathrm{A}) = 12$ & $v(\mathrm{C})-v(\varnothing) = 10$ & \textcolor{orange}{20} \\
\hline $\mathrm{C}, \mathrm{B}, \mathrm{A}$ & $v(\mathrm{C}, \mathrm{B}, \mathrm{A})-v(\mathrm{C}, \mathrm{B}) = 13$ & $v(\mathrm{C}, \mathrm{B})-v(\mathrm{C}) = -3$ & $v(\mathrm{C})-v(\varnothing) = 10$ & \textcolor{orange}{20} \\
\hline Average & $6.3$ & $6.5$ & $8$ & \textcolor{orange}{20} \\
\hline
\end{tabular}
\caption{In a model with 3 features, there are $3! = 6$ possible sequences. We can use the interactions table to derive the marginal contribution of each feature in each sequence. The Shapley values are the averages per column.}
\label{table:marginal_conditional_contribution}
\end{table}

\iffalse
\begin{tabular}{l|l|l|l}
\multicolumn{1}{c|} { Sequence } & \multicolumn{1}{c|} {$x^{(1)}$} & \multicolumn{1}{c} {$x^{(2)}$} & \multicolumn{1}{c} {$x^{(3)}$} \\
\hline$x^{(1)}, x^{(2)}, x^{(3)}$ & $f[x \mid 100]$ & $f[x \mid 110]$ & $f[x \mid 111]$ \\
& $-f[x \mid 000]$ & $-f[x \mid 100]$ & $-f[x \mid 110]$ \\
\hline$x^{(1)}, x^{(3)}, x^{(2)}$ & $f[x \mid 100]$ & $f[x \mid 111]$ & $f[x \mid 101]$ \\
& $-f[x \mid 000]$ & $-f[x \mid 101]$ & $-f[x \mid 100]$ \\
\hline$x^{(2)}, x^{(1)}, x^{(3)}$ & $f[x \mid 110]$ & $f[x \mid 010]$ & $f[x \mid 111]$ \\
& $-f[x \mid 010]$ & $-f[x \mid 000]$ & $-f[x \mid 110]$ \\
\hline$x^{(2)}, x^{(3)}, x^{(1)}$ & $f[x \mid 111]$ & $f[x \mid 010]$ & $f[x \mid 011]$ \\
& $-f[x \mid 011]$ & $-f[x \mid 000]$ & $-f[x \mid 010]$ \\
\hline$x^{(3)}, x^{(1)}, x^{(2)}$ & $f[x \mid 101]$ & $f[x \mid 111]$ & $f[x \mid 001]$ \\
& $-f[x \mid 001]$ & $-f[x \mid 101]$ & $-f[x \mid 000]$ \\
\hline$x^{(3)}, x^{(2)}, x^{(1)}$ & $f[x \mid 111]$ & $f[x \mid 011]$ & $f[x \mid 001]$ \\
& $-f[x \mid 011]$ & $-f[x \mid 001]$ & $-f[x \mid 000]$
\end{tabular}
\fi
%that slide on Interaction Effects. Page 15

\textbf{\underline{Eliminating Redundancy}}: As we can appreciate from the numerical exercise in Tables \ref{table:coalitions} and \ref{table:marginal_conditional_contribution}, some calculations are redundant. For example, the marginal contribution of \emph{C} on sequence \emph{(A, B, C)} must be the same as the marginal contribution of \emph{C} on sequence \emph{(B, A, C)}, because (a) in both cases \emph{C} comes in third position, and (b) the permutations of \emph{(A, B)} do not alter that marginal contribution. The Shapley value of a feature $i$ can be derived as the average contribution of $i$ across all possible coalitions $S$, where $S$ does not include feature $i$:

\begin{align}\label{eq:shapley_main}
    \phi_{i}(v)=\sum_{S \subseteq N \backslash\{i\}} \frac{|S| !(|N|-|S|-1) !}{|N| !}(v(S \cup\{i\})-v(S))
\end{align}

The above equation describes a \emph{coalitional game} (the scenario described previously) with a set \textbf{N} of $|\textbf{N}|$ players. The function $v$ gives the value (payout) for any subset of those players. For example, if $S$ be a subset of \textbf{N}, then \emph{v(S)} gives us the value of that subset. Thus for a coalitional game (\textbf{N}, \emph{v}) Equation \ref{eq:shapley_main} outputs the value for feature \emph{i}, i.e. its the Shapley value.\footnote{Equation \ref{eq:shapley_main} computes the exact Shapley values by grouping the marginal conditional contributions in terms of coalitions $(S)$. For large $N,$ \cite{shap1} and \cite{shap2} have developed fast algorithms for the estimation of $\phi_{i}$} For ease of explanation, we focus our attention on calculating how many of the excess $20$ (thousand) dollars can be attributed to \emph{C}(arol), i.e. calculating the Shapley value for \emph{C}. We then examine the various components of Equation \ref{eq:shapley_main} and study their significance.
\begin{itemize}
    \item If we relate this back to the parameters of the Shapley value formula we have \textbf{N}$ = \{A, B, C\}$ and $i = C$. The highlighted section in Equation \ref{eq:shapley_highlight_0} (derived from Equation \ref{eq:shapley_main} by re-arranging some terms) says that we need to take our group of people and exclude the person that we are focusing on now. Then, we need to consider all of the possible subsets that can be formed. So if we exclude \emph{C} from the group we are left with \emph{{A, B}}. From this remaining group we can form the following subsets (i.e. these are the sets that \emph{S} can take on): $\varnothing, A, B, AB$. In total, we can construct four unique subsets from the remaining team members, including the null set.
    \begin{align}\label{eq:shapley_highlight_0}
    \phi_{i}(v)=\frac{1}{|N|} \sum_{\textcolor{orange}{S \subseteq N \backslash\{i\}}}\left(\begin{array}{c}
    |N|-1 \\
    |S|
    \end{array}\right)^{-1}(v(S \cup\{i\})-v(S))
    \end{align}
    \item Next, we focus on the highlighted term  in Equation \ref{eq:shapley_highlight_1}.
    \begin{align}\label{eq:shapley_highlight_1}
    \phi_{i}(v)=\frac{1}{|N|} \sum_{S \subseteq N \backslash\{i\}}\left(\begin{array}{c}
    |N|-1 \\
    |S|
    \end{array}\right)^{-1}\textcolor{orange}{(v(S \cup\{i\})-v(S))}
    \end{align}
    This refers to the \emph{marginal value} of adding player $i$ to the game. Specifically, we want to see the difference in the money earned daily if we add $C$ to each of our four subsets. We denote these four marginal values as: $\Delta v_{C, \varnothing}, \Delta v_{AC, A}, \Delta v_{BC, B}, \Delta v_{A B C, A B}$. Each of these as a different scenario that we need to observe in order to fairly assess how much \emph{C} contributes to the overall profit. This means that we need to observe how much excess money is produced if everyone is working at their average levels (i.e. the empty set $\varnothing$) and compare it to what happens if we only have \emph{C} working at above or below-average productivity. We also need to observe how much profit is earned by \emph{A} and \emph{B} working simultaneously and compare that to the profit earned by \emph{A} and \emph{B} together with \emph{C}, and so on.
    \item Next, the summation in the Shapley value equation is telling us that we need to add all them together. However, we also need to scale each marginal value before we do that, which we are provided this prescription by the highlighted part in Equation \ref{eq:shapley_highlight_2}.
    \begin{align}\label{eq:shapley_highlight_2}
    \phi_{i}(v)=\frac{1}{|N|} \sum_{S \subseteq N \backslash\{i\}}\textcolor{orange}{\left(\begin{array}{c}
    |N|-1 \\
    |S|
    \end{array}\right)^{-1}}(v(S \cup\{i\})-v(S))
    \end{align}
    It calculates how many permutations of each subset \emph{size} we can have when constructing it out of all remaining team members excluding feature (player) \emph{i}. In other words, given \textbf{$\lvert$N$\rvert$-1} features (players), how many groups of size \textbf{$\lvert$S$\rvert$} can one form with them? We then use this number to divide the marginal contribution of feature (player) $i$ to all groups of size \textbf{$\lvert$S$\rvert$}. For our scenario, we have that \textbf{$\lvert$N$\rvert$-1 = 2}, i.e. these are the remaining team members when we are left with when calculating the Shapley value for \emph{C}. In our case we will use that part of the equation to calculate how many groups we can form of size 0, 1, and 2, since those are only group sizes we can construct with the remaining players. So, for example, if we have that \textbf{$\lvert$S$\rvert$ = 1} then we get that we can construct 2 different groups of this size: \emph{A} and \emph{B}. This means that we should apply the following scaling factors to each of our 4 marginal values: $1\Delta v_{C, \varnothing}, \frac{1}{2}\Delta v_{A C, A}, \frac{1}{2}\Delta v_{B C, B}, 1\Delta v_{A B C, A B}$. By adding this scaling factor we are \emph{averaging out} the effect that the rest of the team members have for each subset size. This means that we are able to capture the \emph{average} marginal contribution of \emph{C} when added to a team of size 0, 1, and 2 \emph{regardless} of the composition of these teams.
    \item Next, we focus on the remaining term, highlighted in Equation \ref{eq:shapley_highlight_3}.
    \begin{align}\label{eq:shapley_highlight_3}
    \phi_{i}(v)=\textcolor{orange}{\frac{1}{|N|}} \sum_{S \subseteq N \backslash\{i\}}\left(\begin{array}{c}
    |N|-1 \\
    |S|
    \end{array}\right)^{-1}(v(S \cup\{i\})-v(S))
    \end{align}
    So far we have averaged out the effects of the other team members for each subset size, allowing us to express how much \emph{C} contributes to groups of size 0, 1, and 2. The final piece of the puzzle is to \emph{average out} the effect of the group size as well, i.e. how much does \emph{C} contribute \emph{regardless} of the size of the team (number of explanatory features in the data set). For our scenario we do this by dividing with 3 since that is the number of different group sizes that we can consider. With this final step, we arrive at the point where can finally compute the Shapley value for \emph{C}. We have observed how much she marginally contributes to all different coalitions the team that can be formed. We have also averaged out the effects of both team member composition as well as team size which finally allows us to compute the Shapley value for \emph{C}:
    \begin{align}\label{eq:shapley_final}
        \phi_{C}(v) &= \frac{1}{3} \sum\left(1\Delta v_{C, \varnothing}, \frac{1}{2} \Delta v_{A C, A}, \frac{1}{2} \Delta v_{B C, B}, 1\Delta v_{ABC, AB}\right) \nonumber \\
        &= \frac{1}{3} \sum\left(1\times 10, \frac{1}{2} \times 6, \frac{1}{2} \times 2, 1\times 10\right) \nonumber \\
        &= 8
    \end{align}
    Comparing Equation \ref{eq:shapley_final} with the Shapley value for \emph{C} in the last row of Table \ref{table:marginal_conditional_contribution}, we see that they match exactly.
\end{itemize}

After calculating the Shapley values for the rest of the players, we can then determine their individual contributions to the excess 20 (thousand) dollars earned today, allowing us to fairly divide the bonus amongst all team members:
\begin{align}
    20 &=v(\{A, B, C\}) \nonumber \\
    &=\phi_{A}(v)+\phi_{B}(v)+\phi_{C}(v)
\end{align}
This is also evident from the last row of Table \ref{table:marginal_conditional_contribution}, where the Shapley values for \emph{A}, \emph{B}, and \emph{C} add to the total of 20.

\textbf{\underline{Properties}}:
As they apply to local explanations of predictions from machine learning models, Shapley values have some nice uniqueness guarantees \citep{shapley}. As described above, Shapley values are computed by introducing each feature, one at at time, into a conditional
expectation function of the model's output, $v_{x}(S) \approx E[v(x) | x_{S}]$, and attributing the change produced at each step to the feature that was introduced; then averaging this process over all possible feature orderings. Shapley values are provably the only possible method in the broad class of \emph{additive feature attribution methods} \cite{shap1} that simultaneously satisfy four important properties of \emph{fair} credit-attribution: \emph{efficiency}, \emph{symmetry}, \emph{missingness}, and \emph{additivity}.
\begin{enumerate}
    \item \emph{Efficiency/ Local Accuracy}: The sum of the Shapley values of all features (players) equals the value of the total coalition. That is, the value of $20$ in the bottom-rightmost corner in Table \ref{table:coalitions} is the same as the $20$ in the bottom-rightmost cell in Table \ref{table:marginal_conditional_contribution}.
    \item \emph{Missingness}: If a feature (player) never adds any marginal value regardless of the coalition, its payoff (Shapley value $\phi_{i}(v)$) is 0. That is, if: $v(S \cup \{i\}) = v(S) \forall $ subsets of features \emph{S}, then $\phi_{i}(f) = 0$.
    \item \emph{Symmetry}: All players have a fair chance to join the game. That’s why Table \ref{table:coalitions} lists all the permutations of the players. If two players always add the same marginal value to any subset to which they are added, their payoff portion should be the same. In other words, $\phi_{i} = \phi_{j}$ IIF feature \emph{i} and feature \emph{j} contribute equally to all possible coalitions.
    \item \emph{Additivity}: A function with combined outputs has as Shapley values the sum of the constituent ones. For any pair of games $v, w$, $\phi(v+w) = \phi(v)+\phi(w)$, where $(v+w)(S)=v(S)+w(S) \forall S$. This property allows us to combine Shapley values from $n$ different models, thus enabling interpretability for ensembles. 
\end{enumerate}

\iffalse
\textbf{\underline{Interaction Effects}}: An interaction effect occurs when the effect of one variable on the output depends on the value of another variable. The estimation of interaction effects requires having an accurate attribution of the individual effect of the variables involved. This makes Shapley values particularly useful for estimating these. For $i \neq j$:
\begin{align}
    \phi_{i, j}=\sum_{S \subseteq(X \backslash\{i, j\})} \frac{|S| !(|N|-|S|-2) !}{2(|N|-1) !} \delta_{i, j}(S), \nonumber \\
    \delta_{i, j}(S)=v(S \cup\{i, j\})-v(S \cup\{i\})-v(S \cup\{j\})+v(S)
\end{align}
\fi

\iffalse
%https://towardsdatascience.com/explain-your-model-with-the-shap-values-bc36aac4de3d
Lloyd Shapley came up with this solution concept for a cooperative game in 1953. Shapley wants to calculate the contribution of each player in a coalition game. Assume there are N players and S is a subset of the N players. Let v(S) be the total value of the S players. When player i join the S players, Player i’s marginal contribution is v(S union {i})- v(S). If we take the average of the contribution over the possible different permutations in which the coalition can be formed, we get the right contribution of player i:
\fi

\iffalse
\begin{align}
    \phi_{i}(v)=\frac{1}{|N|} \sum_{S \subseteq N \backslash\{i\}}\left(\begin{array}{c}
|N|-1 \\
|S|
\end{array}\right)^{-1}(v(S \cup\{i\})-v(S))
\end{align}
\fi

\end{document}